\documentclass[aps, prx,superscriptaddress,reprint,showpacs,longbibliography]{revtex4-2}
\usepackage{color}
\usepackage[normalem]{ulem}  
\usepackage{xcolor}          
\usepackage{graphicx,textcomp,amssymb,amsmath,dcolumn}
\usepackage[colorlinks=false, pdfborder={0 0 0}]{hyperref}
\usepackage{siunitx}
\usepackage{comment}
\usepackage{physics}
\usepackage{siunitx}
\usepackage{adjustbox}

\usepackage{romannum}

\begin{document}

\title{Tunable spin--orbit splitting in bilayer graphene/WSe\textsubscript{2} quantum devices} 

\author{Jonas D. Gerber}
\email{gerberjo@phys.ethz.ch}
\affiliation{Solid State Physics Laboratory, ETH Z\"urich, 8093 Z\"urich, Switzerland}
\author{Efe Ersoy}%
\affiliation{Solid State Physics Laboratory, ETH Z\"urich, 8093 Z\"urich, Switzerland}
\author{Michele Masseroni}%
\affiliation{Solid State Physics Laboratory, ETH Z\"urich, 8093 Z\"urich, Switzerland}
\author{Markus Niese}
\affiliation{Solid State Physics Laboratory, ETH Z\"urich, 8093 Z\"urich, Switzerland}
\author{Michael Laumer}
    \affiliation{Institut f\"ur Theoretische Physik, Universit\"at Regensburg, D-93040 Regensburg, Germany}
 \author{Artem O. Denisov}
 \affiliation{Solid State Physics Laboratory, ETH Z\"urich, 8093 Z\"urich, Switzerland}
  \author{Hadrien Duprez}
   \affiliation{Solid State Physics Laboratory, ETH Z\"urich, 8093 Z\"urich, Switzerland}
  \author{Wister Wei Huang}
     \affiliation{Solid State Physics Laboratory, ETH Z\"urich, 8093 Z\"urich, Switzerland}
   \author{Christoph Adam}
     \affiliation{Solid State Physics Laboratory, ETH Z\"urich, 8093 Z\"urich, Switzerland}
   \author{Lara Ostertag}
     \affiliation{Solid State Physics Laboratory, ETH Z\"urich, 8093 Z\"urich, Switzerland}
\author{Chuyao Tong}
     \affiliation{Solid State Physics Laboratory, ETH Z\"urich, 8093 Z\"urich, Switzerland}
\author{Takashi Taniguchi}
	\affiliation{Research Center for Materials Nanoarchitectonics, National Institute for Materials Science,  1-1 Namiki, Tsukuba 305-0044, Japan}
\author{Kenji Watanabe}
	\affiliation{Research Center for Electronic and Optical Materials, National Institute for Materials Science, 1-1 Namiki, Tsukuba 305-0044, Japan}

\author{Vladimir I. Fal’ko}
    \affiliation{National Graphene Institute, University of Manchester,  Manchester M13 9PL, UK}
\author{Thomas Ihn}
    \affiliation{Solid State Physics Laboratory, ETH Z\"urich, 8093 Z\"urich, Switzerland}
\author{Klaus Ensslin}
    \affiliation{Solid State Physics Laboratory, ETH Z\"urich, 8093 Z\"urich, Switzerland}
\author{Angelika Knothe}
    \affiliation{Institut f\"ur Theoretische Physik, Universit\"at Regensburg, D-93040 Regensburg, Germany}
\date{\today}

\begin{abstract}

Abstract: Bilayer graphene (BLG)–based quantum devices represent a promising platform for emerging technologies such as quantum computing and spintronics. However, their intrinsically weak spin–orbit coupling (SOC) 
complicates spin and valley manipulation. Integrating BLG with transition metal dichalcogenides (TMDs) enhances SOC via proximity effects. While this enhancement has been demonstrated in 2D-layered structures,  1D and 0D-nanostructures in BLG/TMD remain unrealized, with open questions regarding SOC strength, and tunability. Here, we investigate quantum point contacts and quantum dots in two BLG/WSe\textsubscript{2} heterostructures with different stacking orders. Across multiple devices, we reproducibly demonstrate spin--orbit splitting up to \SI{1.5}{meV}---more than one order of magnitude higher than in pristine BLG. Furthermore, we show that the induced SOC can be tuned in situ from its maximum value to near-complete suppression via the perpendicular electric field.
This enhancement and in situ tunability establish SOC as a control mechanism for dynamic spin and valley manipulation.


\end{abstract}
\maketitle

With its tunable band gap \cite{Oostinga2008}, high carrier mobility \cite{Wang2013}, and a nuclear-spin-free environment \cite{Trauzettel2007}, bilayer graphene (BLG) is a promising material for advanced quantum devices in spintronics, valleytronics, and quantum computation. Recent experiments have demonstrated long spin and valley relaxation times in BLG-based quantum dots \cite{Denisov2025,Garreis2024, gachter_single_2022, Banszerus2022, banszerus2024phononlimited}, underscoring its potential for spin qubits. For these applications,  strong spin--orbit coupling (SOC) can be advantageous, enabling efficient spin control via electric dipole spin resonance \cite{Hendrickx2020} and playing a key role in spin- and valleytronics, including spin--orbit valves \cite{GmitraFabian2017} and spin--valley filters \cite{Rycerz2007, Sui2015}. However, in pristine BLG, SOC is intrinsically weak, with a spin--orbit splitting ($\Delta_\mathrm{SO}$) of only \qtyrange[range-phrase=~--~]{40}{80}{\micro eV} \cite{SichauSOCMicrowave, Banszerus_SOC_QPC,Banszerus2021, Kurzmann2021_Kondo,Duprez2024}, arising from Kane--Mele \cite{PhysRevLett.KanMeleSOC} and Bychkov--Rashba mechanisms \cite{Konschuh2007}. Furthermore, its limited in situ tunability  \cite{Banszerus2021,Banszerus_SOC_QPC} poses additional challenges.

A promising approach to enhance SOC in BLG-based quantum devices is proximity coupling to TMDs with strong intrinsic SOC. This combination has been shown to significantly increase SOC in two-dimensional bulk BLG, resulting in spin--orbit splittings of up to several meV \cite{Gmitra2016, GmitraFabian2017, David2019, Li2019,Avsar2014,Wang2015,Yang2016,Volkl2017,Wakamura2018,Zihlmann2018,Island2019,Wang2016, Masseroni2024, zhang2024, Seiler2024, Tiwari2022, Holleis2024, Fulop2021, Rao2023}. 
Experimental estimates of SOC strength in BLG/TMD heterostructures were first obtained using traditional transport techniques, such as weak anti-localization measurements \cite{Wang2015,Yang2016,Volkl2017,Wakamura2018,Zihlmann2018, Fulop2021}  and  Shubnikov-de-Haas oscillations \cite{Wang2016, Masseroni2024, zhang2024, Tiwari2022, Holleis2024}.

Unlike transport techniques, extracting SOC from confined quantum devices such as quantum dots (QDs) and quantum point contacts (QPCs) is more direct, model-independent, and less affected by disorder, allowing precise determination of $\Delta_\mathrm{SO}$. These devices probe energy gaps near the band edge, a regime typically inaccessible in 2D transport measurements. QPCs serve as a robust tool for detecting degeneracy lifting, while QDs enable precise spin–orbit gap measurements. Together, they provide a comprehensive view of spin--orbit effects, with their sensitivity to layer polarization enabling layer-resolved probing.

This work presents high-precision $\Delta_\mathrm{SO}$ measurements in two BLG/WSe\textsubscript{2} heterostructures using QPCs and QDs. We identify spin--valley--Zeeman SOC (also referred to as Ising SOC in the literature) as the dominant mechanism causing this splitting. Additionally, we demonstrate quantized conductance and Coulomb blockade in BLG/TMD systems. Our experimental data closely align with single-particle calculations, indicating a strong understanding of electronic states in confined BLG/TMD structures. By systematically varying the displacement field in devices with different stacking order types, we achieve in situ tuning of spin--orbit splitting over more than an order of magnitude. This tunability is essential for developing adaptive and flexible quantum devices for applications in quantum computing, spintronics, and valleytronics.
\\


\begin{figure}
	\includegraphics{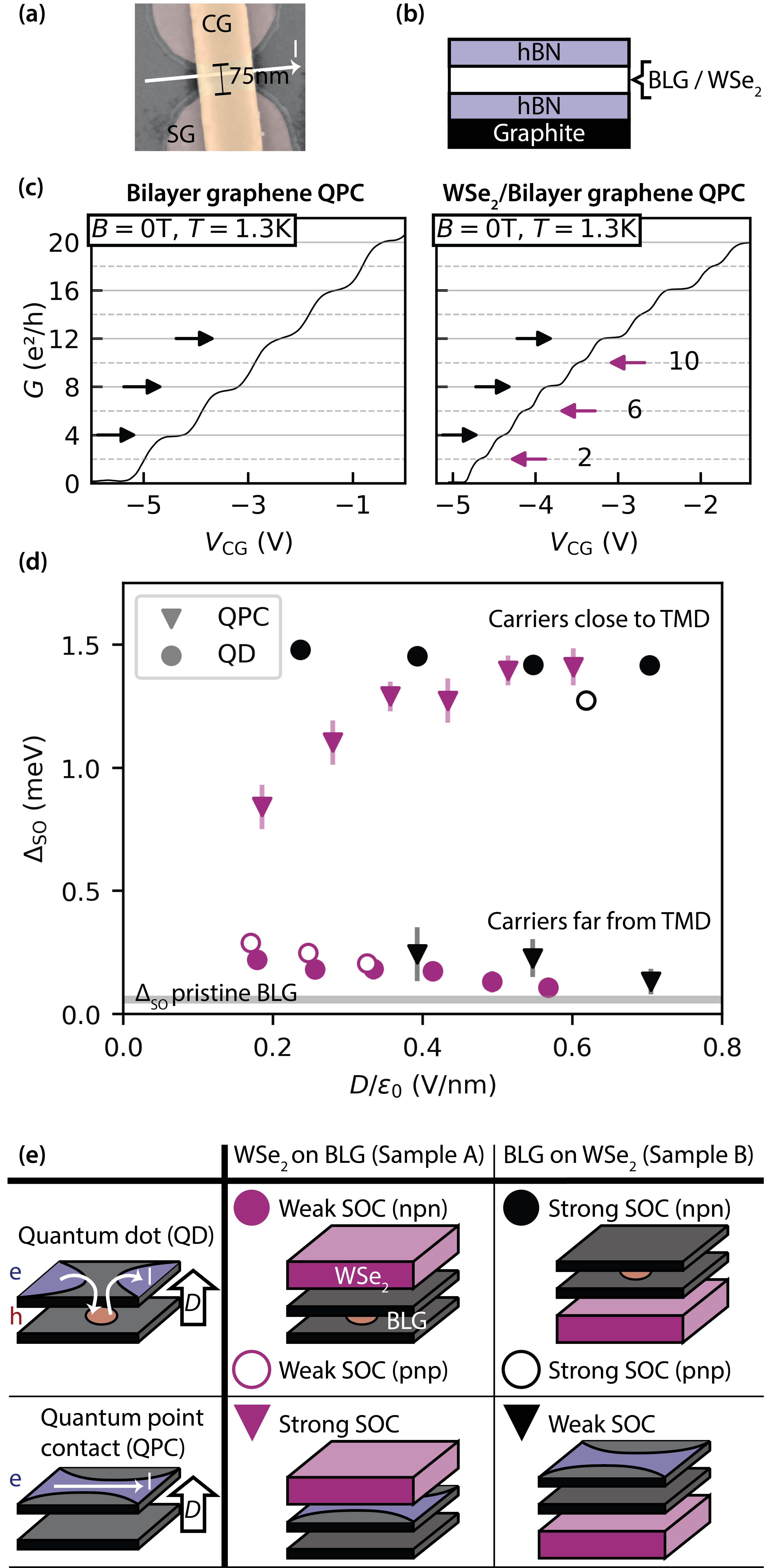}
	\caption{(a) False-color SEM image showing the top view of the measured devices. (b) Schematic cross-sectional view of the BLG/TMD heterostructures. (c) Quantized conductance of QPCs in pristine BLG and BLG/WSe\textsubscript{2}, corrected for parasitic resistances (see SI B2). Black arrows indicate fourfold-degenerate plateaus, while the magenta arrows highlight SOC-induced degeneracy lifting. (d) Extracted spin--orbit splitting of two heterostructures (magenta/black) as a function of displacement field applied beneath the split gates. The gray band marks the typical range of $\Delta_\mathrm{SO}$ observed in pristine BLG quantum devices. The data exhibit excellent agreement with expectations based on layer polarization, as summarized in (e). Open symbols correspond to measurements taken with reversed displacement field polarity relative to the convention defined in (e).}
   
	\label{Fig1}
\end{figure}

Using the gate structure shown in the false-color SEM top view (Fig.~\ref{Fig1}a), we electrostatically define QDs and QPCs. The two split gates (SG), separated by a lithographic width of \SI{75}{nm}, work in combination with the graphite back gate to open a band gap and create a 1D confinement in bilayer graphene. A channel gate (CG), positioned above the SGs and insulated by a layer of Al\textsubscript{2}O\textsubscript{3}, controls the local electrostatic potential within the channel. By tuning the channel gate voltage ($V_\mathrm{CG}$), we can define either QPCs or QDs. Notably, QDs in this design rely on pn-junction tunneling barriers, allowing the formation of only p-type dots with n-type leads and vice versa \cite{Eich2018, Banszerus2018}. A cross-sectional view of the heterostructure is shown in Fig.~\ref{Fig1}b. In this study, we investigate quantum devices in two heterostructures: WSe\textsubscript{2}-on-BLG (sample A, with a WSe\textsubscript{2}/BLG twist angle of \SI{0 (2)}{\degree}) and BLG-on-WSe\textsubscript{2} (sample B, \SI{4 (2)}{\degree} twist angle). Details on sample fabrication and twist angle determination are provided in the Supporting Information (SI) A.

 As the first compelling evidence of enhanced proximity-induced SOC, Fig.~\ref{Fig1}c compares the quantized conductance of a QPC in a pristine BLG with that in a WSe\textsubscript{2}/BLG heterostructure. Both measurements were performed at a temperature of $T=\SI{1.3}{K}$ and zero magnetic field, with corrections applied for parasitic resistances as detailed in SI B2. The pristine BLG QPC shows quantized conductance steps in units of $4e^2/h$, reflecting the fourfold degeneracy of its energy levels. This behavior arises because the intrinsic $\Delta_\mathrm{SO}$ is smaller than the thermal energy $k_\mathrm{B}T$, consistent with previous studies \cite{Kraft2018, Overweg2018Nano, Banszerus_SOC_QPC, Lee2020, Overweg2018PRL}. In contrast, the WSe\textsubscript{2}/BLG heterostructure exhibits conductance steps in units of $2e^2/h$ (magenta arrows in Fig.~\ref{Fig1}c), indicating that the fourfold degeneracy has been lifted into two pairs (2+2), with a splitting energy significantly exceeding $ k_\mathrm{B}T$.

Figure~\ref{Fig1}d presents the central result of this work: a summary of the experimentally extracted spin--orbit splittings for the first modes in all measured QPC devices (triangles) and the first single-particle levels in all QD devices (circles). All measurements were conducted at a base temperature of \SI{10}{mK}, with the extraction procedure detailed in subsequent chapters. Figure~\ref{Fig1}e provides a legend to Fig.~\ref{Fig1}d, indicating the stacking order of the two heterostructures: sample A (purple symbols) and sample B (black symbols).

We begin by discussing measurements taken at a displacement field that yields n-type QD leads and n-type QPC channels, corresponding to carrier polarization in the top layer of BLG (indicated in blue in Fig.~\ref{Fig1}e). The layer polarization of carriers \cite{Fogler2010} plays a critical role: in sample A, where carriers reside in the layer adjacent to the WSe\textsubscript{2}, the QPC exhibits a large $\Delta_\mathrm{SO}$ (filled purple triangles in Fig.~\ref{Fig1}d). In contrast, in sample B, where the carriers occupy the remote layer, $\Delta_\mathrm{SO}$ is significantly reduced (filled black triangles). A complementary trend is observed in the QD measurements. In sample A, the p-type QD forms in the layer remote from the WSe\textsubscript{2} layer, resulting in a small $\Delta_\mathrm{SO}$ (filled purple circles). Conversely, in sample B, the QD forms adjacent to the WSe\textsubscript{2}, leading to a significantly larger $\Delta_\mathrm{SO}$ (filled black circles).  

Reversing the sign of the displacement field results in p-type leads, with holes once again polarized in the top layer. Consequently, for an n-type quantum dot, carriers remain polarized in the bottom layer in both samples A and B. As a result, the QD in sample A continues to exhibit weak SOC (open purple circles in Fig.~\ref{Fig1}d), while the QD in sample B maintains strong SOC (open black circles). The layer polarization can be continuously tuned in all devices, enabling modulation of $\Delta_\mathrm{SO}$ as a function of the displacement field applied beneath the split-gates, as shown in Fig.~\ref{Fig1}d. As the displacement field decreases, the layer polarization is reduced, leading to a more symmetric distribution of the electronic wavefunction across both layers. This results in a convergence of SOC values between the strong and weak SOC regimes. This trend is clearly observed in both QDs and QPCs in sample A (magenta symbols). For sample B, data at low-displacement fields are absent due to the loss of quantum confinement.
Minor \(\Delta_{\mathrm{SO}}\) differences between the datasets arise from strain or twist-angle variations between samples~\cite{David2019, Li2019, Naimer2021, Zollner2023, zhang2024}. In addition, slight sample-to-sample variations in the effective displacement field exist, as discussed in SI~B5.

Owing to the similarly low twist angles in samples A and B, the measured SOC values are consistent with recent transport experiments in BLG/WSe\textsubscript{2} heterostructures \cite{zhang2024}. At large displacement fields, the enhanced spin--orbit splitting $\Delta_\mathrm{SO}$ saturates between \num{1.3} and \SI{1.5}{meV}. By switching the layer polarization in situ, \( \Delta_\mathrm{SO} \) can be tuned down to $ \num{100} - \SI{300}{\micro\electronvolt}$. Both the maximum and minimum values significantly exceed the typical \( \Delta_\mathrm{SO} \) in pristine BLG quantum devices---indicated by the gray band in Fig.~\ref{Fig1}d---demonstrating the additive effect of proximity-induced SOC on top of the intrinsic Kane--Mele SOC in BLG. Moreover, the observed in situ tunability of \( \Delta_\mathrm{SO} \) via layer polarization aligns with the expected short-range nature of the orbital overlap mechanism responsible for proximity-induced SOC~\cite{Gmitra2016,Fulop2021}.
While WSe\textsubscript{2} is commonly used to induce enhanced SOC in BLG, other TMDs can produce a comparable increase in $\Delta_\mathrm{SO}$, as demonstrated by data from a MoS\textsubscript{2}/BLG quantum device presented in SI E2.\\


\begin{figure}
	\includegraphics{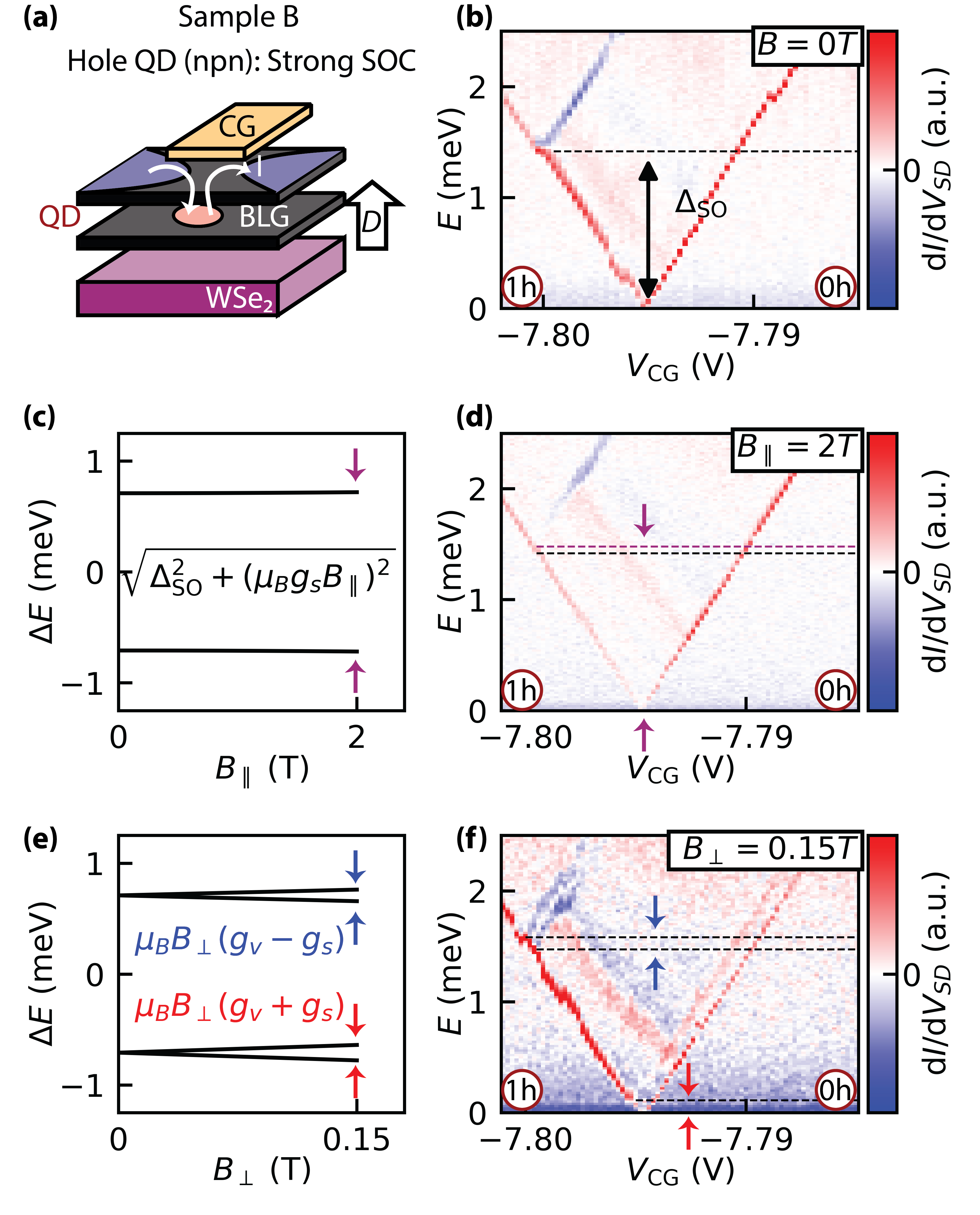}
	\caption{(a) Schematic of the measured npn-type QD in the BLG/WSe\textsubscript{2} sample. (b)  
    Finite-bias spectroscopy at the 0h $\rightarrow$ 1h charge transition at zero magnetic field. The intersection of the excited state line with the Coulomb diamond edges (black dashed line) yields a spin--orbit gap of $\Delta_\mathrm{SO}=\SI{1.42(2)}{meV}$. (c) Calculated evolution of quantum states in an in-plane magnetic field, using the extracted $\Delta_\mathrm{SO}$. (d) Corresponding finite-bias measurement at $B_\parallel=\SI{2}{T}$. The excited-state energy (magenta dashed line) shows only a minor shift relative to the zero-field gap (black dashed line). (e)  Calculated energy evolution in a perpendicular magnetic field, using $\Delta_\mathrm{SO}$ and $g_v=14$, predicting the characteristic splitting of the two Kramers pairs. (f) Experimental finite-bias measurement at $B_\perp=\SI{0.15}{T}$, in agreement with the predicted splitting.    }
	\label{Fig2}
\end{figure}

This section focuses on characterizing proximity-induced SOC in quantum dot devices.
We begin with bias spectroscopy measurements on a npn-QD in sample B (BLG/WSe\textsubscript{2}), where strong SOC is expected due to the proximity of the QD to the TMD layer (Fig.~\ref{Fig2}a). The quantum dot potential is tuned via the channel gate voltage $V_\mathrm{CG}$.
Fintie-bias spectroscopy at the 0h $\rightarrow$ 1h transition (Fig.~\ref{Fig2}b) reveals a clear spin--orbit splitting between the ground- and first excited state, with $\Delta_\mathrm{SO}=\SI{1.42(2)}{meV}$. This value corresponds to the black circle in Fig.~\ref{Fig1}d at a displacement field of $D=\SI{0.7}{\volt/nm}$.

To verify that the observed excited state corresponds to the spin--orbit split-off Kramers pair, we investigate its evolution under an in-plane magnetic field ($B_\parallel$). In this confined QD regime, Rashba-type spin--orbit coupling is expected to be strongly suppressed compared to the 2D case \cite{Kormanyos2014}. Under the assumption of spin--valley--Zeeman SOC  (see SI C for details) as the dominant mechanism, the expected energy splitting follows $\Delta E = \sqrt{\Delta_\mathrm{SO}^2 + (g_\mathrm{S}\mu_\mathrm{B}B_\parallel)^2}$, as illustrated in Fig.~\ref{Fig2}c.
At \( B_\parallel = \SI{2}{\tesla} \), this model predicts a modest increase of \SI{19}{\micro eV} relative to the zero-field value. In the experiment (Fig.~\ref{Fig2}d), the measured splitting at $B_\parallel=\SI{2}{T}$ (magenta dashed line) shows a small increase of \SI{60(30)}{\micro eV} compared to the zero-field spin--orbit gap (black dashed line), consistent with expectations. The discrepancy is attributed to a slight perpendicular field component due to an unavoidable sample tilt. Importantly, the observed field dependence is far weaker than that expected from a linear spin--Zeeman splitting with $g_s=2$, which would result in a shift of \SI{230}{\micro eV}. 

Additional confirmation of the spin--orbit nature is obtained by applying a perpendicular magnetic field ($B_\perp$). Assuming predominantly out-of-plane SOC and using the previously extracted values $\Delta_\mathrm{SO}=\SI{1.42(2)}{meV}$ and $g_\mathrm{v} \approx 14$, we predict the characteristic magnetic-field evolution of the Kramers pairs shown in Fig.~\ref{Fig2}e. This prediction is consistent with finite-bias spectroscopy data presented in Fig.~\ref{Fig2}f.  In this device, increasing $B_\perp$ significantly suppressed the transport current, which limits the precision of the extracted $g_v$.
Nonetheless, the magnetic field dependence observed across measurements aligns well with a single-particle model incorporating proximity-induced SOC, supporting the identification of the excited state as a spin--orbit split-off level. The full evolution of the QD states for this sample is provided in SI E5.\\


\begin{figure}
	\includegraphics{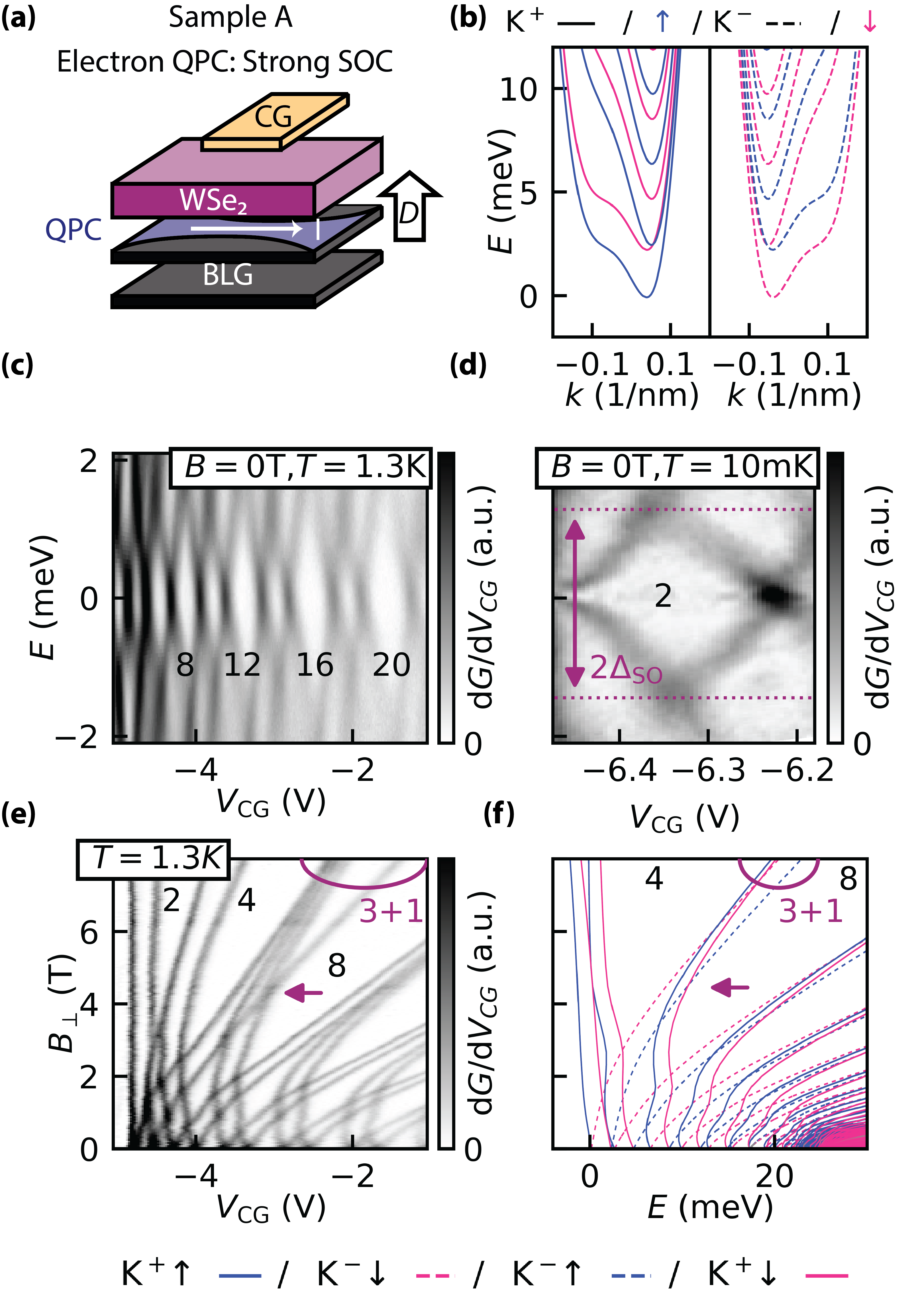}
	\caption{(a)  Schematic of the electron QPC in the WSe\textsubscript{2}/BLG heterostructure. (b)  Single-particle band structure of a 30\,nm wide QPC on BLG/WSe\textsubscript{2}, calculated using the DFT SOC parameters from Ref.~\cite{Seiler2024}. (c) Bias spectroscopy of the QPC at $B=0$ and $T=\SI{1.3}{K}$. The alternating white diamond-shaped regions correspond to conductance plateaus, with values indicated in units of $e^2/h$. (d) High-resolution
    bias spectroscopy of the first $G=2e^2/h$ plateau, measured at \SI{10}{mK}. The diamond height directly yields $2\Delta_\mathrm{SO}$.   (e) Transconductance $dG/dV_\mathrm{CG}$ as a function of $V_\mathrm{CG}$ and $B_\perp$, measured at T\,=\,\SI{1.3}{K}. Black labels indicate corresponding conductance quantum numbers. (f) Calculated single-particle magnetic subband evolution for a \SI{30}{nm}-wide QPC in BLG/WSe\textsubscript{2}, showing excellent agreement with the experimental data across both low and high magnetic fields. The characteristic ``3+1" feature, resulting from spin--valley--Zeeman coupling, is highlighted with the magenta circle. The magenta arrow marks the crossing between the \( 1\mathrm{K}^-\uparrow \)--\( 3\mathrm{K}^+\downarrow \) states, in (e) and (f). This ``3+1" feature is discussed more in detail in the Supporting Information E6.}
	\label{Fig3}
\end{figure}

We now turn to the analysis of the QPC in sample A (BLG/WSe\textsubscript{2} heterostructure, Fig.~\ref{Fig3}a) to extract both the magnitude and nature of the SOC. Figure~\ref{Fig3}b presents the calculated single-particle band structure of a 30\,nm wide QPC in BLG/WSe\textsubscript{2}, based on density functional theory (DFT) SOC parameters from Ref.~\cite{Seiler2024}. The lowest-energy subband, formed by the $\mathrm{K}^+\uparrow$/$\mathrm{K}^-\downarrow$ states, is separated by the spin–orbit gap $\Delta_\mathrm{SO}$ from $\mathrm{K}^+\downarrow$/$\mathrm{K}^-\uparrow$ states.

This gap is experimentally determined using bias spectroscopy (Fig.~\ref{Fig3}c). The black numbers indicate the conductance values in units of $e^2/h$. The corresponding zero-bias conductance trace $G(V_\mathrm{CG})$ is shown in the right panel of Fig.~\ref{Fig1}c. To enhance energy resolution, we remeasured the first spin--orbit split plateau at $2e^2/h$ in a dilution refrigerator with a base temperature of \SI{10}{mK} (Fig.~\ref{Fig3}d). The height of the corresponding diamond directly yields $2\Delta_\mathrm{SO}$, from which we extract $\Delta_\mathrm{SO}=\SI{1.37(8)}{meV}$. This data point corresponds to the magenta triangle in Fig.~\ref{Fig1}d at $D=\SI{0.6}{\volt/nm}$. SI B4 describes the procedure used to determine the corresponding energy scale.


To characterize the quantum states, we analyze the magnetic depopulation of magnetoelectric subbands by measuring $dG/dV_\mathrm{CG}$ as a function of $B_\perp$ (Fig.~\ref{Fig3}e). The conductance steps appear as dark lines, each of which splits into two at low $B_\perp$, consistent with the valley--Zeeman effect. This pronounced splitting confirms that the degenerate states at $B=0$ originate from opposite valley flavors. As the field increases, lines corresponding to states with the same valley and subband index evolve in parallel but remain offset due to spin--orbit splitting. While the spin--Zeeman effect is unresolved at low magnetic fields, opposite spin flavors lead to the formation of a ``3+1" feature at high magnetic fields, as discussed in SI E6. Unlike in pristine BLG (see SI E3), the enhanced spin–orbit coupling combined with the valley--Zeeman effect enables all four states to be individually resolved.

To further elucidate the interplay between subband spacing and SOC, we compare the measured spectrum (Fig.~\ref{Fig3}e) and single-particle calculations (Fig.~\ref{Fig3}f). The modeling of the electrostatically defined QPC in bilayer graphene follows previous approaches for pristine BLG \cite{Knothe2018, Overweg2018PRL, Lee2020}, with additional details provided in SI C. Using the Hamiltonian outlined therein, we compute the QPC subband spectrum shown in Fig.~\ref{Fig3}f. The calculations use the DFT SOC parameters for BLG/WSe\textsubscript{2} from Ref. \cite{Seiler2024} and a channel width of $L=\SI{30}{nm}$, estimated from the measured level spacing in Fig.~\ref{Fig3}e. The discrepancy between this value and the lithographically defined width ($L=\SI{75}{nm}$) arises from stray electric fields near the split gates, which effectively narrow the electronic channel. The agreement between the measured subband evolution and the theoretical spectrum is excellent across both low and high magnetic fields. Importantly, this agreement depends sensitively on the choice of SOC parameters in the model. We do not reproduce the measured magnetic field pattern for parameters strongly deviating from the DFT parameters of Ref.~\cite{Seiler2024}. In SI D, we explore the influence of various SOC parameters and show that Rashba-type SOC has minimal impact on the magnetic field dependence of the subbands. These findings confirm that the observed \( \Delta_\mathrm{SO} \) arises predominantly from proximity-induced spin--valley--Zeeman SOC.
\\


\begin{figure}
        \includegraphics{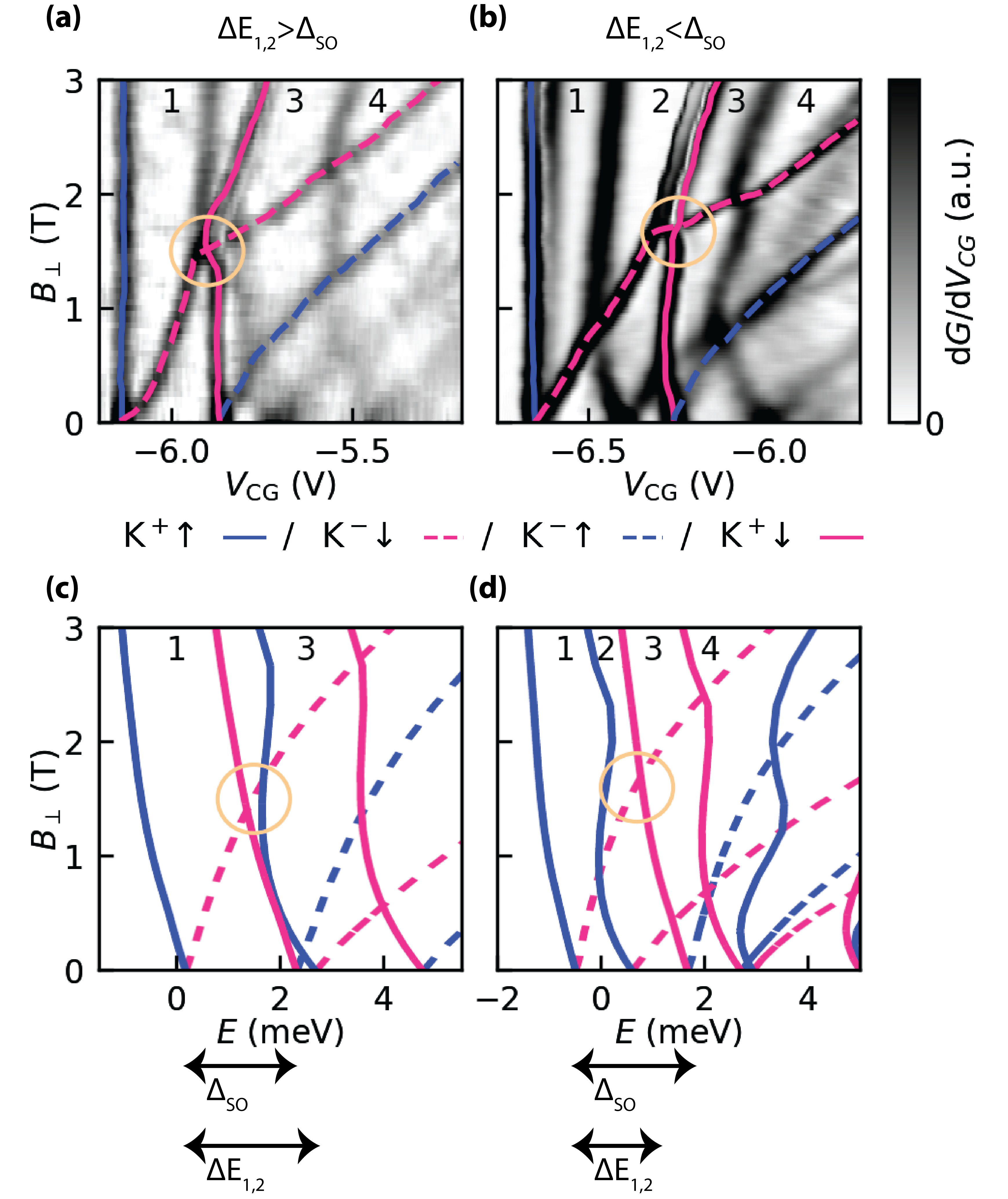}
	\caption{(a), (b) Experimentally measured transconductance versus $B_\perp$ of two QPCs in a WSe\textsubscript{2}/BLG heterostructure (sample A) at \SI{10}{mK}. The left QPC (a) is the same device shown in Fig.~\ref{Fig3}. The right QPC (b) is located on the same heterostructure stack but exhibits a different subband spectrum, likely due to differences in applied gate voltages and possible device-to-device variations.  (c), (d) Single-particle calculations of QPC energy levels in a BLG/WSe\textsubscript{2} heterostructure using DFT SOC parameters from \cite{Seiler2024}, with a QPC width of $L=\SI{30}{nm}$ (left) and $L=\SI{50}{nm}$ (right). The measured subband evolution, with the states of the first subband highlighted, shows good agreement with theoretical calculations.  A notable deviation from the theoretical prediction is marked by an orange circle, indicating a strong bending of the $1\mathrm{K}^-\downarrow$ mode when crossing the $1\mathrm{K}^+\downarrow$ mode.   
}
	\label{Fig4}
\end{figure}

In this section, we examine the magnetic subband evolution in greater detail and present evidence for an exchange-enhanced \( g \)-factor, demonstrating that this platform is well suited for exploring many-body physics. We remeasure the QPC in sample A (WSe\textsubscript{2}/BLG, Fig.~\ref{Fig4}a)  at a temperature of \SI{10}{mK} and compare its magneto-transconductance to that of a second QPC fabricated on the same sample (Fig.~\ref{Fig4}b). In both plots, the evolution of the first subband states is color-highlighted, with the subband assignment confirmed by high-field measurements (SI E6). Despite identical lithographic widths, the subband spectra in Figs.~\ref{Fig4}a,b differ significantly. In particular, the subband spacing in Fig.~\ref{Fig4}a is markedly larger than in Fig.~\ref{Fig4}b, arising from a higher applied split-gate voltage in Fig.~\ref{Fig4}a and possible device-to-device variations of the electrostatic landscape.

To interpret the experimental observations, we compare the data to single-particle calculations for QPCs with different channel widths, adjusted to reproduce the measured subband spacings while keeping the SOC parameters fixed. In Fig.~\ref{Fig4}c, for a channel width of \SI{30}{nm}, the calculated subband spacing $\Delta E_{1,2}$ exceeds the spin--orbit-induced splitting $\Delta_\mathrm{SO}$ and vice versa for the \SI{50}{nm} wide QPC (Fig.~\ref{Fig4}d).

A prominent deviation between the experimental data and the single-particle calculations is the pronounced nonlinearity observed in the magnetic field evolution of the subbands.  In both measured QPCs the \(1\mathrm{K}^-\downarrow\) subband exhibits a pronounced bending after crossing the \(1\mathrm{K}^+\downarrow\) state, as highlighted by the orange circle. We attribute this feature to an enhancement of the effective g-factor, $g^\ast$, arising from exchange interactions. Enhancements of \(g^\ast\), have been reported in various low-dimensional systems~\cite{Fang1968, Pallecchi2002, Gall2022} and are commonly attributed to spin-based exchange interactions~\cite{Janak1969, Ando1974}. In our case, however, the crossing between the \(1\mathrm{K}^-\downarrow\) and \(1\mathrm{K}^+\downarrow\) states involves a change in valley occupancy. This suggests that interaction effects in our system are caused by more complex spin--valley interactions.

While a complete theoretical description of the effect is beyond the scope of this work, our results highlight the interplay of spin, valley, and subband index in shaping the magnetic response of QPCs in graphene-based heterostructures. Owing to the resolution of individual subbands, such behavior becomes experimentally accessible in BLG/TMD systems, establishing them as a promising platform for future investigations of correlated electronic phenomena.\\


We have presented direct energy spectroscopy measurements of a tunable spin--valley--Zeeman spin--orbit coupling in QPCs and QDs based on BLG/WSe\textsubscript{2} heterostructures.
Our results demonstrate that the SOC in BLG can be enhanced from below \SI{80}{\micro eV} to $\SI{1.5}{meV}$ via proximity to a TMD layer.
This strong SOC regime, reached at large displacement fields, remains robust in the range of \SI{1.3}{meV} to \SI{1.5}{meV} across device types (QPCs and QDs) and carrier polarities (electrons and holes), establishing BLG/TMD heterostructures as a versatile platform for engineering strong SOC.
We further show that the spin--orbit gap $\Delta_\mathrm{SO}$ is highly tunable---by over an order of magnitude---via the displacement field, which controls the layer polarization in BLG.
This tunability opens the door to QD architectures with multiple gates, such as those demonstrated in Refs.~\cite{Eich2018, Banszerus2018}, that can switch between strong-SOC (nn'n) and weak-SOC (npn) QD regimes in situ, without reversing the displacement field. The ability to modulate SOC  within a single device allows direct comparison of quantum phenomena under distinct SOC conditions, effectively adding a new dimension of control. 
While our measurements primarily reveal a spin--valley--Zeeman SOC, the detection and fine control of Rashba SOC in these confined hybrid systems remain an open question for future investigation.\\

Our experimental results show excellent agreement with single-particle modeling across a broad range of parameters. Given the full lifting of degeneracies in the subband spectra, remaining deviations can be attributed to many-body interactions, indicating that this platform is well suited for exploring correlated phenomena.\\

The combination of high material quality, gate-tunable confinement, and proximity-induced SOC makes BLG/WSe\textsubscript{2} heterostructures a compelling candidate for next-generation spintronic, valleytronic, and quantum information devices. In particular, the ability to switch between strong and weak SOC regimes in situ provides a powerful lever for tailoring and optimizing device functionality.\\

\textbf{\Large Supporting Information}
Supporting Information. Details on device
fabrication, measurement setup, data analysis, model de-
tails, and additional measurement data.\\

\textbf{\Large Acknowledgments}
We thank Peter Märki, Thomas Bähler, as well as the staff of the ETH-cleanroom FIRST for their technical support. We acknowledge support from the European Graphene Flagship Core3 Project, Swiss National Science Foundation via NCCR Quantum Science and Technology, and H2020 European Research Council (ERC) Synergy Grant under Grant Agreement 95154.
K.W. and T.T. acknowledge support from the JSPS KAKENHI (Grant Numbers 21H05233 and 23H02052), the CREST (JPMJCR24A5), JST and World Premier International Research Center Initiative (WPI), MEXT, Japan. A.K. acknowledges support from the Deutsche Forschungsgemeinschaft (DFG, German Research Foundation) within DFG Individual grants KN 1383/4,  KN 1383/7, and SFB 1277 (Project-ID 314695032), as well as the Academic Research Sabbatical program of the University of Regensburg. We thank Christoph Schulz, Jaroslav Fabian, and Klaus Zollner for helpful discussions.\\


\clearpage
\newpage
\textbf{\Large References}

%



\clearpage
\newpage
\twocolumngrid 

\onecolumngrid

\begin{center}
    \LARGE \textbf{Supporting Information}\\[0.5cm] 
    
    \large{\textbf{Tunable spin--orbit splitting in bilayer graphene/WSe\textsubscript{2} quantum devices}}\\[0.5cm]

    \normalsize Jonas D. Gerber\textsuperscript{1,*}, Efe Ersoy\textsuperscript{1}, Michele Masseroni\textsuperscript{1}, Markus Niese\textsuperscript{1}, Michael Laumer\textsuperscript{2}, Artem O. Denisov\textsuperscript{1},\\ Hadrien Duprez\textsuperscript{1}, Wister Wei Huang\textsuperscript{1}, Christoph Adam\textsuperscript{1}, Lara Ostertag\textsuperscript{1}, Chuyao Tong\textsuperscript{1},\\ Takashi Taniguchi\textsuperscript{3}, Kenji Watanabe\textsuperscript{4}, Vladimir I. Fal’ko\textsuperscript{5}, Thomas Ihn\textsuperscript{1}, Klaus Ensslin\textsuperscript{1}, Angelika Knothe\textsuperscript{2}\\[0.2cm]
    \small
    \textit{\textsuperscript{1}Solid State Physics Laboratory, ETH Zürich, 8093 Zürich, Switzerland\\
    \textsuperscript{2}Institut für Theoretische Physik, Universität Regensburg, D-93040 Regensburg, Germany\\
    \textsuperscript{3}Research Center for Materials Nanoarchitectonics, National Institute for Materials Science,  1-1 Namiki, Tsukuba 305-0044, Japan\\
    \textsuperscript{4}Research Center for Electronic and Optical Materials, National Institute for Materials Science, 1-1 Namiki, Tsukuba 305-0044, Japan\\
    \textsuperscript{5}National Graphene Institute, University of Manchester, Manchester M13 9PL, UK}\\
    *Email: gerberjo@phys.ethz.ch
\end{center}

\vspace{1cm}

\twocolumngrid

\appendix
\renewcommand\thefigure{\thesection.\arabic{figure}}  
\renewcommand\thetable{\thesection.\arabic{table}} 

\setcounter{figure}{0}    
\setcounter{table}{0}


\section{Device fabrication}\label{app:DevFab}

The 2D materials used in this study were prepared by mechanical exfoliation of bulk crystals. For the WSe\textsubscript{2} and MoS\textsubscript{2} layers, we used material from \textit{HQ graphene}. The corresponding heterostructures were assembled using a polymer-based dry transfer technique. The stacks presented in the main text have the following layer thicknesses.

\begin{table}[h]
    \centering
    \renewcommand{\arraystretch}{1.2}
    \begin{tabular}{c|c|c}
        \hline
        \textbf{Layer Nr}  & \textbf{WSe\textsubscript{2}-on-BLG (A)} & \textbf{BLG-on-WSe\textsubscript{2} (B)} \\
        \hline
        1 & hBN: \SI{21}{nm} & hBN: \SI{23}{nm} \\
        2 & WSe\textsubscript{2}: 5 layers & BLG \\
        3 & BLG  & WSe\textsubscript{2}: 3 layers \\
        4 & hBN: \SI{85}{nm} & hBN: \SI{42}{nm} \\
        5 & Graphite: \SI{3}{nm} & Graphite: \SI{25}{nm} \\
        \hline
    \end{tabular}
    \caption{Layer composition and thicknesses from top to bottom for samples A and B.}
    \label{tab:stack_thickness}
\end{table}

 Using atomic force microscopy (AFM), the relative twist angle between WSe\textsubscript{2} and BLG was determined to be \SI{0 (2)}{\degree} for the WSe\textsubscript{2}-on-BLG stack (sample A) and \SI{4 (2)}{\degree} for the BLG-on-WSe\textsubscript{2} heterostructure (sample B). It is important to note that exfoliated edges in hexagonal materials predominantly exhibit zig-zag or armchair orientations, leading to an intrinsic \SI{0}{\degree}/\SI{30}{\degree} uncertainty. Since we observe a significant enhancement of the spin--orbit gap $\Delta_\mathrm{SO}$, we conclude that the twist angle in both devices is close to  \SI{0}{\degree}. This is supported by both theoretical \cite{David2019, Li2019, Naimer2021, Zollner2023} and experimental \cite{zhang2024} studies, which indicate a strong reduction of  $\lambda_\mathrm{VZ}$ near \SI{30}{\degree}.
 
 We used standard electron-beam lithography (EBL) and metal deposition techniques for device fabrication. 3/20\,nm Cr/Au split gates (SGs)  were formed first, creating a \SI{75}{nm} wide channel. Ohmic contacts were formed by a CHF\textsubscript{3}/O\textsubscript{2} etch followed by 5/50\,nm Cr/Au metal deposition. Finally, a \SI{200}{nm} wide and 10/90\,nm high Cr/Au channel gate is deposited on top of a \SI{20}{nm} ALD grown Al\textsubscript{2}O\textsubscript{3} layer.

\section{Measurement setup and Data Analysis}\label{app:MeasurmentSetUp}

\subsection{Measurement setup} \label{sub:MeasurmentSetUp}

\begin{figure}[hbt!]
	\includegraphics{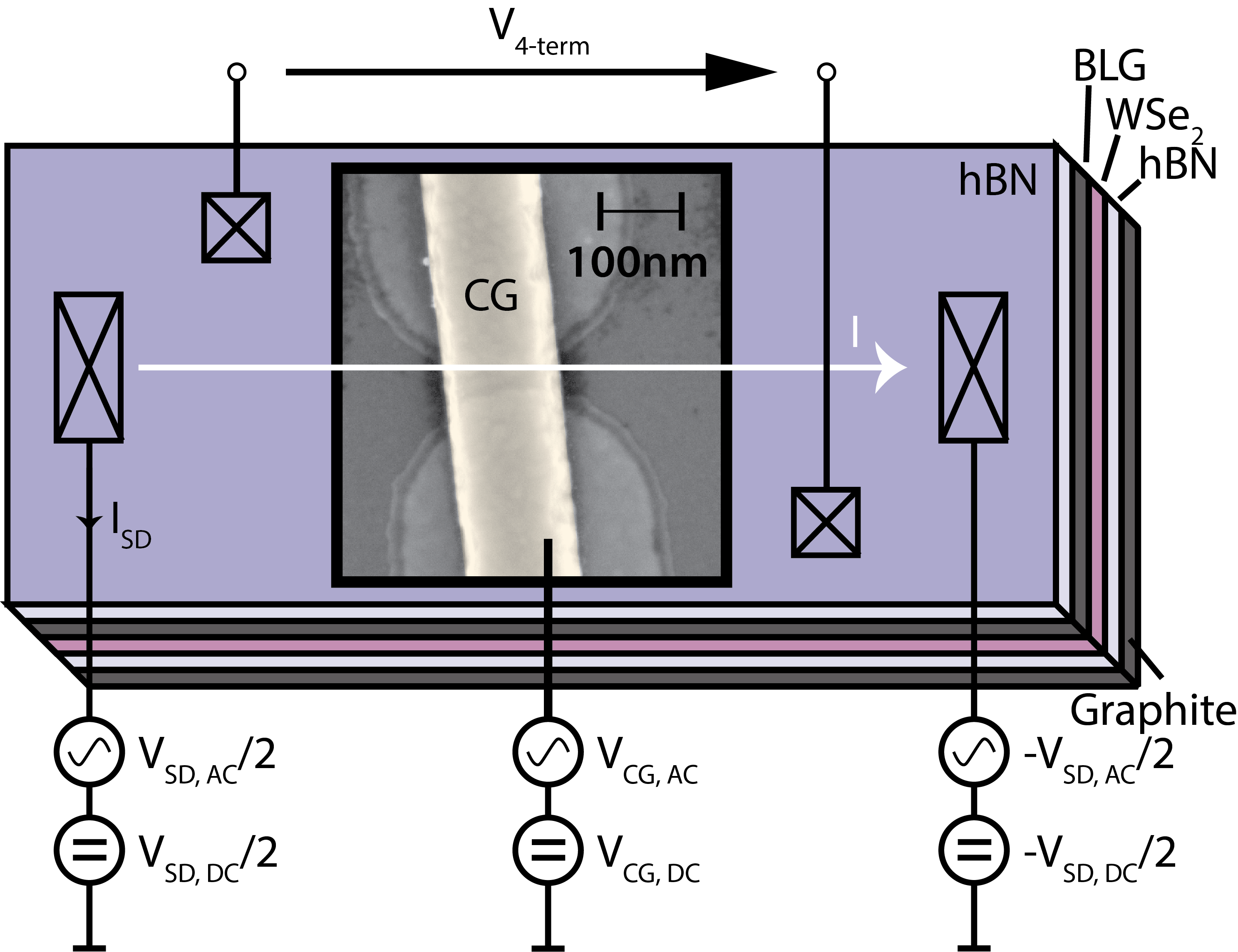}
	\caption{Measurement setup using an AC and DC voltage applied source-drain bias, measuring the current. A DC voltage is applied to the channel gate with the option of additionally applying an AC bias. For QPC measurement, the setup allows four-terminal voltage measurements. }
	\label{Fig:SupSteup}
\end{figure}

Measurements were conducted both in a variable temperature insert (\SI{1.3}{K} base temperature) and a dilution refrigerator (\SI{10}{mK} base temperature). We use the measurement setup shown in Fig.~\ref{Fig:SupSteup}. 

A small AC voltage on a DC voltage was applied to the source-drain (SD) leads. The resulting current $I_\mathrm{SD}$ (referred to as $I$) was measured using an IV converter. The conductance shown throughout this work was obtained from the differential current response. Unless otherwise stated, all data presented in the figures were acquired using this AC measurement technique.

This lock-in technique reduces noise, which is crucial for accurately extracting excited state energies. All initial measurements were performed in a two-terminal configuration. In high-resistance measurements, such as in QDs, this method introduces a small, universal offset in $dI/dV_\mathrm{SD}$, which depends on $V_\mathrm{SD, AC}$ and $f_\mathrm{SD, AC}$. We correct for this offset in Figs.~2b,d,f and Figs.~\ref{Fig:SupParaSpin}b,c,d. Moreover, this two-terminal setup was used for QPC measurements at $T = \SI{1.3}{K}$ (Figs.~3c,e), where the line resistances of roughly \SI{100}{\ohm} are negligible.

Additional contacts on the sample allow for a four-terminal voltage $V_\mathrm{4-term}$ measurement. This four-terminal setup was used for the $T = \SI{10}{mK}$ transconductance measurements (Figs.~4a,b) to eliminate artifacts from the \SI{10}{k\ohm} filtering resistances at these low-temperatures. To present accurate conductance values for the conductance traces in Fig.~1c, we further correct for parasitic resistances, as discussed in Appendix~\ref{sub:ParRes}.

For the QPC bias spectroscopy (Figs.~3c,d), an additional AC voltage was applied to the channel gate, superimposed on the DC voltage. This approach optimizes the measurement of transconductance ($dG/dV_\mathrm{CG}$), as explained in more detail in Appendix~\ref{sub: transconducatnce}.

\subsection{Correcting for parasitic resistances in QPCs}\label{sub:ParRes}

\begin{figure}[h!]
	\includegraphics{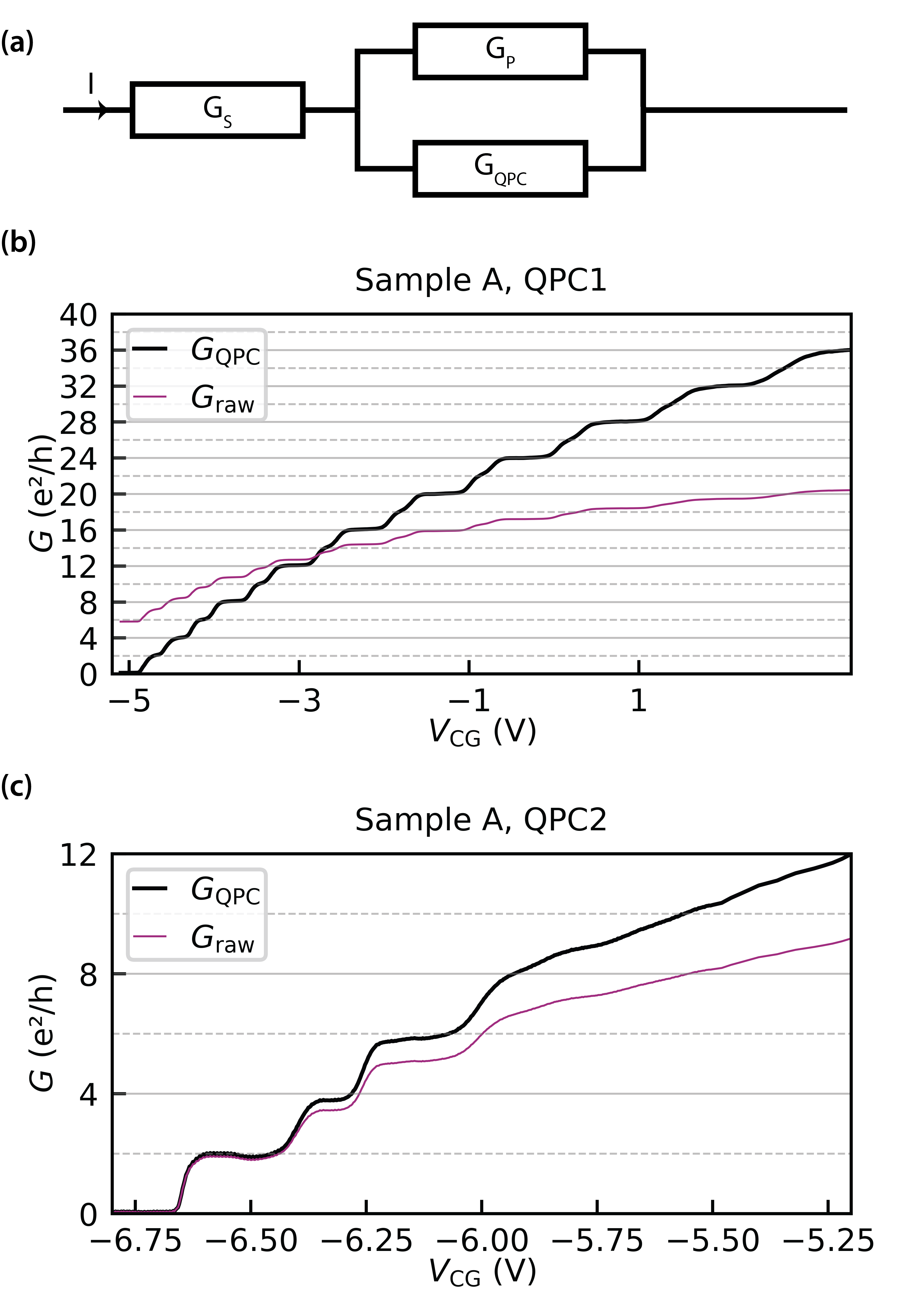}
	\caption{\textbf{(a)} Equivalent circuit illustrating the measured conductance $G_\mathrm{raw} = I/V_\mathrm{4-term}$ and its dependence on the parasitic resistances $G_\mathrm{S}$ and $G_\mathrm{P}$. The correct QPC conductance $G_\mathrm{QPC}$ is extracted using these parameters. \textbf{(b)} Conductance of QPC1 (same as in the main study) showing both the uncorrected ($G_\mathrm{raw}$) and corrected conductance ($G_\mathrm{QPC}$) with $G_\mathrm{S}=39.1e^2/h$ and  $G_\mathrm{P}=6.7e^2/h$ at $T=\SI{1.3}{K}$ and $B=\SI{0}{T}$. \textbf{(c)}  $G_\mathrm{raw}$ and $G_\mathrm{QPC}$ of QPC2 on the same stack with $G_\mathrm{S}=39.1e^2/h$ and  $G_\mathrm{P}=0$ at $T=\SI{10}{mK}$ and $B=\SI{0}{T}$.}
	\label{Fig:SupRes}
\end{figure}

In this chapter, we discuss the correction for parasitic resistances in order to recover the conductance plateaus of the QPCs.

Despite employing a four-terminal setup, the measured conductance, defined as $G_\mathrm{raw} = I/V_\mathrm{4-term}$, does not directly correspond to the intrinsic QPC conductance ($G_\mathrm{QPC}$). This discrepancy arises because both the QPC and the ungated BLG contribute to a voltage drop between the four-terminal contacts. In the low-resistance regime of the QPC, the series conductance ($G_\mathrm{S}$) of the ungated BLG is of a similar magnitude than the actual QPC conductance $G_\mathrm{QPC}$. Consequently, $G_\mathrm{S}$ must be considered to obtain precise conductance values, particularly at high mode numbers.

Additionally, some devices have a parallel conductance path ($G_\mathrm{P}$) alongside the QPC. If this parallel conductance is spatially separate from the channel and remains constant across all mode numbers, we can subtract it from the measurement. The intrinsic QPC conductance $G_\mathrm{QPC}$ is then obtained using the equivalent circuit shown in Fig.~\ref{Fig:SupRes}a with the following correction formula:

 \begin{equation}
G_\mathrm{QPC}=\frac{G_\mathrm{raw}(G_\mathrm{S}+G_\mathrm{P})-G_\mathrm{S}G_\mathrm{P}}{G_\mathrm{S}-G_\mathrm{raw}}.
\end{equation} \label{eq:GQPC}

 By using these parameters $G_\mathrm{S}$ and $G_\mathrm{P}$, we corrected the conductance traces in Fig.~1c by $G_\mathrm{S}=600e^2/h$ and $G_\mathrm{P}=8.5e^2/h$ for the BLG QPC and $G_\mathrm{S}=39.1e^2/h$ and  $G_\mathrm{P}=6.75e^2/h$ for the WSe\textsubscript{2}/BLG QPC.
 Figure~\ref{Fig:SupRes}b illustrates this correction for the WSe\textsubscript{2}/BLG QPC on sample A, where more than 10 conductance plateaus were recovered using just two parasitic resistance parameters, demonstrating the validity of the model. The data from this QPC, labeled QPC1, is the one shown in Fig.~1c, Figs.~3c,d,e, and Fig.~4a.
 
 The correctness of the extracted conductance plateaus is supported by the measured magnetic field pattern in Fig.~3c. For BLG (Appendix~\ref{sub:BLGQPC}), this magnetic field pattern resembles the ones in Refs. \cite{Kraft2018, Overweg2018Nano, Overweg2018PRL, Banszerus_SOC_QPC, Lee2020}.

To further validate the correction procedure, we present additional data from a second QPC (QPC2, shown in Fig.~\ref{Fig:SupRes}c and Fig.~4b) on the same WSe\textsubscript{2}/BLG heterostructure (sample A) as QPC1.
This QPC2 differs from QPC1 in that it does not possess a parallel conductance channel ($G_\mathrm{P} = 0$). Here, the two-fold degeneracy lifting due to enhanced SOC is already evident in the raw data, with the lowest plateau occurring at $G_\mathrm{QPC} \approx G_\mathrm{raw} \approx 2e^2/h$. At higher mode numbers, applying the same series conductance correction
($G_\mathrm{S} = 39.1e^2/h$) as in QPC1 yields quantized conductance in steps of $2e^2/h$. Since QPC1 and QPC2 are connected in parallel within the same four-terminal contact configuration, they share the same ungated graphene leads in series, and thus the same $G_\mathrm{S}$. This further validates that $G_\mathrm{S}$ is a global sample property,
while $G_\mathrm{P}$ remains device-specific.
 

\subsection{Transconductance measurement using two-frequency lock-in detection}\label{sub: transconducatnce}

The transconductance $dG/dV_\mathrm{CG}=d^2I/dV_\mathrm{SD}/dV_\mathrm{CG}$ in the $V_\mathrm{CG}$-$B$ plots (Fig.~3e and Figs.~4a,b) is computed numerically. This approach works well since data points used for differentiation are recorded sequentially.\\

In contrast, the measurement procedure for bias spectroscopy (Figs.~3c,d) differs, as here, we continuously sweep the source-drain voltage $V_\mathrm{SD}$ while stepping the channel gate voltage $V_\mathrm{CG}$. Taking a numerical derivative in the $V_\mathrm{CG}$ direction in this case introduces significant sensitivity to noise and drift because data points are differentiated that were not recorded consecutively. To avoid these issues, we measure in a two-frequency lock-in amplifier setup, as illustrated in Fig.~\ref{Fig:SupSteup}, with a source-drain AC frequency of $f_\mathrm{SD}=\SI{130}{Hz}$ and a channel gate frequency of $f_\mathrm{CG}=\SI{35}{Hz}$. The resulting source-drain current $I$ can be expressed as:

 \begin{equation}
 \begin{split}
   I(V_\mathrm{SD}, V_\mathrm{SD})\approx I_0+\frac{\partial I}{\partial V_\mathrm{SD}}dV_\mathrm{SD} + \frac{\partial I}{\partial V_\mathrm{CG}}dV_\mathrm{CG} +\\
   \frac{\partial ^2 I}{\partial V_\mathrm{SD}\partial V_\mathrm{CG}}dV_\mathrm{SD}dV_\mathrm{CG} + \frac{\partial^2 I}{2\partial V_\mathrm{SD}^2}dV_\mathrm{SD}^2+ \frac{\partial^2 I}{2\partial V_\mathrm{CG}^2}dV_\mathrm{CG}^2
   \end{split}
\end{equation}

The first lock-in amplifier ($f_\mathrm{SD}$), with a time constant of \SI{10}{ms}, outputs a signal proportional to  
\begin{equation}
\frac{\partial I}{\partial V_\mathrm{SD}} dV_\mathrm{SD}
\end{equation}
which remains constant over time, as well as  
\begin{equation}
\frac{\partial^2 I}{\partial V_\mathrm{SD} \partial V_\mathrm{CG}} dV_\mathrm{SD} dV_\mathrm{CG},
\end{equation}
which is modulated by the frequency $f_\mathrm{CG}$. The second lock-in amplifier ($f_\mathrm{CG}$) then directly extracts the transconductance $dG/dV_\mathrm{CG}$.

\subsection{Converting voltages into energy scales}\label{sub:convertEnergy}

For carrier transport through the high-resistance quantum dot, nearly the entire source-drain voltage drops across the dot. As a result, changes in source-drain voltage directly translate to changes in the quantum dot energy levels, such that $\Delta V_\mathrm{SD}=\Delta E/e$. The spin--orbit splitting is directly extracted from the intersection of the excited state line with the edge of the Coulomb diamond, fitted by two straight lines.

For the lower-resistive QPC, parasitic resistances must be considered, especially in the low-resistance regime. First, we determine the series resistance $G_\mathrm{S}$, as shown in the previous subsection. The voltage drop across the QPC — and thus the corresponding energy is then obtained by correcting for the series resistance contribution using $\Delta E = e(V_\mathrm{4-term} - I/G_\mathrm{s})$. We extract $\Delta _\mathrm{SO}$ by fitting the diamond edges with four individual lines and dividing its height by two.

\subsection{Characterizing the displacement field}\label{sub:DisplacementField}
All displacement field values presented in the main text correspond to the field underneath the split gates in operation. This allows us to use the analytical simple plate capacitor model instead of accounting for the geometry of stray field components from the channel and split gates, which would be necessary for calculating potential landscape in the channel. Since the spin--orbit gap $\Delta_\mathrm{SO}$ is extracted always for the first carrier/subband, the displacement field in the channel is close to pinch-off at the bend-edge and thus closely matches the value beneath the split gates.

Additionally, because the channel gate manipulates both the density and the displacement field during a measurement, the displacement field beneath the split gates provides a more consistent metric for comparison.

Its value is defined by:

 \begin{equation}
D=1/2\,(C_\mathrm{B}V_\mathrm{BG} - C_\mathrm{T}V_\mathrm{SG}) + D_0
\label{Eq_DField}
\end{equation}

with $V_\mathrm{BG}$ as back gate and $V_\mathrm{SG}$ as split gate voltage. To calculate the capacitances  $C_\mathrm{T}$ and $C_\mathrm{B}$, we use the thicknesses provided in Appendix~\ref{app:DevFab} with $\epsilon_r=3.24$ for hBN \cite{MasseroniphD2024} and $\epsilon_r=6$ for WSe\textsubscript{2} \cite{Hou2022}. Due to the relatively large ungated regions compared to the top gate area, these ungated regions dominate the measured resistance. This effect becomes particularly problematic near the charge neutrality point of the gated region---where precise data is required to extract $D_0$---as the resistance is dominated by the ungated regions. As a result, an exact value of $D_0$ could not be determined and is therefore set to $D_0=0$.

Hence, while the data points within each sample follow a reproducible trend, points from different samples that are labeled with the same displacement field may correspond to slightly different actual displacement values, due to possible sample-to-sample variations in $D_0$.

\clearpage
\newpage

\section{Modelling the proximitized BLG channel}
\label{app:Model}
To theoretically describe QPCs in proximitized BLG/TMD heterostructures, we work in the frame of the four-band model, including the effects of confinement and proximity-induced SOC. The full Hamiltonian in valley $K^{\xi}$ with $\xi\pm 1$ reads
\begin{equation}
H_{BLG}^{\xi}+H_Z+H_{SOC}^{\xi}.
\label{eqn:Hamiltonian}
\end{equation}
The first term \cite{mccann2013, Knothe2018, Overweg2018PRL},
\begin{align}
\nonumber &H^{\xi}_{BLG}= \sigma_0 \otimes\\
\xi
&\setlength{\arraycolsep}{-5pt} \begin{pmatrix} 
 \xi U(x)-\frac{1}{2}\Delta(x) & v_3\pi & 0 &v \pi^{\dagger}\\
 v_3 \pi^{\dagger}& \xi U(x)+\frac{1}{2}\Delta(x) & v\pi &0\\
 0 & v\pi^{\dagger} & \xi U(x)+\frac{1}{2}\Delta(x) & \xi \gamma_1\\
 v\pi & 0 & \xi \gamma_1 & \xi U(x)-\frac{1}{2}\Delta(x)
\end{pmatrix},
\label{eqn:HBLG}
\end{align}
captures the confinement by a confinement potential U(x) and spatially modulated gap $\Delta(x)$,
\begin{equation}
U(x)=\frac{U_0}{\cosh{\frac{x}{L}}}, \hskip20pt \Delta(x)=\Delta_0-\beta\frac{\Delta_0}{\cosh{\frac{x}{L}}},
\label{eqn:UD}
\end{equation}
with $\beta=0.2$ and confinement depth $U_0=-20$ meV, and width $L$ chosen to match the parameters of the experimental probes. Similar, smooth potential landscapes have been used previously to successfully describe electrostatically confined channels in BLG \cite{Knothe2018, Overweg2018PRL, Lee2020}. Furthermore, $\pi=p_x+ip_y$, $\pi^{\dagger}=p_x-ip_y$, with $\mathbf{p}=-i\hbar\nabla-\frac{e}{c}\mathbf{A}$,    $v=1.0228*10^6$ m/s, $v_3=1.2299*10^5$ m/s, and  $\gamma_1=381$ meV.~

The second term, $H_Z=-\frac{1}{2}g_S\mu_B B\sigma_z \otimes \sigma_0 \otimes \sigma_0$, with Bohr magneton $\mu_B$ and spin g-factor $g_S=2$, describes spin Zeeman coupling.  

The last term describes the proximity-induced SOC \cite{Zollner2021},
\begin{align} &H_{SOC}^{\xi=+1}= \sigma_z \otimes
    \begin{pmatrix}
        -  \lambda_I^{A_1}&0&0&0\\
        0&   \lambda_I^{B_2} &0&0\\
        0&0& -  \lambda_I^{A_2}&0\\
        0&0&0&   \lambda_I^{B_1}
    \end{pmatrix}\\&+
    \sigma_x \otimes
    \setlength{\arraycolsep}{-5pt}\begin{pmatrix}
        0&0&0&-i \lambda_{R_1} (1-s)\\
        0&0&i \lambda_{R_2} (1+s)&0\\
        0& -i\lambda_{R_2} (1-s)& 0&0\\
        i \lambda_{R_1} (1+s)& 0&0&0
    \end{pmatrix},
    \label{eqn:HSOC+}
\end{align}
and
\begin{align} &H_{SOC}^{\xi=-1}= \sigma_z \otimes
    \begin{pmatrix}
        -  \lambda_I^{B_2}&0&0&0\\
        0&   \lambda_I^{A_1} &0&0\\
        0&0& -  \lambda_I^{B_1}&0\\
        0&0&0&   \lambda_I^{A_2}
    \end{pmatrix}\\&+
    \sigma_x \otimes
    \setlength{\arraycolsep}{-5pt}\begin{pmatrix}
        0&0&0&i \lambda_{R_2} (1-s)\\
        0&0&-i \lambda_{R_1} (1+s)&0\\
        0& i\lambda_{R_1} (1-s)& 0&0\\
        -i \lambda_{R_2} (1+s)& 0&0&0
    \end{pmatrix}.
     \label{eqn:HSOC-}
\end{align}
In \eqref{eqn:HSOC+}, \eqref{eqn:HSOC-}, $\lambda_I$ measures the strength of the intrinsic SOC and $\lambda_{R}$ of Rashba SOC, respectively, and $s=\pm1$ labels the spin states $\uparrow = 1$ and $\downarrow = -1$. We work in the basis $\Phi_{K^+}=(\Psi_{A1}\uparrow,\Psi_{B2}\uparrow,\Psi_{A2}\uparrow,\Psi_{B1}\uparrow, \Psi_{A1}\downarrow,\Psi_{B2}\downarrow,\Psi_{A2}\downarrow,\Psi_{B1}\downarrow)$ or $\Phi_{K^-}=(\Psi_{B2}\uparrow,\Psi_{A1}\uparrow,\Psi_{B1}\uparrow,\Psi_{A2}\uparrow , \Psi_{B2}\downarrow,\Psi_{A1}\downarrow,\Psi_{B1}\downarrow,\Psi_{A2}\downarrow)$. Note that our convention of labeling the graphene sublattices and layers differs from that of previous works, and to reach consistency with the proximity-induced SOC parameters presented in Refs.~\cite{Zollner2021, Seiler2024} one has to interchange sublattices A and B as well as graphene layers 1 and 2. The spin--valley--Zeeman SOC parameter is related to the parameters above as \cite{Naimer2021}
\begin{equation}
 \lambda_{VZ} =  \frac{\lambda_I^B - \lambda_I^A}{2}   .
\end{equation}

To calculate the confined subband spectra and states (as the ones shown in Fig.~3f and Figs.~4c,d of the main text), we numerically diagonalize the Hamiltonian in \eqref{eqn:HBLG} in a suitable basis of confined states. We follow the procedure described in Refs.~\cite{Knothe2018, Overweg2018PRL}, using harmonic oscillator wave functions in the confinement direction across the channel axis, $\Tilde{x}=x\sin{\theta} + y\cos{\theta}$ and assuming free propagation of the electrons along the channel axis, $\tilde{y}=x\cos{\theta}-y\sin{\theta}$, where the angle $\theta$ interpolates between orientation of the channel axis along the armchair $(\theta=0)$ and zigzag $(\theta=\frac{\pi}{2})$ direction of the graphene lattice. We refer the reader to Refs.~\cite{Knothe2018, Overweg2018PRL} for details about the numerical implementation.

We extract the subband edges for different magnetic field values for deducing the magnetic field patterns as shown in Fig.~4. These points correspond to the energies where we expect steps in the quantized conductance.

\section{Impact of the SOC parameters on the calculated spectrum}\label{app:Rashba}
\begin{figure*}[htb!]
	\includegraphics[width=\textwidth]{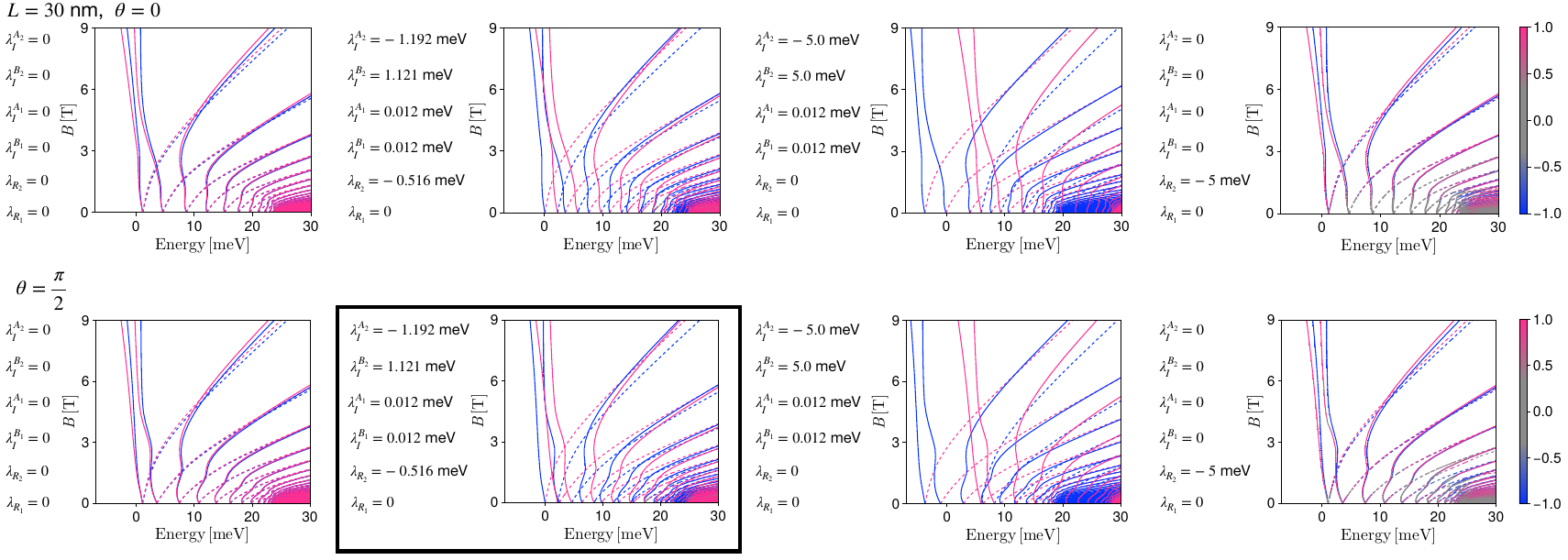}
	\caption{Calculated magnetic field patterns for different SOC parameters and channel orientations for width $L=30$ nm and $\Delta=50$ meV. The boxed plot corresponds to the one shown in the main text, with parameters obtained from DFT in \cite{Seiler2024}, showing the best agreement with experimental data. Solid and dashed lines distinguish between the $K^+$ and $K^-$ valley. The color scale quantifies the spin polarization of the bands.}
	\label{fig:SupModelL30}
\end{figure*}
\begin{figure*}[htb!]
	\includegraphics[width=\textwidth]{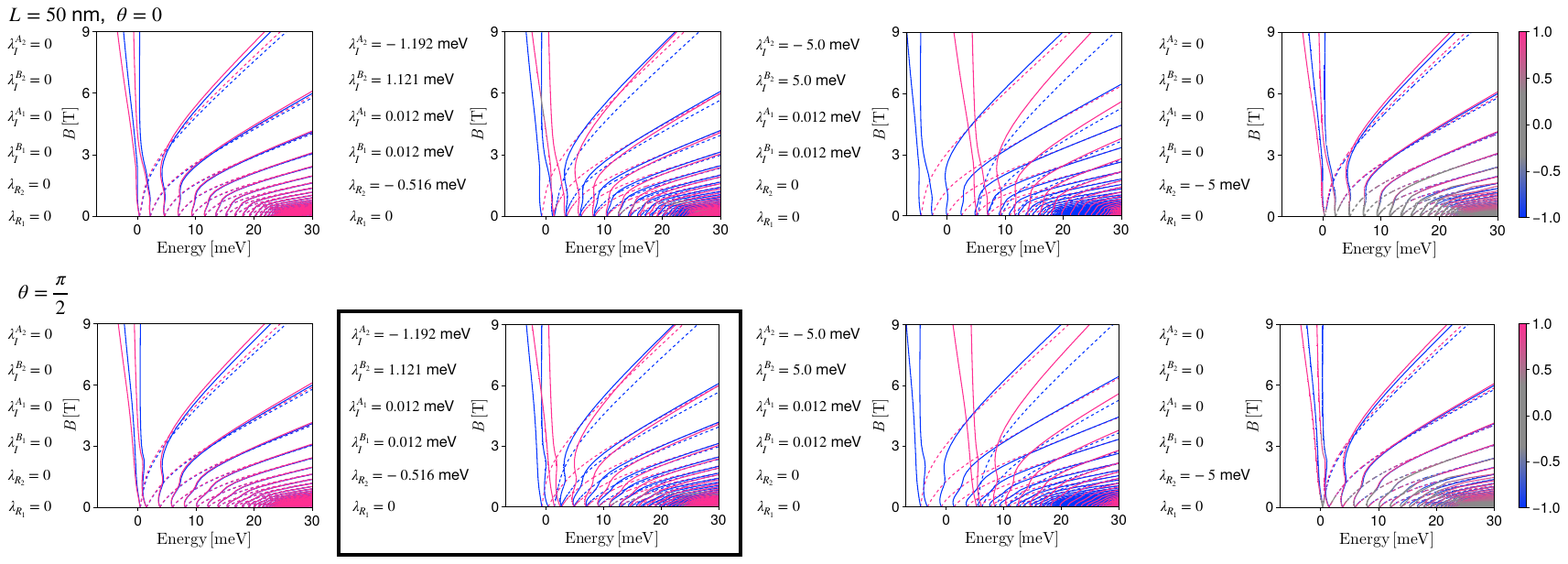}
	\caption{Calculated magnetic field patterns for different SOC parameters and channel orientations for width $L=50$ nm and $\Delta=50$ meV. The boxed plot corresponds to the one shown in the main text, with parameters obtained from DFT in \cite{Seiler2024}, showing the best agreement with experimental data. Solid and dashed lines distinguish between the $K^+$ and $K^-$ valley. The color scale quantifies the spin polarization of the bands.}
	\label{fig:SupModelL50}
\end{figure*}
We discuss how choosing different SOC parameters in the theoretical model changes the calculated magnetic field patterns compared to experimental observations. In Figs.~\ref{fig:SupModelL30} and \ref{fig:SupModelL50}, we show the magnetic field patterns obtained from calculations with different SOC parameters, $\lambda_I^{A_1}, \lambda_I^{A_2}, \lambda_I^{B_1}, \lambda_I^{B_2}, \lambda_{R_1}, \lambda_{R_2}$, ranging from no SOC to strong intrinsic or Rashba SOC, respectively. This analysis allows the following conclusions about the role of the SOC parameters: First, we observe that the intrinsic SOC $\lambda_I$ is indeed responsible for the characteristic 3+1 pattern observed in the experiment (compare first and second column in Figs.~\ref{fig:SupModelL30} and \ref{fig:SupModelL50}). However, the value of $\lambda_I$ cannot be much larger than the subband spacing to obtain this pattern (compare columns two and three). Further, the Rashba parameters have negligible influence on the magnetic field patterns since it does not induce noticeable splittings at the band minima near the $K$-points, even for large values of $\lambda_R$ (fourth column in Figs.~\ref{fig:SupModelL30} and \ref{fig:SupModelL50}). We note that Rashba SOC does, however, affect the spin polarization of the subbands. In the case of the relatively small value 
$\lambda_{R_2}=-0.5$ meV predicted by DFT \cite{Seiler2024}, we obtain spin polarizations above $0.99$ for all data points (data presented in the main text, Fig.~3). Increasing $\lambda_{R_2}$ decreases the degree of spin polarization. For $\lambda_{R_2}=-5$ meV in the fourth column of Figs.~\ref{fig:SupModelL30} and \ref{fig:SupModelL50}, we find the spin polarizations vary significantly depending on the band index and the strength of the external magnetic field.

\clearpage
\newpage
\section{Additional measurement data}\label{app:add_measurement}
\subsection{PNP dot with weak spin-orbit coupling}

\begin{figure}[hbt!]
	\includegraphics{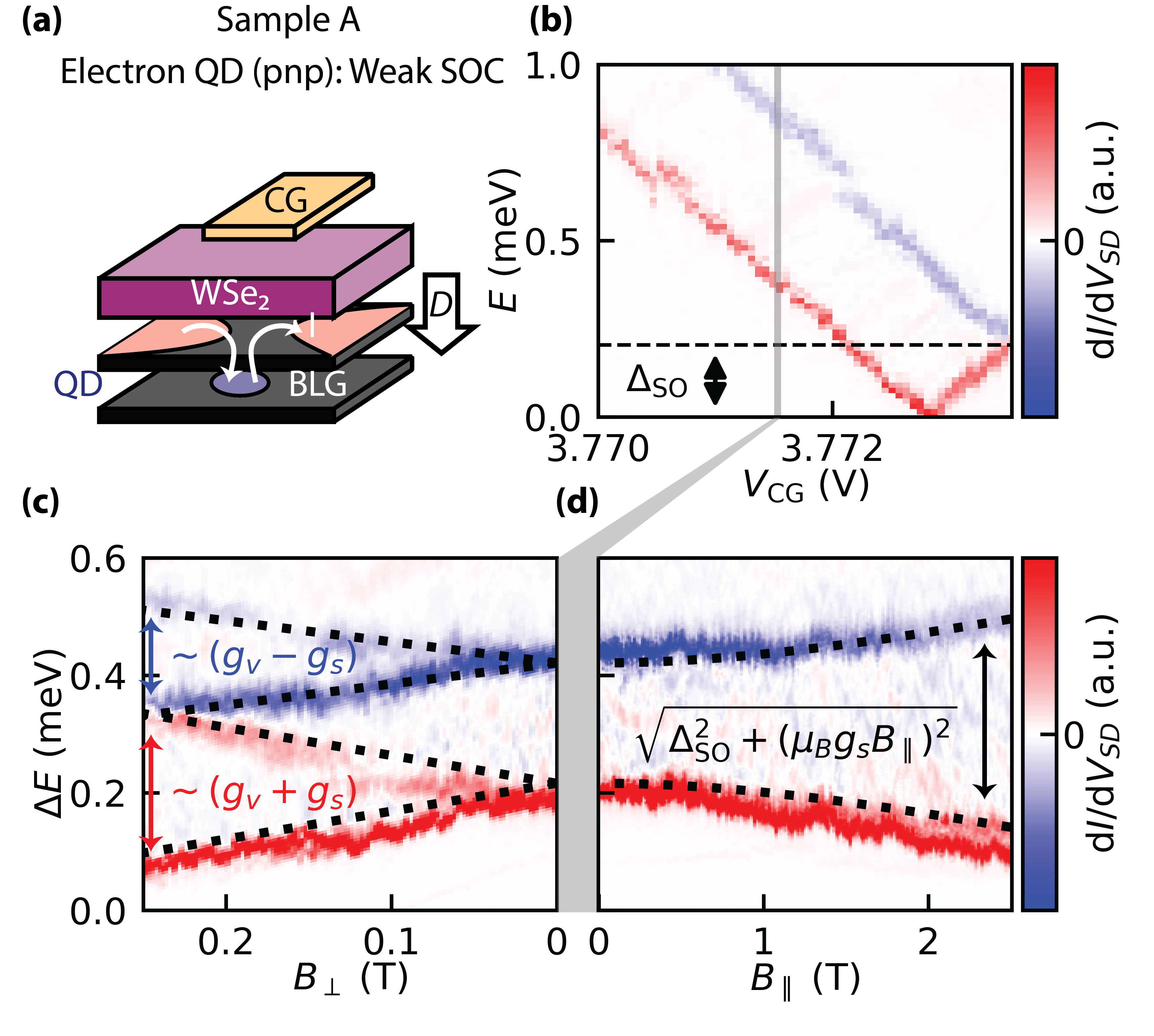}
	\caption{\textbf{(a)} Schematic of the pnp-type quantum dot in sample A (WSe\textsubscript{2}/BLG). 
\textbf{(b)} Finite-bias spectroscopy of the first charge carrier at \( B_\perp = 0 \) and \( T = \SI{10}{mK} \). 
\textbf{(c)}, \textbf{(d)} Evolution of the QD energy levels in perpendicular and parallel magnetic fields, respectively. The data are overlaid with theoretical curves calculated using the extracted spin–orbit splitting \( \Delta_\mathrm{SO} \), valley \( g \)-factor \( g_\mathrm{v} = 14.4 \), and spin \( g \)-factor \( g_\mathrm{s} = 2 \). The energy levels were extracted from bias spectroscopy performed at varying magnetic fields, as indicated by the gray line. }
	\label{Fig:SupParaSpin}
\end{figure}

To clearly demonstrate the expected quadratic dependence on magnetic field, we turn to measurements in the weak SOC regime, conducted on the pnp-type quantum dot in sample A (WSe\textsubscript{2}/BLG), shown in Fig.~\ref{Fig:SupParaSpin}a. Finite-bias spectroscopy (Fig.~\ref{Fig:SupParaSpin}b) yields a spin--orbit splitting of $\Delta_\mathrm{SO}=$\SI{0.21(2)}{meV}.
Measurements in a perpendicular magnetic field reveal a clear splitting of each Kramers pair, consistent with the valley--Zeeman effect (Fig.~\ref{Fig:SupParaSpin}c). By analyzing the energy separation within each pair---given by \( (g_\mathrm{v} + g_\mathrm{s})\mu_BB_\perp \) and \( (g_\mathrm{v} - g_\mathrm{s})\mu_BB_\perp \)---we extract valley and spin \( g \)-factors of $ g_\mathrm{v} = $\SI{14.4(3)} and $g_\mathrm{s} = $\SI{2.0(3)}, respectively. The data are overlaid with the theoretically expected evolution in \( B_\perp \) (dotted line), calculated using these \( g \)-factors and the zero-field spin--orbit splitting \( \Delta_\mathrm{SO} = \SI{0.21}{meV} \) from Fig.~\ref{Fig:SupParaSpin}b. The agreement between theory and experiment confirms that each Kramers pair comprises two states with opposite valley and spin quantum numbers.

Compared to the quantum dot presented in Fig.~2, the reduced SOC in this regime allows for a more unambiguous observation of the expected quadratic dependence of the energy splitting on \( B_\parallel \). This behavior is clearly visible in Fig.~\ref{Fig:SupParaSpin}d, confirming that the two Kramers pairs arise from opposite spin orientations coupled via an out-of-plane spin--orbit field. The data are overlaid with the theoretically expected evolution, calculated using the extracted spin--orbit gap $\Delta_\mathrm{SO}$ and the spin $g$-factor $g_s = 2$. The slight deviation from the quadratic trend at higher fields is attributed to a small \( B_\perp \) component resulting from the residual tilt of the sample. 

Overall, the measurements closely follow the expected quadratic evolution of the spin--orbit gap with in-plane magnetic field. In contrast, an orbital excited state would exhibit a $B_\parallel$ dependence identical to the ground state, which shares the same spin configuration. The distinct responses to both in-plane and out-of-plane magnetic fields therefore provide unambiguous evidence that the measured excitation corresponds to the spin--orbit splitting of the first charge carrier. The observed magnetic field dependence is fully consistent with previous studies on pristine bilayer graphene QDs \cite{Banszerus2021,Kurzmann2021_Kondo,Duprez2024}, with the notable distinction of a significantly enhanced spin–orbit gap in the present devices.


\subsection{MoS\textsubscript{2}/BLG reference QPC}\label{sub:Mos2}

\begin{figure}[hbt!]
	\includegraphics{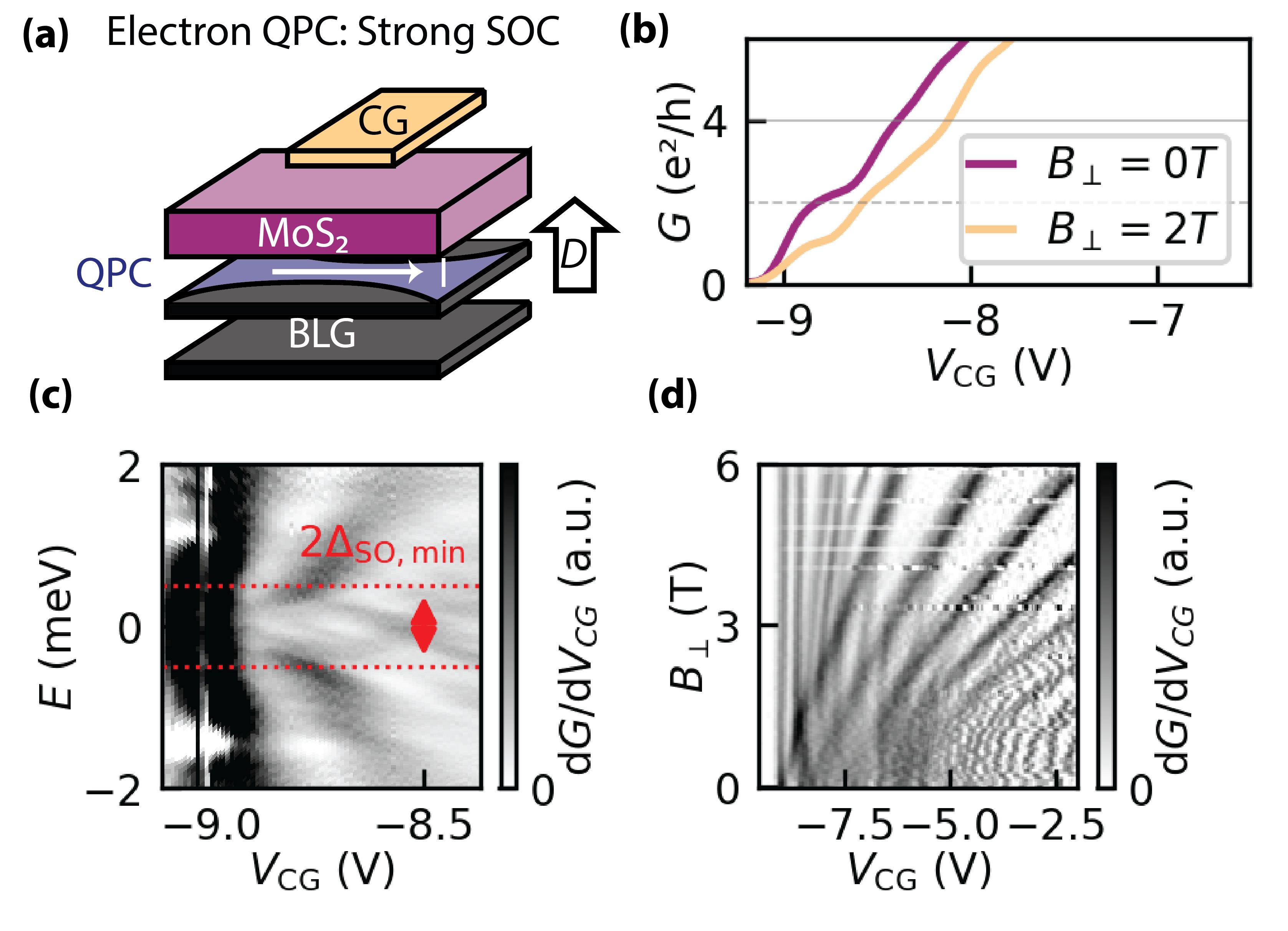}
	\caption{\textbf{(a)} Schematic side view of the MoS\textsubscript{2}/BLG device. \textbf{(b)} Four-terminal conductance traces at $B_\perp=\SI{0}{T}$ and $B_\perp=\SI{2}{T}$. \textbf{(c)} Bias spectroscopy of the  $G=2e^2/h$-step, indicating a $\Delta_\mathrm{SO}\,>\,\SI{500}{\micro eV}$. \textbf{(d)} Transconductance as function of $B_\perp$ and $V_\mathrm{CG}$.  }
	\label{SupMoS2}
\end{figure}

We demonstrate that replacing WSe\textsubscript{2} with MoS\textsubscript{2} in the same device geometry (Fig.~\ref{SupMoS2}a) leads to a similar enhanced and tunable SOC. We measure a clear $G=2e^2/h$ plateau in Fig.~\ref{SupMoS2}b (uncorrected conductance), indicating a lifted degeneracy, caused by an enhanced SOC. This plateau further splits at a relatively small $B_\perp$ due to the valley--Zeeman effect (Fig.~\ref{SupMoS2}b). Fig.~\ref{SupMoS2}c presents the energy gap between the two lowest states. Due to a low subband spacing in the measured QPC, direct state assignment in the $B_\perp$ transconductance data is not possible (Fig.~\ref{SupMoS2}d). Thus, we interpret the measured gap of $\SI{500}{\micro eV}$ as a lower bound for $\Delta_\mathrm{SO}$. Measurements were taken at $T=\SI{10}{mK}$.

\subsection{Bilayer graphene reference QPC}\label{sub:BLGQPC}

\begin{figure}[hbt!]
	\includegraphics{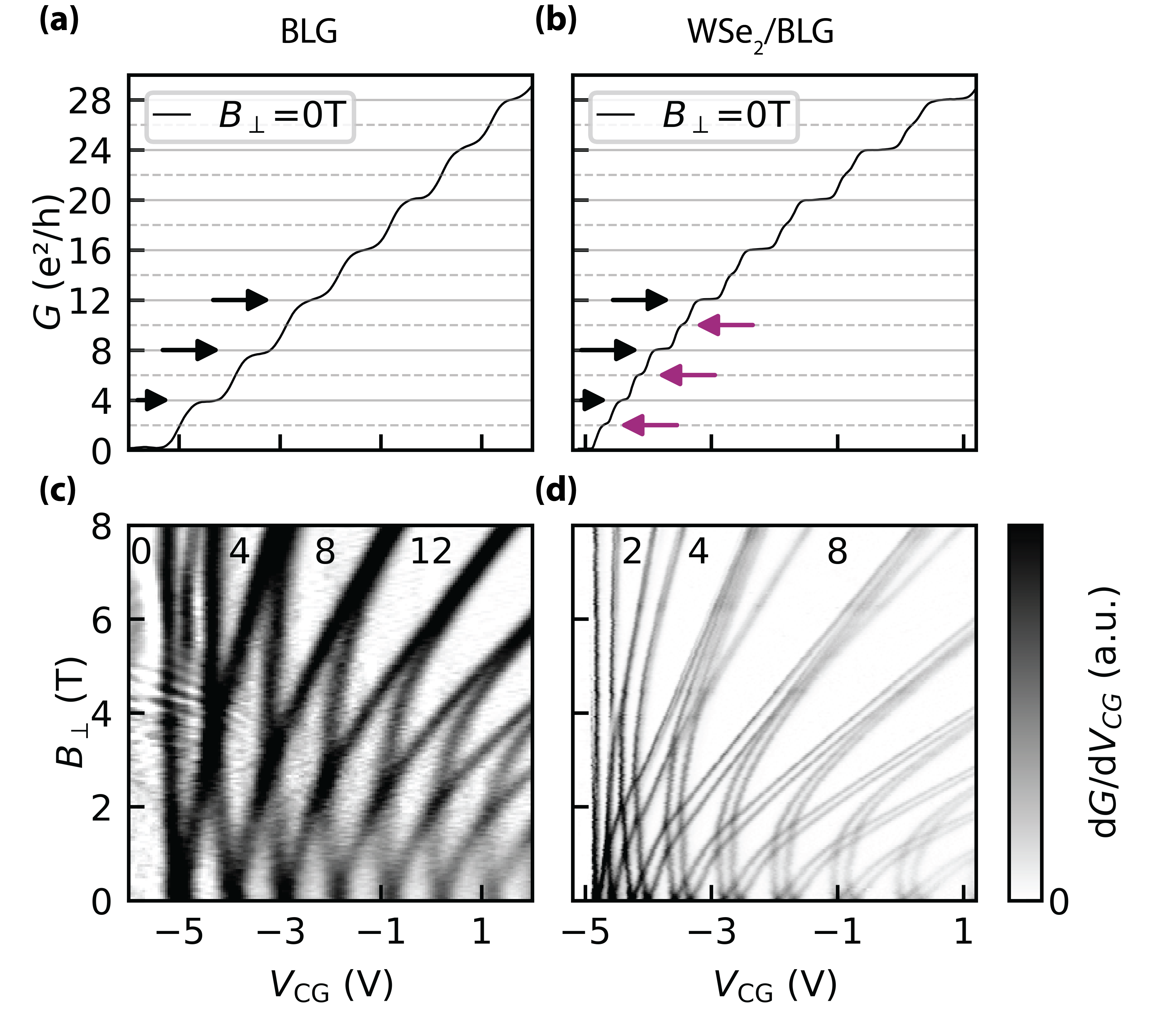}
	\caption{\textbf{(a), (b)} Four-terminal conductance $G$ as a function of the channel gate voltage $V_\mathrm{CG}$ (T\,=\,\SI{1.3}{K}) for BLG \textbf{(a)} and WSe\textsubscript{2}/BLG \textbf{(b)}. The traces have been corrected from parasitic resistances.\textbf{(c), (d)} Two-terminal transconductance $dG/dV_\mathrm{CG}$ as function of $V_\mathrm{CG}$ and $B_\perp$ at T\,=\,\SI{1.3}{K} for BLG \textbf{(c)} and WSe\textsubscript{2}/BLG \textbf{(d)}. The black numbers in the plot represent the quantum numbers of the respective conductance plateaus.   }
	\label{Fig:SupBLG}
\end{figure}

In this section, we present the magnetic field-dependent transconductance of the reference BLG QPC, whose conductance trace was shown in Fig.~1c, and compare it to the corresponding measurements for the WSe\textsubscript{2}/BLG QPC. The BLG QPC has a \SI{100}{nm} channel width - in contrast to the \SI{75}{nm} channel width used for all other QPCs in this study.

Figures~\ref{Fig:SupBLG}a,b display the corrected conductance plateaus of the BLG and WSe\textsubscript{2}/BLG QPC, while Figs.~\ref{Fig:SupBLG}c,d show the respective transconductance $dG/dV_\mathrm{CG}$ as a function of $B_\perp$. In the pristine BLG QPC, increasing $B_\perp$ splits the subbands due to the valley--Zeeman effect, forming two valley-polarized pairs: ($\mathrm{K}^+\downarrow$, $\mathrm{K}^+\uparrow$) and ($\mathrm{K}^-\uparrow$, $\mathrm{K}^-\downarrow$) (Fig.~1c). At sufficiently high magnetic fields, the magnetic length becomes smaller than the channel width, leading to the formation of Landau levels and an apparent convergence of the $(n+2)\mathrm{K^+}$ and $n\mathrm{K^-}$ states. Similar behavior in BLG QPCs has been reported in Refs. \cite{Kraft2018, Overweg2018Nano, Banszerus_SOC_QPC, Lee2020, Overweg2018PRL}.

The magnetotransport data for the WSe\textsubscript{2}/BLG QPC (Fig.~\ref{Fig:SupBLG}d) strongly resembles that of pristine BLG  (Fig.~\ref{Fig:SupBLG}c), with one crucial distinction: in BLG/WSe\textsubscript{2}, two parallel evolving states within the same subband and same valley quantum number remain well-resolved due to the enhanced spin--orbit splitting. In contrast, in pristine BLG, these states are too close in energy to be resolved separately.

Additionally, we note that the quality of the pristine BLG QPC device appears to be lower compared to the WSe\textsubscript{2}/BLG QPC. While we do not attribute this to a fundamental effect, this observation suggests that introducing a TMD layer does not reduce device quality more than typical sample-to-sample variations in BLG-based quantum devices.

\subsection{Spin--orbit gap measurements across all device configurations}\label{sub:BiasCollection}

\begin{figure*}[tb]
	\includegraphics{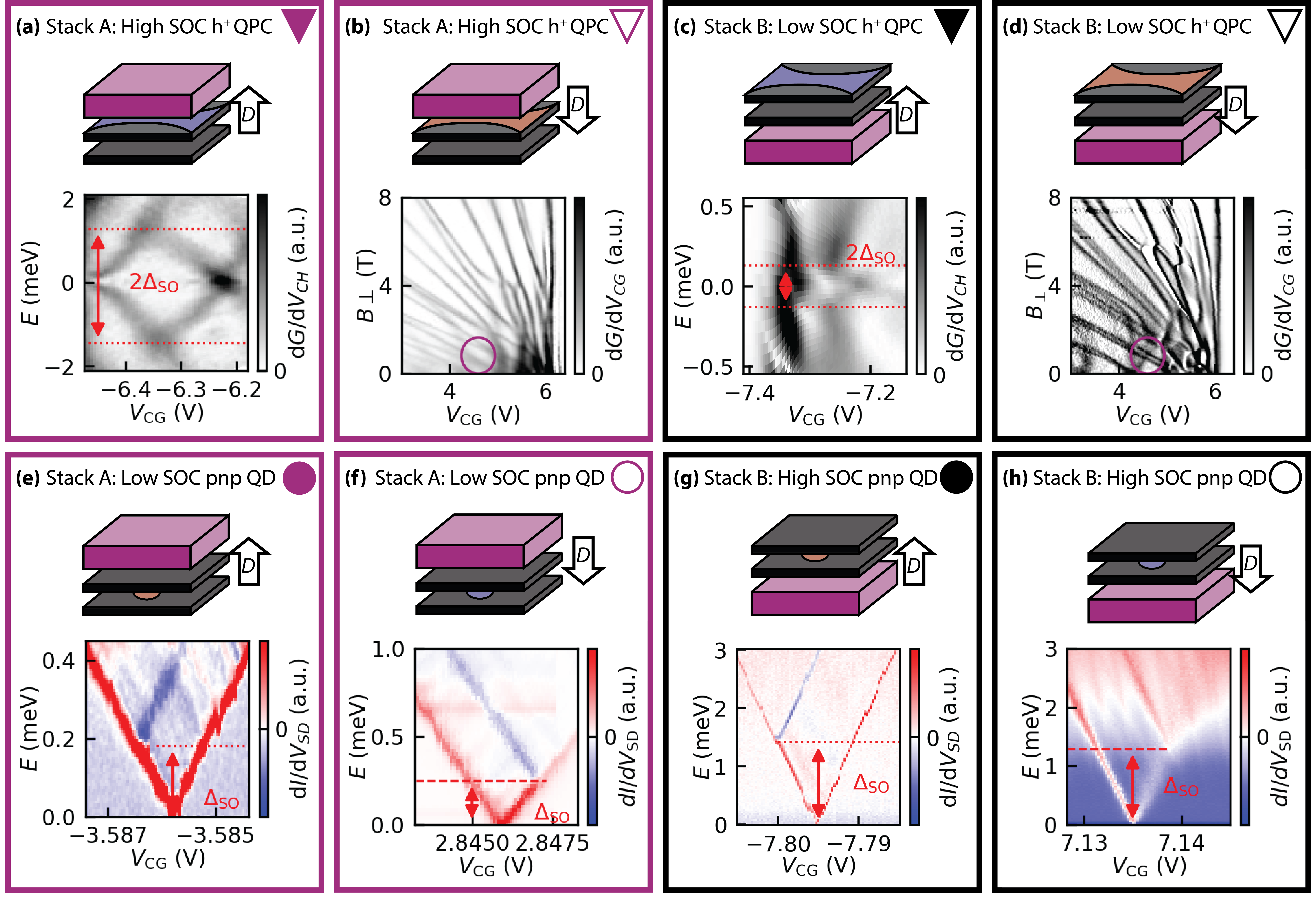}
	\caption{Exemplary bias spectroscopy across all configurations at \SI{10}{mK}, showing the extracted spin--orbit gap $\Delta_\mathrm{SO}$ and a schematic electrostatic illustration. For the p-type QPC in both samples, $\Delta_\mathrm{SO}$ could not be determined due to a low subband spacing. We instead present magneto transconductance data for \textbf{(b)} and \textbf{(d)} verifying the expected SOC trend qualitatively when comparing the $3\mathrm{K}^+\downarrow$ and  $3\mathrm{K}^+\uparrow$ states in this system highlighted by the magenta circle.}
	\label{Fig:SupRevD}
\end{figure*}

In the following table (Fig.~\ref{Fig:SupRevD}), we present representative bias spectroscopy measurements for extracting the spin--orbit gaps $\Delta_\mathrm{SO}$ across all eight possible configurations. The measured values show excellent agreement with layer polarization predictions, which are also included in the figure.

For the p-type QPCs (Fig.~\ref{Fig:SupRevD}b, d), an exact $\Delta_\mathrm{SO}$ could not be determined, as the subband spacing of the lowest modes is smaller than the spin–orbit gap, making an unambiguous assignment of low-energy subband states at zero magnetic field impossible. However, strong SOC is evident from the highly broken degeneracy. The magenta circle highlights the states $3\mathrm{K}^+\downarrow$ and $3\mathrm{K}^+\uparrow$, which originate from different energies at $B = \SI{0}{T}$ in the high-SOC case (Fig.~\ref{Fig:SupRevD}b), whereas in the low-SOC scenario (Fig.~\ref{Fig:SupRevD}d), they originate from nearly the same zero-magnetic-field energy.


\subsection{State Evolution in a Strong SOC Quantum Dot under Magnetic Fields}\label{sub:QDhighSOCStateEvo}

In this chapter, we present magnetic field measurements corresponding to the high SOC quantum dot in sample B (WSe\textsubscript{2}/BLG), whose results were shown in Fig.~2. Since the excited state energies of interest are relatively high, we cut the Coulomb diamond at a constant energy of \( E = \SI{1.9}{meV} \) and sweep the $V_\mathrm{CG}$-axis in a magnetic field. The measurements are overlaid with the expected behavior using the extracted lever arm.

From the \( B_\perp \) data, it is evident that the current through the quantum dot is strongly suppressed above \( B_\perp = \SI{0.15}{T} \), reaching the noise level. As a result, a more precise extraction of the valley \( g_v \)-factor was not possible.

The \( B_\parallel \) data, on the other hand, clearly follows the expected trend. However, a quadratic dependence is not distinctly resolved due to the large SOC.

Overall, the states follow the single-particle predictions well.

\begin{figure}[hbt!]
	\includegraphics{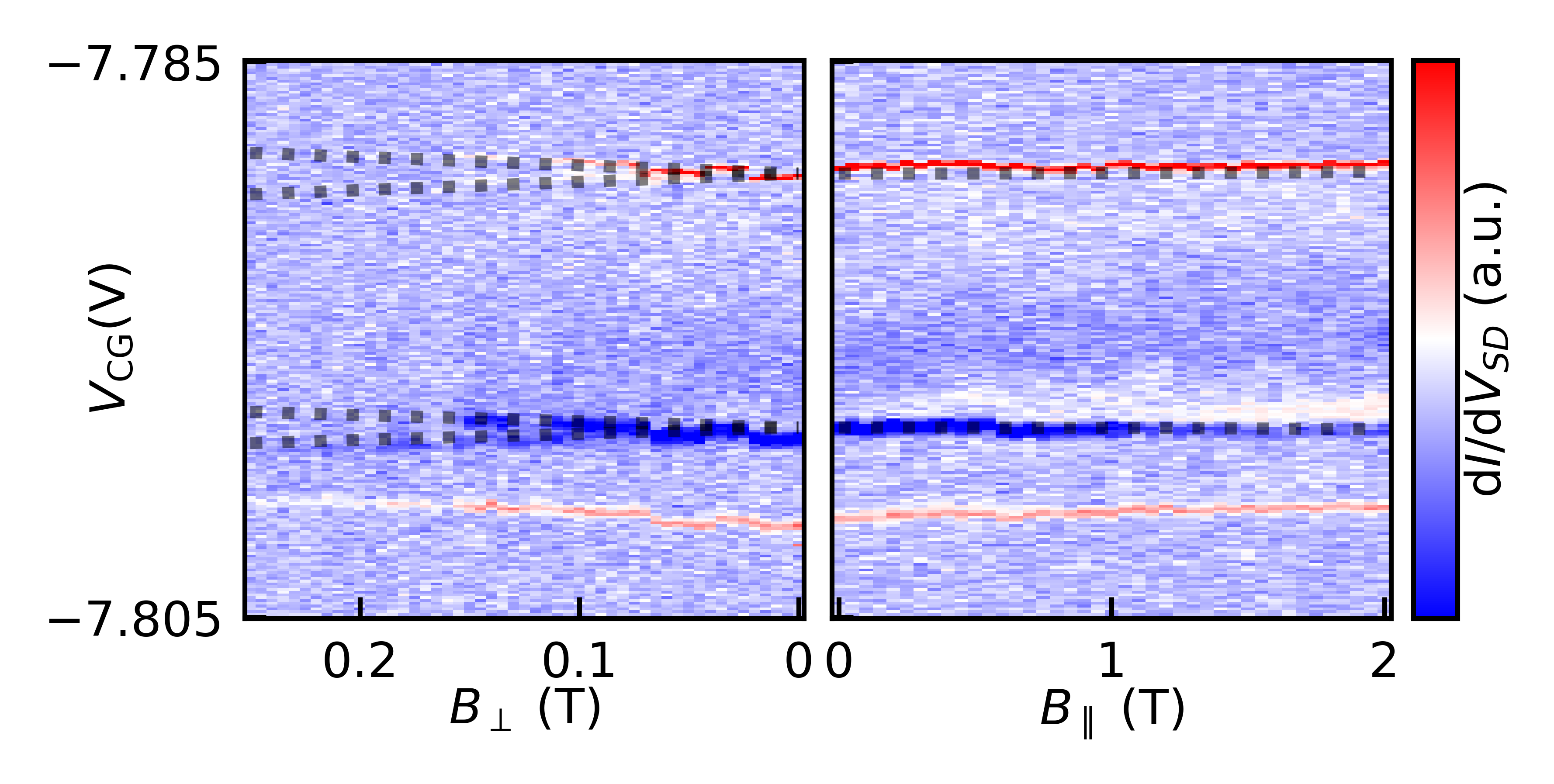}
	\caption{Magnetic field evolution of the quantum dot states in sample B (as shown in Fig.~2), measured in both \( B_\perp \) (left) and \( B_\parallel \) (right). The state evolution is overlaid with a gray dotted line, showing the expected behavior based on the extracted values \( \Delta_\mathrm{SO} \), \( g_\mathrm{v} \approx 14 \), and \( g_\mathrm{s} = 2 \). }
	\label{Fig:QDhighSOCStateEvo}
\end{figure}

\subsection{High magnetic field transport data of  high-SOC QPCs at  \SI{10}{mK}}\label{sub:HighBFieldQPC} 

\begin{figure}[hbt!]
	\includegraphics{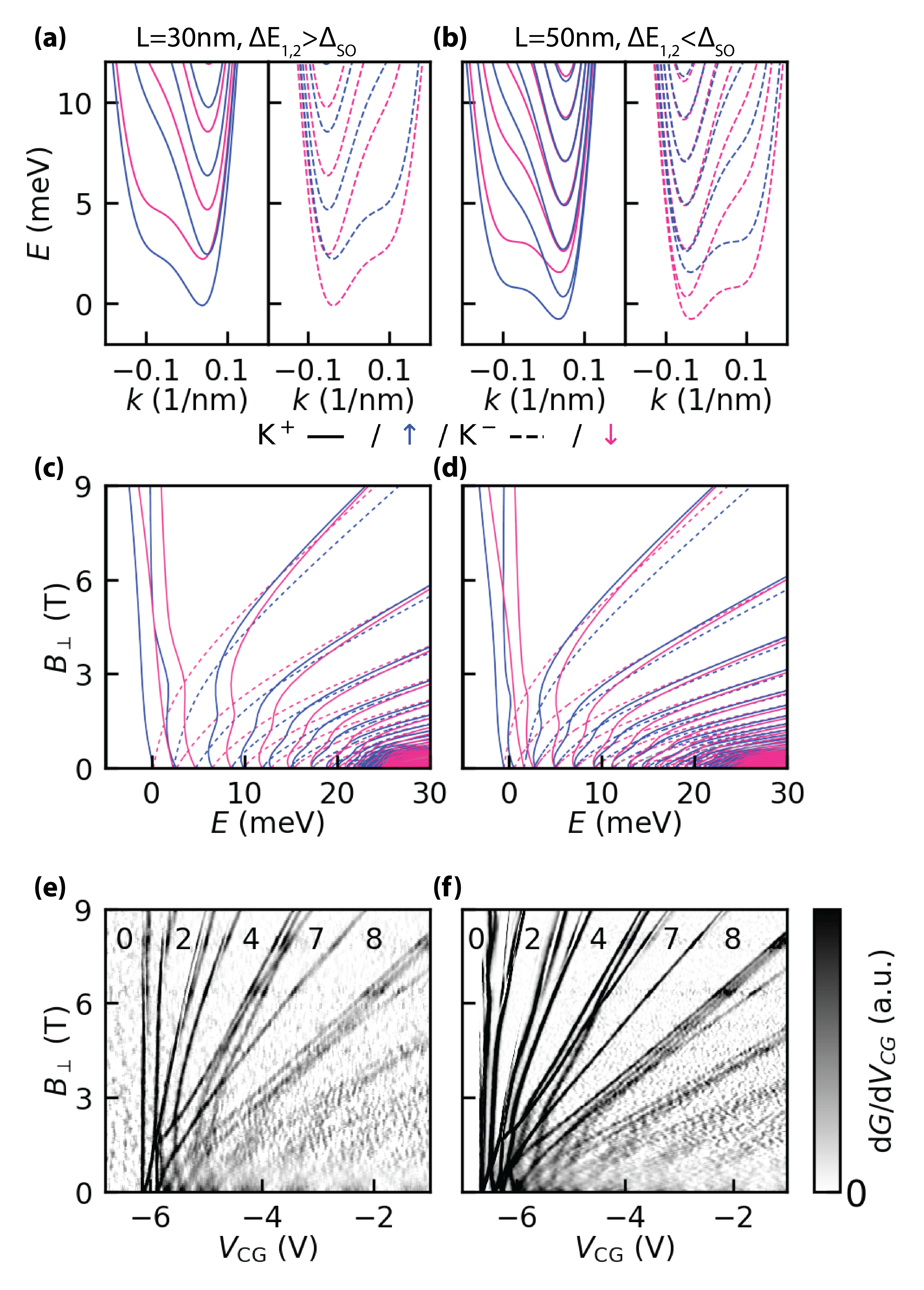}
	\caption{\textbf{(a)}, \textbf{(b)} Calculated band structure for WSe\textsubscript{2}/BLG QPCs with channel widths of \SI{30}{nm} and \SI{50}{nm}, respectively. \textbf{(c)}, \textbf{(d)} Corresponding single-particle calculations of subband evolution in a perpendicular magnetic field.  \textbf{(e)}, \textbf{(f)} Experimentally measured magnetotransport data at high magnetic fields and \SI{10}{mK}, with zoomed-in sections shown in Fig.~4a, b. Although both QPCs have the same \SI{75}{nm} lithographic width, they exhibit different subband spacings, qualitatively matching the single-particle calculations in \textbf{(c)} and \textbf{(d)}. Possible explanations for this behavior are discussed in the main text.  }
	\label{Fig:SupHighmagnetic}
\end{figure}

In this section, we present the high-magnetic-field transconductance data from Figs.~4a,b, and compare it with theoretical predictions. The corresponding band structures for channel widths of \SI{30}{nm} and \SI{50}{nm} are shown in Figs.~\ref{Fig:SupHighmagnetic}a,b. Figs.~\ref{Fig:SupHighmagnetic}c,d depict the single-particle calculations, which qualitatively match the measured magnetotransport data in Figs.~\ref{Fig:SupHighmagnetic}e,f. By analyzing the measured magnetic field dependence, we can confidently assign all observed states in the measurements, confirming that the subband spacing in Fig.~\ref{Fig:SupHighmagnetic}f is smaller than in Fig.~\ref{Fig:SupHighmagnetic}e.

The \SI{10}{mK} measurement temperature improves the resolution of each subband state. Both experimental data and theoretical predictions consistently show that the previously observed 3+1 feature evolves into a ``2+1+1" splitting at higher magnetic fields.   Additionally, we observe that in a high magnetic field the $2\mathrm{K}^-\uparrow\ $ splits off from the $2\mathrm{K}^-\downarrow\ $/$4\mathrm{K}^+$ branch in a similar way to the $1\mathrm{K}^-\uparrow\ $ state from $1\mathrm{K}^-\downarrow\ $/$3\mathrm{K}^+$. This increased magnetic field for the second split-off is also captured in the model. Furthermore, the model correctly predicts the trend, that the magnetic split-off field decreases for configurations with lower subband spacing.

This ``3+1" feature enables an unambiguous assignment of all observed quantum states. This characteristic feature, evident in both experiment and theory, arises from the energetic separation of the \( 1\mathrm{K}^-\uparrow \) state from the closely spaced \( 1\mathrm{K}^-\downarrow \)/\( 3\mathrm{K}^+ \) branch. This pattern results from the interplay between SOC and the spin--Zeeman effect in opposite valleys. In the absence of SOC, the states would organize into two Zeeman-split pairs, as shown in SI~\ref{app:Rashba} Figs.~\ref{fig:SupModelL30} and \ref{fig:SupModelL50}. However, the presence of strong SOC induces opposite spin splittings in opposite valleys. For the 1$\mathrm{K}^-$ states, SOC and the spin--Zeeman effect act in the same direction, thereby enhancing the energy separation between 1$\mathrm{K}^-\uparrow$ and 1$\mathrm{K}^-\downarrow$, leading to a pronounced spin split-off state at high magnetic field. In contrast, for the 3$\mathrm{K}^+$ states, SOC counteracts the spin--Zeeman effect, which reduces the energy gap between 3$\mathrm{K}^+\uparrow$ and 3$\mathrm{K}^+\downarrow$ as the field increases---ultimately giving rise to the observed 3+1 state pattern. The observation of this feature in both calculated and measured magnetic field spectra provides compelling evidence for opposite spin orientations in states associated with the same subband and valley index. The appearance of this feature in higher branches allows for a recursive assignment of quantum numbers to all observed states. By tracing the \( \mathrm{K}^- \) and \( \mathrm{K}^+ \) states back to zero magnetic field, we identify the ground states as \( 1\mathrm{K}^+\uparrow \)/\( 1\mathrm{K}^-\downarrow \), and the excited states as \( 1\mathrm{K}^-\uparrow \)/\( 1\mathrm{K}^+\downarrow \), in agreement with theoretical calculations. The energy gap between these states thus corresponds to the spin--orbit gap, characterized by spin--valley--Zeeman SOC.

Overall, the strong agreement between theory and experiment highlights the model's robustness and a strong understanding of this system.

\begin{thebibliography}{57}%
\makeatletter
\providecommand \@ifxundefined [1]{%
 \@ifx{#1\undefined}
}%
\providecommand \@ifnum [1]{%
 \ifnum #1\expandafter \@firstoftwo
 \else \expandafter \@secondoftwo
 \fi
}%
\providecommand \@ifx [1]{%
 \ifx #1\expandafter \@firstoftwo
 \else \expandafter \@secondoftwo
 \fi
}%
\providecommand \natexlab [1]{#1}%
\providecommand \enquote  [1]{``#1''}%
\providecommand \bibnamefont  [1]{#1}%
\providecommand \bibfnamefont [1]{#1}%
\providecommand \citenamefont [1]{#1}%
\providecommand \href@noop [0]{\@secondoftwo}%
\providecommand \href [0]{\begingroup \@sanitize@url \@href}%
\providecommand \@href[1]{\@@startlink{#1}\@@href}%
\providecommand \@@href[1]{\endgroup#1\@@endlink}%
\providecommand \@sanitize@url [0]{\catcode `\\12\catcode `\$12\catcode `\&12\catcode `\#12\catcode `\^12\catcode `\_12\catcode `\%12\relax}%
\providecommand \@@startlink[1]{}%
\providecommand \@@endlink[0]{}%
\providecommand \url  [0]{\begingroup\@sanitize@url \@url }%
\providecommand \@url [1]{\endgroup\@href {#1}{\urlprefix }}%
\providecommand \urlprefix  [0]{URL }%
\providecommand \Eprint [0]{\href }%
\providecommand \doibase [0]{https://doi.org/}%
\providecommand \selectlanguage [0]{\@gobble}%
\providecommand \bibinfo  [0]{\@secondoftwo}%
\providecommand \bibfield  [0]{\@secondoftwo}%
\providecommand \translation [1]{[#1]}%
\providecommand \BibitemOpen [0]{}%
\providecommand \bibitemStop [0]{}%
\providecommand \bibitemNoStop [0]{.\EOS\space}%
\providecommand \EOS [0]{\spacefactor3000\relax}%
\providecommand \BibitemShut  [1]{\csname bibitem#1\endcsname}%
\let\auto@bib@innerbib\@empty
\bibitem [{\citenamefont {Oostinga}\ \emph {et~al.}(2008)\citenamefont {Oostinga}, \citenamefont {Heersche}, \citenamefont {Liu}, \citenamefont {Morpurgo},\ and\ \citenamefont {Vandersypen}}]{Oostinga2008}%
  \BibitemOpen
  \bibfield  {author} {\bibinfo {author} {\bibfnamefont {J.~B.}\ \bibnamefont {Oostinga}}, \bibinfo {author} {\bibfnamefont {H.~B.}\ \bibnamefont {Heersche}}, \bibinfo {author} {\bibfnamefont {X.}~\bibnamefont {Liu}}, \bibinfo {author} {\bibfnamefont {A.~F.}\ \bibnamefont {Morpurgo}},\ and\ \bibinfo {author} {\bibfnamefont {L.~M.~K.}\ \bibnamefont {Vandersypen}},\ }\bibfield  {title} {\bibinfo {title} {Gate-induced insulating state in bilayer graphene devices},\ }\href {https://doi.org/10.1038/nmat2082} {\bibfield  {journal} {\bibinfo  {journal} {Nature Materials}\ }\textbf {\bibinfo {volume} {7}},\ \bibinfo {pages} {151} (\bibinfo {year} {2008})}\BibitemShut {NoStop}%
\bibitem [{\citenamefont {Wang}\ \emph {et~al.}(2013)\citenamefont {Wang}, \citenamefont {Meric}, \citenamefont {Huang}, \citenamefont {Gao}, \citenamefont {Gao}, \citenamefont {Tran}, \citenamefont {Taniguchi}, \citenamefont {Watanabe}, \citenamefont {Campos}, \citenamefont {Muller}, \citenamefont {Guo}, \citenamefont {Kim}, \citenamefont {Hone}, \citenamefont {Shepard},\ and\ \citenamefont {Dean}}]{Wang2013}%
  \BibitemOpen
  \bibfield  {author} {\bibinfo {author} {\bibfnamefont {L.}~\bibnamefont {Wang}}, \bibinfo {author} {\bibfnamefont {I.}~\bibnamefont {Meric}}, \bibinfo {author} {\bibfnamefont {P.~Y.}\ \bibnamefont {Huang}}, \bibinfo {author} {\bibfnamefont {Q.}~\bibnamefont {Gao}}, \bibinfo {author} {\bibfnamefont {Y.}~\bibnamefont {Gao}}, \bibinfo {author} {\bibfnamefont {H.}~\bibnamefont {Tran}}, \bibinfo {author} {\bibfnamefont {T.}~\bibnamefont {Taniguchi}}, \bibinfo {author} {\bibfnamefont {K.}~\bibnamefont {Watanabe}}, \bibinfo {author} {\bibfnamefont {L.~M.}\ \bibnamefont {Campos}}, \bibinfo {author} {\bibfnamefont {D.~A.}\ \bibnamefont {Muller}}, \bibinfo {author} {\bibfnamefont {J.}~\bibnamefont {Guo}}, \bibinfo {author} {\bibfnamefont {P.}~\bibnamefont {Kim}}, \bibinfo {author} {\bibfnamefont {J.}~\bibnamefont {Hone}}, \bibinfo {author} {\bibfnamefont {K.~L.}\ \bibnamefont {Shepard}},\ and\ \bibinfo {author} {\bibfnamefont {C.~R.}\ \bibnamefont {Dean}},\ }\bibfield  {title} {\bibinfo {title} {One-dimensional
  electrical contact to a two-dimensional material},\ }\href {https://doi.org/10.1126/science.1244358} {\bibfield  {journal} {\bibinfo  {journal} {Science}\ }\textbf {\bibinfo {volume} {342}},\ \bibinfo {pages} {614} (\bibinfo {year} {2013})}\BibitemShut {NoStop}%
\bibitem [{\citenamefont {Trauzettel}\ \emph {et~al.}(2007)\citenamefont {Trauzettel}, \citenamefont {Bulaev}, \citenamefont {Loss},\ and\ \citenamefont {Burkard}}]{Trauzettel2007}%
  \BibitemOpen
  \bibfield  {author} {\bibinfo {author} {\bibfnamefont {B.}~\bibnamefont {Trauzettel}}, \bibinfo {author} {\bibfnamefont {D.~V.}\ \bibnamefont {Bulaev}}, \bibinfo {author} {\bibfnamefont {D.}~\bibnamefont {Loss}},\ and\ \bibinfo {author} {\bibfnamefont {G.}~\bibnamefont {Burkard}},\ }\bibfield  {title} {\bibinfo {title} {Spin qubits in graphene quantum dots},\ }\href {https://doi.org/10.1038/nphys544} {\bibfield  {journal} {\bibinfo  {journal} {Nature Physics}\ }\textbf {\bibinfo {volume} {3}},\ \bibinfo {pages} {192} (\bibinfo {year} {2007})}\BibitemShut {NoStop}%
\bibitem [{\citenamefont {Denisov}\ \emph {et~al.}(2025)\citenamefont {Denisov}, \citenamefont {Reckova}, \citenamefont {Cances}, \citenamefont {Ruckriegel}, \citenamefont {Masseroni}, \citenamefont {Adam}, \citenamefont {Tong}, \citenamefont {Gerber}, \citenamefont {Huang}, \citenamefont {Watanabe}, \citenamefont {Taniguchi}, \citenamefont {Ihn}, \citenamefont {Ensslin},\ and\ \citenamefont {Duprez}}]{Denisov2025}%
  \BibitemOpen
  \bibfield  {author} {\bibinfo {author} {\bibfnamefont {A.~O.}\ \bibnamefont {Denisov}}, \bibinfo {author} {\bibfnamefont {V.}~\bibnamefont {Reckova}}, \bibinfo {author} {\bibfnamefont {S.}~\bibnamefont {Cances}}, \bibinfo {author} {\bibfnamefont {M.~J.}\ \bibnamefont {Ruckriegel}}, \bibinfo {author} {\bibfnamefont {M.}~\bibnamefont {Masseroni}}, \bibinfo {author} {\bibfnamefont {C.}~\bibnamefont {Adam}}, \bibinfo {author} {\bibfnamefont {C.}~\bibnamefont {Tong}}, \bibinfo {author} {\bibfnamefont {J.~D.}\ \bibnamefont {Gerber}}, \bibinfo {author} {\bibfnamefont {W.~W.}\ \bibnamefont {Huang}}, \bibinfo {author} {\bibfnamefont {K.}~\bibnamefont {Watanabe}}, \bibinfo {author} {\bibfnamefont {T.}~\bibnamefont {Taniguchi}}, \bibinfo {author} {\bibfnamefont {T.}~\bibnamefont {Ihn}}, \bibinfo {author} {\bibfnamefont {K.}~\bibnamefont {Ensslin}},\ and\ \bibinfo {author} {\bibfnamefont {H.}~\bibnamefont {Duprez}},\ }\bibfield  {title} {\bibinfo {title} {Spin--valley protected kramers pair in bilayer graphene},\
  }\bibfield  {journal} {\bibinfo  {journal} {Nature Nanotechnology}\ }\href {https://doi.org/10.1038/s41565-025-01858-8} {10.1038/s41565-025-01858-8} (\bibinfo {year} {2025})\BibitemShut {NoStop}%
\bibitem [{\citenamefont {Garreis}\ \emph {et~al.}(2024)\citenamefont {Garreis}, \citenamefont {Tong}, \citenamefont {Terle}, \citenamefont {Ruckriegel}, \citenamefont {Gerber}, \citenamefont {G{\"a}chter}, \citenamefont {Watanabe}, \citenamefont {Taniguchi}, \citenamefont {Ihn}, \citenamefont {Ensslin},\ and\ \citenamefont {Huang}}]{Garreis2024}%
  \BibitemOpen
  \bibfield  {author} {\bibinfo {author} {\bibfnamefont {R.}~\bibnamefont {Garreis}}, \bibinfo {author} {\bibfnamefont {C.}~\bibnamefont {Tong}}, \bibinfo {author} {\bibfnamefont {J.}~\bibnamefont {Terle}}, \bibinfo {author} {\bibfnamefont {M.~J.}\ \bibnamefont {Ruckriegel}}, \bibinfo {author} {\bibfnamefont {J.~D.}\ \bibnamefont {Gerber}}, \bibinfo {author} {\bibfnamefont {L.~M.}\ \bibnamefont {G{\"a}chter}}, \bibinfo {author} {\bibfnamefont {K.}~\bibnamefont {Watanabe}}, \bibinfo {author} {\bibfnamefont {T.}~\bibnamefont {Taniguchi}}, \bibinfo {author} {\bibfnamefont {T.}~\bibnamefont {Ihn}}, \bibinfo {author} {\bibfnamefont {K.}~\bibnamefont {Ensslin}},\ and\ \bibinfo {author} {\bibfnamefont {W.~W.}\ \bibnamefont {Huang}},\ }\bibfield  {title} {\bibinfo {title} {Long-lived valley states in bilayer graphene quantum dots},\ }\href {https://doi.org/10.1038/s41567-023-02334-7} {\bibfield  {journal} {\bibinfo  {journal} {Nature Physics}\ }\textbf {\bibinfo {volume} {20}},\ \bibinfo {pages} {428} (\bibinfo
  {year} {2024})}\BibitemShut {NoStop}%
\bibitem [{\citenamefont {G\"achter}\ \emph {et~al.}(2022)\citenamefont {G\"achter}, \citenamefont {Garreis}, \citenamefont {Gerber}, \citenamefont {Ruckriegel}, \citenamefont {Tong}, \citenamefont {Kratochwil}, \citenamefont {de~Vries}, \citenamefont {Kurzmann}, \citenamefont {Watanabe}, \citenamefont {Taniguchi}, \citenamefont {Ihn}, \citenamefont {Ensslin},\ and\ \citenamefont {Huang}}]{gachter_single_2022}%
  \BibitemOpen
  \bibfield  {author} {\bibinfo {author} {\bibfnamefont {L.~M.}\ \bibnamefont {G\"achter}}, \bibinfo {author} {\bibfnamefont {R.}~\bibnamefont {Garreis}}, \bibinfo {author} {\bibfnamefont {J.~D.}\ \bibnamefont {Gerber}}, \bibinfo {author} {\bibfnamefont {M.~J.}\ \bibnamefont {Ruckriegel}}, \bibinfo {author} {\bibfnamefont {C.}~\bibnamefont {Tong}}, \bibinfo {author} {\bibfnamefont {B.}~\bibnamefont {Kratochwil}}, \bibinfo {author} {\bibfnamefont {F.~K.}\ \bibnamefont {de~Vries}}, \bibinfo {author} {\bibfnamefont {A.}~\bibnamefont {Kurzmann}}, \bibinfo {author} {\bibfnamefont {K.}~\bibnamefont {Watanabe}}, \bibinfo {author} {\bibfnamefont {T.}~\bibnamefont {Taniguchi}}, \bibinfo {author} {\bibfnamefont {T.}~\bibnamefont {Ihn}}, \bibinfo {author} {\bibfnamefont {K.}~\bibnamefont {Ensslin}},\ and\ \bibinfo {author} {\bibfnamefont {W.~W.}\ \bibnamefont {Huang}},\ }\bibfield  {title} {\bibinfo {title} {Single-shot spin readout in graphene quantum dots},\ }\href {https://doi.org/10.1103/PRXQuantum.3.020343}
  {\bibfield  {journal} {\bibinfo  {journal} {PRX Quantum}\ }\textbf {\bibinfo {volume} {3}},\ \bibinfo {pages} {020343} (\bibinfo {year} {2022})}\BibitemShut {NoStop}%
\bibitem [{\citenamefont {Banszerus}\ \emph {et~al.}(2022)\citenamefont {Banszerus}, \citenamefont {Hecker}, \citenamefont {M{\"o}ller}, \citenamefont {Icking}, \citenamefont {Watanabe}, \citenamefont {Taniguchi}, \citenamefont {Volk},\ and\ \citenamefont {Stampfer}}]{Banszerus2022}%
  \BibitemOpen
  \bibfield  {author} {\bibinfo {author} {\bibfnamefont {L.}~\bibnamefont {Banszerus}}, \bibinfo {author} {\bibfnamefont {K.}~\bibnamefont {Hecker}}, \bibinfo {author} {\bibfnamefont {S.}~\bibnamefont {M{\"o}ller}}, \bibinfo {author} {\bibfnamefont {E.}~\bibnamefont {Icking}}, \bibinfo {author} {\bibfnamefont {K.}~\bibnamefont {Watanabe}}, \bibinfo {author} {\bibfnamefont {T.}~\bibnamefont {Taniguchi}}, \bibinfo {author} {\bibfnamefont {C.}~\bibnamefont {Volk}},\ and\ \bibinfo {author} {\bibfnamefont {C.}~\bibnamefont {Stampfer}},\ }\bibfield  {title} {\bibinfo {title} {Spin relaxation in a single-electron graphene quantum dot},\ }\href {https://doi.org/10.1038/s41467-022-31231-5} {\bibfield  {journal} {\bibinfo  {journal} {Nature Communications}\ }\textbf {\bibinfo {volume} {13}},\ \bibinfo {pages} {3637} (\bibinfo {year} {2022})}\BibitemShut {NoStop}%
\bibitem [{\citenamefont {Banszerus}\ \emph {et~al.}(2025)\citenamefont {Banszerus}, \citenamefont {Hecker}, \citenamefont {Wang}, \citenamefont {M\"oller}, \citenamefont {Watanabe}, \citenamefont {Taniguchi}, \citenamefont {Burkard}, \citenamefont {Volk},\ and\ \citenamefont {Stampfer}}]{banszerus2024phononlimited}%
  \BibitemOpen
  \bibfield  {author} {\bibinfo {author} {\bibfnamefont {L.}~\bibnamefont {Banszerus}}, \bibinfo {author} {\bibfnamefont {K.}~\bibnamefont {Hecker}}, \bibinfo {author} {\bibfnamefont {L.}~\bibnamefont {Wang}}, \bibinfo {author} {\bibfnamefont {S.}~\bibnamefont {M\"oller}}, \bibinfo {author} {\bibfnamefont {K.}~\bibnamefont {Watanabe}}, \bibinfo {author} {\bibfnamefont {T.}~\bibnamefont {Taniguchi}}, \bibinfo {author} {\bibfnamefont {G.}~\bibnamefont {Burkard}}, \bibinfo {author} {\bibfnamefont {C.}~\bibnamefont {Volk}},\ and\ \bibinfo {author} {\bibfnamefont {C.}~\bibnamefont {Stampfer}},\ }\bibfield  {title} {\bibinfo {title} {Phonon-limited valley lifetimes in single-particle bilayer graphene quantum dots},\ }\href {https://doi.org/10.1103/dkgn-pfjb} {\bibfield  {journal} {\bibinfo  {journal} {Phys. Rev. B}\ }\textbf {\bibinfo {volume} {112}},\ \bibinfo {pages} {035409} (\bibinfo {year} {2025})}\BibitemShut {NoStop}%
\bibitem [{\citenamefont {Hendrickx}\ \emph {et~al.}(2020)\citenamefont {Hendrickx}, \citenamefont {Franke}, \citenamefont {Sammak}, \citenamefont {Scappucci},\ and\ \citenamefont {Veldhorst}}]{Hendrickx2020}%
  \BibitemOpen
  \bibfield  {author} {\bibinfo {author} {\bibfnamefont {N.~W.}\ \bibnamefont {Hendrickx}}, \bibinfo {author} {\bibfnamefont {D.~P.}\ \bibnamefont {Franke}}, \bibinfo {author} {\bibfnamefont {A.}~\bibnamefont {Sammak}}, \bibinfo {author} {\bibfnamefont {G.}~\bibnamefont {Scappucci}},\ and\ \bibinfo {author} {\bibfnamefont {M.}~\bibnamefont {Veldhorst}},\ }\bibfield  {title} {\bibinfo {title} {Fast two-qubit logic with holes in germanium},\ }\href {https://doi.org/10.1038/s41586-019-1919-3} {\bibfield  {journal} {\bibinfo  {journal} {Nature}\ }\textbf {\bibinfo {volume} {577}},\ \bibinfo {pages} {487} (\bibinfo {year} {2020})}\BibitemShut {NoStop}%
\bibitem [{\citenamefont {Gmitra}\ and\ \citenamefont {Fabian}(2017)}]{GmitraFabian2017}%
  \BibitemOpen
  \bibfield  {author} {\bibinfo {author} {\bibfnamefont {M.}~\bibnamefont {Gmitra}}\ and\ \bibinfo {author} {\bibfnamefont {J.}~\bibnamefont {Fabian}},\ }\bibfield  {title} {\bibinfo {title} {Proximity effects in bilayer graphene on monolayer {${\mathrm{WSe}}_{2}$}: Field-effect spin valley locking, spin-orbit valve, and spin transistor},\ }\href {https://doi.org/10.1103/PhysRevLett.119.146401} {\bibfield  {journal} {\bibinfo  {journal} {Phys. Rev. Lett.}\ }\textbf {\bibinfo {volume} {119}},\ \bibinfo {pages} {146401} (\bibinfo {year} {2017})}\BibitemShut {NoStop}%
\bibitem [{\citenamefont {Rycerz}\ \emph {et~al.}(2007)\citenamefont {Rycerz}, \citenamefont {Tworzyd{\l}o},\ and\ \citenamefont {Beenakker}}]{Rycerz2007}%
  \BibitemOpen
  \bibfield  {author} {\bibinfo {author} {\bibfnamefont {A.}~\bibnamefont {Rycerz}}, \bibinfo {author} {\bibfnamefont {J.}~\bibnamefont {Tworzyd{\l}o}},\ and\ \bibinfo {author} {\bibfnamefont {C.~W.~J.}\ \bibnamefont {Beenakker}},\ }\bibfield  {title} {\bibinfo {title} {Valley filter and valley valve in graphene},\ }\href {https://doi.org/10.1038/nphys547} {\bibfield  {journal} {\bibinfo  {journal} {Nature Physics}\ }\textbf {\bibinfo {volume} {3}},\ \bibinfo {pages} {172} (\bibinfo {year} {2007})}\BibitemShut {NoStop}%
\bibitem [{\citenamefont {Sui}\ \emph {et~al.}(2015)\citenamefont {Sui}, \citenamefont {Chen}, \citenamefont {Ma}, \citenamefont {Shan}, \citenamefont {Tian}, \citenamefont {Watanabe}, \citenamefont {Taniguchi}, \citenamefont {Jin}, \citenamefont {Yao}, \citenamefont {Xiao},\ and\ \citenamefont {Zhang}}]{Sui2015}%
  \BibitemOpen
  \bibfield  {author} {\bibinfo {author} {\bibfnamefont {M.}~\bibnamefont {Sui}}, \bibinfo {author} {\bibfnamefont {G.}~\bibnamefont {Chen}}, \bibinfo {author} {\bibfnamefont {L.}~\bibnamefont {Ma}}, \bibinfo {author} {\bibfnamefont {W.-Y.}\ \bibnamefont {Shan}}, \bibinfo {author} {\bibfnamefont {D.}~\bibnamefont {Tian}}, \bibinfo {author} {\bibfnamefont {K.}~\bibnamefont {Watanabe}}, \bibinfo {author} {\bibfnamefont {T.}~\bibnamefont {Taniguchi}}, \bibinfo {author} {\bibfnamefont {X.}~\bibnamefont {Jin}}, \bibinfo {author} {\bibfnamefont {W.}~\bibnamefont {Yao}}, \bibinfo {author} {\bibfnamefont {D.}~\bibnamefont {Xiao}},\ and\ \bibinfo {author} {\bibfnamefont {Y.}~\bibnamefont {Zhang}},\ }\bibfield  {title} {\bibinfo {title} {Gate-tunable topological valley transport in bilayer graphene},\ }\href {https://doi.org/10.1038/nphys3485} {\bibfield  {journal} {\bibinfo  {journal} {Nature Physics}\ }\textbf {\bibinfo {volume} {11}},\ \bibinfo {pages} {1027} (\bibinfo {year} {2015})}\BibitemShut {NoStop}%
\bibitem [{\citenamefont {Sichau}\ \emph {et~al.}(2019)\citenamefont {Sichau}, \citenamefont {Prada}, \citenamefont {Anlauf}, \citenamefont {Lyon}, \citenamefont {Bosnjak}, \citenamefont {Tiemann},\ and\ \citenamefont {Blick}}]{SichauSOCMicrowave}%
  \BibitemOpen
  \bibfield  {author} {\bibinfo {author} {\bibfnamefont {J.}~\bibnamefont {Sichau}}, \bibinfo {author} {\bibfnamefont {M.}~\bibnamefont {Prada}}, \bibinfo {author} {\bibfnamefont {T.}~\bibnamefont {Anlauf}}, \bibinfo {author} {\bibfnamefont {T.~J.}\ \bibnamefont {Lyon}}, \bibinfo {author} {\bibfnamefont {B.}~\bibnamefont {Bosnjak}}, \bibinfo {author} {\bibfnamefont {L.}~\bibnamefont {Tiemann}},\ and\ \bibinfo {author} {\bibfnamefont {R.~H.}\ \bibnamefont {Blick}},\ }\bibfield  {title} {\bibinfo {title} {Resonance microwave measurements of an intrinsic spin-orbit coupling gap in graphene: A possible indication of a topological state},\ }\href {https://doi.org/10.1103/PhysRevLett.122.046403} {\bibfield  {journal} {\bibinfo  {journal} {Phys. Rev. Lett.}\ }\textbf {\bibinfo {volume} {122}},\ \bibinfo {pages} {046403} (\bibinfo {year} {2019})}\BibitemShut {NoStop}%
\bibitem [{\citenamefont {Banszerus}\ \emph {et~al.}(2020)\citenamefont {Banszerus}, \citenamefont {Frohn}, \citenamefont {Fabian}, \citenamefont {Somanchi}, \citenamefont {Epping}, \citenamefont {M\"uller}, \citenamefont {Neumaier}, \citenamefont {Watanabe}, \citenamefont {Taniguchi}, \citenamefont {Libisch}, \citenamefont {Beschoten}, \citenamefont {Hassler},\ and\ \citenamefont {Stampfer}}]{Banszerus_SOC_QPC}%
  \BibitemOpen
  \bibfield  {author} {\bibinfo {author} {\bibfnamefont {L.}~\bibnamefont {Banszerus}}, \bibinfo {author} {\bibfnamefont {B.}~\bibnamefont {Frohn}}, \bibinfo {author} {\bibfnamefont {T.}~\bibnamefont {Fabian}}, \bibinfo {author} {\bibfnamefont {S.}~\bibnamefont {Somanchi}}, \bibinfo {author} {\bibfnamefont {A.}~\bibnamefont {Epping}}, \bibinfo {author} {\bibfnamefont {M.}~\bibnamefont {M\"uller}}, \bibinfo {author} {\bibfnamefont {D.}~\bibnamefont {Neumaier}}, \bibinfo {author} {\bibfnamefont {K.}~\bibnamefont {Watanabe}}, \bibinfo {author} {\bibfnamefont {T.}~\bibnamefont {Taniguchi}}, \bibinfo {author} {\bibfnamefont {F.}~\bibnamefont {Libisch}}, \bibinfo {author} {\bibfnamefont {B.}~\bibnamefont {Beschoten}}, \bibinfo {author} {\bibfnamefont {F.}~\bibnamefont {Hassler}},\ and\ \bibinfo {author} {\bibfnamefont {C.}~\bibnamefont {Stampfer}},\ }\bibfield  {title} {\bibinfo {title} {Observation of the spin-orbit gap in bilayer graphene by one-dimensional ballistic transport},\ }\href
  {https://doi.org/10.1103/PhysRevLett.124.177701} {\bibfield  {journal} {\bibinfo  {journal} {Phys. Rev. Lett.}\ }\textbf {\bibinfo {volume} {124}},\ \bibinfo {pages} {177701} (\bibinfo {year} {2020})}\BibitemShut {NoStop}%
\bibitem [{\citenamefont {Banszerus}\ \emph {et~al.}(2021)\citenamefont {Banszerus}, \citenamefont {M{\"o}ller}, \citenamefont {Steiner}, \citenamefont {Icking}, \citenamefont {Trellenkamp}, \citenamefont {Lentz}, \citenamefont {Watanabe}, \citenamefont {Taniguchi}, \citenamefont {Volk},\ and\ \citenamefont {Stampfer}}]{Banszerus2021}%
  \BibitemOpen
  \bibfield  {author} {\bibinfo {author} {\bibfnamefont {L.}~\bibnamefont {Banszerus}}, \bibinfo {author} {\bibfnamefont {S.}~\bibnamefont {M{\"o}ller}}, \bibinfo {author} {\bibfnamefont {C.}~\bibnamefont {Steiner}}, \bibinfo {author} {\bibfnamefont {E.}~\bibnamefont {Icking}}, \bibinfo {author} {\bibfnamefont {S.}~\bibnamefont {Trellenkamp}}, \bibinfo {author} {\bibfnamefont {F.}~\bibnamefont {Lentz}}, \bibinfo {author} {\bibfnamefont {K.}~\bibnamefont {Watanabe}}, \bibinfo {author} {\bibfnamefont {T.}~\bibnamefont {Taniguchi}}, \bibinfo {author} {\bibfnamefont {C.}~\bibnamefont {Volk}},\ and\ \bibinfo {author} {\bibfnamefont {C.}~\bibnamefont {Stampfer}},\ }\bibfield  {title} {\bibinfo {title} {Spin-valley coupling in single-electron bilayer graphene quantum dots},\ }\href {https://doi.org/10.1038/s41467-021-25498-3} {\bibfield  {journal} {\bibinfo  {journal} {Nature Communications}\ }\textbf {\bibinfo {volume} {12}},\ \bibinfo {pages} {5250} (\bibinfo {year} {2021})}\BibitemShut {NoStop}%
\bibitem [{\citenamefont {Kurzmann}\ \emph {et~al.}(2021)\citenamefont {Kurzmann}, \citenamefont {Kleeorin}, \citenamefont {Tong}, \citenamefont {Garreis}, \citenamefont {Knothe}, \citenamefont {Eich}, \citenamefont {Mittag}, \citenamefont {Gold}, \citenamefont {de~Vries}, \citenamefont {Watanabe}, \citenamefont {Taniguchi}, \citenamefont {Fal'ko}, \citenamefont {Meir}, \citenamefont {Ihn},\ and\ \citenamefont {Ensslin}}]{Kurzmann2021_Kondo}%
  \BibitemOpen
  \bibfield  {author} {\bibinfo {author} {\bibfnamefont {A.}~\bibnamefont {Kurzmann}}, \bibinfo {author} {\bibfnamefont {Y.}~\bibnamefont {Kleeorin}}, \bibinfo {author} {\bibfnamefont {C.}~\bibnamefont {Tong}}, \bibinfo {author} {\bibfnamefont {R.}~\bibnamefont {Garreis}}, \bibinfo {author} {\bibfnamefont {A.}~\bibnamefont {Knothe}}, \bibinfo {author} {\bibfnamefont {M.}~\bibnamefont {Eich}}, \bibinfo {author} {\bibfnamefont {C.}~\bibnamefont {Mittag}}, \bibinfo {author} {\bibfnamefont {C.}~\bibnamefont {Gold}}, \bibinfo {author} {\bibfnamefont {F.~K.}\ \bibnamefont {de~Vries}}, \bibinfo {author} {\bibfnamefont {K.}~\bibnamefont {Watanabe}}, \bibinfo {author} {\bibfnamefont {T.}~\bibnamefont {Taniguchi}}, \bibinfo {author} {\bibfnamefont {V.}~\bibnamefont {Fal'ko}}, \bibinfo {author} {\bibfnamefont {Y.}~\bibnamefont {Meir}}, \bibinfo {author} {\bibfnamefont {T.}~\bibnamefont {Ihn}},\ and\ \bibinfo {author} {\bibfnamefont {K.}~\bibnamefont {Ensslin}},\ }\bibfield  {title} {\bibinfo {title} {Kondo effect and
  spin--orbit coupling in graphene quantum dots},\ }\href {https://doi.org/10.1038/s41467-021-26149-3} {\bibfield  {journal} {\bibinfo  {journal} {Nature Communications}\ }\textbf {\bibinfo {volume} {12}},\ \bibinfo {pages} {6004} (\bibinfo {year} {2021})}\BibitemShut {NoStop}%
\bibitem [{\citenamefont {Duprez}\ \emph {et~al.}(2024)\citenamefont {Duprez}, \citenamefont {Cances}, \citenamefont {Omahen}, \citenamefont {Masseroni}, \citenamefont {Ruckriegel}, \citenamefont {Adam}, \citenamefont {Tong}, \citenamefont {Garreis}, \citenamefont {Gerber}, \citenamefont {Huang}, \citenamefont {G{\"a}chter}, \citenamefont {Watanabe}, \citenamefont {Taniguchi}, \citenamefont {Ihn},\ and\ \citenamefont {Ensslin}}]{Duprez2024}%
  \BibitemOpen
  \bibfield  {author} {\bibinfo {author} {\bibfnamefont {H.}~\bibnamefont {Duprez}}, \bibinfo {author} {\bibfnamefont {S.}~\bibnamefont {Cances}}, \bibinfo {author} {\bibfnamefont {A.}~\bibnamefont {Omahen}}, \bibinfo {author} {\bibfnamefont {M.}~\bibnamefont {Masseroni}}, \bibinfo {author} {\bibfnamefont {M.~J.}\ \bibnamefont {Ruckriegel}}, \bibinfo {author} {\bibfnamefont {C.}~\bibnamefont {Adam}}, \bibinfo {author} {\bibfnamefont {C.}~\bibnamefont {Tong}}, \bibinfo {author} {\bibfnamefont {R.}~\bibnamefont {Garreis}}, \bibinfo {author} {\bibfnamefont {J.~D.}\ \bibnamefont {Gerber}}, \bibinfo {author} {\bibfnamefont {W.}~\bibnamefont {Huang}}, \bibinfo {author} {\bibfnamefont {L.}~\bibnamefont {G{\"a}chter}}, \bibinfo {author} {\bibfnamefont {K.}~\bibnamefont {Watanabe}}, \bibinfo {author} {\bibfnamefont {T.}~\bibnamefont {Taniguchi}}, \bibinfo {author} {\bibfnamefont {T.}~\bibnamefont {Ihn}},\ and\ \bibinfo {author} {\bibfnamefont {K.}~\bibnamefont {Ensslin}},\ }\bibfield  {title} {\bibinfo {title}
  {Spin-valley locked excited states spectroscopy in a one-particle bilayer graphene quantum dot},\ }\href {https://doi.org/10.1038/s41467-024-54121-4} {\bibfield  {journal} {\bibinfo  {journal} {Nature Communications}\ }\textbf {\bibinfo {volume} {15}},\ \bibinfo {pages} {9717} (\bibinfo {year} {2024})}\BibitemShut {NoStop}%
\bibitem [{\citenamefont {Kane}\ and\ \citenamefont {Mele}(2005)}]{PhysRevLett.KanMeleSOC}%
  \BibitemOpen
  \bibfield  {author} {\bibinfo {author} {\bibfnamefont {C.~L.}\ \bibnamefont {Kane}}\ and\ \bibinfo {author} {\bibfnamefont {E.~J.}\ \bibnamefont {Mele}},\ }\bibfield  {title} {\bibinfo {title} {Quantum spin hall effect in graphene},\ }\href {https://doi.org/10.1103/PhysRevLett.95.226801} {\bibfield  {journal} {\bibinfo  {journal} {Phys. Rev. Lett.}\ }\textbf {\bibinfo {volume} {95}},\ \bibinfo {pages} {226801} (\bibinfo {year} {2005})}\BibitemShut {NoStop}%
\bibitem [{\citenamefont {Konschuh}\ \emph {et~al.}(2012)\citenamefont {Konschuh}, \citenamefont {Gmitra}, \citenamefont {Kochan},\ and\ \citenamefont {Fabian}}]{Konschuh2007}%
  \BibitemOpen
  \bibfield  {author} {\bibinfo {author} {\bibfnamefont {S.}~\bibnamefont {Konschuh}}, \bibinfo {author} {\bibfnamefont {M.}~\bibnamefont {Gmitra}}, \bibinfo {author} {\bibfnamefont {D.}~\bibnamefont {Kochan}},\ and\ \bibinfo {author} {\bibfnamefont {J.}~\bibnamefont {Fabian}},\ }\bibfield  {title} {\bibinfo {title} {Theory of spin-orbit coupling in bilayer graphene},\ }\href {https://doi.org/10.1103/PhysRevB.85.115423} {\bibfield  {journal} {\bibinfo  {journal} {Phys. Rev. B}\ }\textbf {\bibinfo {volume} {85}},\ \bibinfo {pages} {115423} (\bibinfo {year} {2012})}\BibitemShut {NoStop}%
\bibitem [{\citenamefont {Gmitra}\ \emph {et~al.}(2016)\citenamefont {Gmitra}, \citenamefont {Kochan}, \citenamefont {H\"ogl},\ and\ \citenamefont {Fabian}}]{Gmitra2016}%
  \BibitemOpen
  \bibfield  {author} {\bibinfo {author} {\bibfnamefont {M.}~\bibnamefont {Gmitra}}, \bibinfo {author} {\bibfnamefont {D.}~\bibnamefont {Kochan}}, \bibinfo {author} {\bibfnamefont {P.}~\bibnamefont {H\"ogl}},\ and\ \bibinfo {author} {\bibfnamefont {J.}~\bibnamefont {Fabian}},\ }\bibfield  {title} {\bibinfo {title} {Trivial and inverted dirac bands and the emergence of quantum spin hall states in graphene on transition-metal dichalcogenides},\ }\href {https://doi.org/10.1103/PhysRevB.93.155104} {\bibfield  {journal} {\bibinfo  {journal} {Phys. Rev. B}\ }\textbf {\bibinfo {volume} {93}},\ \bibinfo {pages} {155104} (\bibinfo {year} {2016})}\BibitemShut {NoStop}%
\bibitem [{\citenamefont {David}\ \emph {et~al.}(2019)\citenamefont {David}, \citenamefont {Rakyta}, \citenamefont {Korm\'anyos},\ and\ \citenamefont {Burkard}}]{David2019}%
  \BibitemOpen
  \bibfield  {author} {\bibinfo {author} {\bibfnamefont {A.}~\bibnamefont {David}}, \bibinfo {author} {\bibfnamefont {P.}~\bibnamefont {Rakyta}}, \bibinfo {author} {\bibfnamefont {A.}~\bibnamefont {Korm\'anyos}},\ and\ \bibinfo {author} {\bibfnamefont {G.}~\bibnamefont {Burkard}},\ }\bibfield  {title} {\bibinfo {title} {Induced spin-orbit coupling in twisted graphene--transition metal dichalcogenide heterobilayers: Twistronics meets spintronics},\ }\href {https://doi.org/10.1103/PhysRevB.100.085412} {\bibfield  {journal} {\bibinfo  {journal} {Phys. Rev. B}\ }\textbf {\bibinfo {volume} {100}},\ \bibinfo {pages} {085412} (\bibinfo {year} {2019})}\BibitemShut {NoStop}%
\bibitem [{\citenamefont {Li}\ and\ \citenamefont {Koshino}(2019)}]{Li2019}%
  \BibitemOpen
  \bibfield  {author} {\bibinfo {author} {\bibfnamefont {Y.}~\bibnamefont {Li}}\ and\ \bibinfo {author} {\bibfnamefont {M.}~\bibnamefont {Koshino}},\ }\bibfield  {title} {\bibinfo {title} {Twist-angle dependence of the proximity spin-orbit coupling in graphene on transition-metal dichalcogenides},\ }\href {https://doi.org/10.1103/PhysRevB.99.075438} {\bibfield  {journal} {\bibinfo  {journal} {Phys. Rev. B}\ }\textbf {\bibinfo {volume} {99}},\ \bibinfo {pages} {075438} (\bibinfo {year} {2019})}\BibitemShut {NoStop}%
\bibitem [{\citenamefont {Avsar}\ \emph {et~al.}(2014)\citenamefont {Avsar}, \citenamefont {Tan}, \citenamefont {Taychatanapat}, \citenamefont {Balakrishnan}, \citenamefont {Koon}, \citenamefont {Yeo}, \citenamefont {Lahiri}, \citenamefont {Carvalho}, \citenamefont {Rodin}, \citenamefont {O'Farrell}, \citenamefont {Eda}, \citenamefont {Castro~Neto},\ and\ \citenamefont {{\"O}zyilmaz}}]{Avsar2014}%
  \BibitemOpen
  \bibfield  {author} {\bibinfo {author} {\bibfnamefont {A.}~\bibnamefont {Avsar}}, \bibinfo {author} {\bibfnamefont {J.~Y.}\ \bibnamefont {Tan}}, \bibinfo {author} {\bibfnamefont {T.}~\bibnamefont {Taychatanapat}}, \bibinfo {author} {\bibfnamefont {J.}~\bibnamefont {Balakrishnan}}, \bibinfo {author} {\bibfnamefont {G.~K.~W.}\ \bibnamefont {Koon}}, \bibinfo {author} {\bibfnamefont {Y.}~\bibnamefont {Yeo}}, \bibinfo {author} {\bibfnamefont {J.}~\bibnamefont {Lahiri}}, \bibinfo {author} {\bibfnamefont {A.}~\bibnamefont {Carvalho}}, \bibinfo {author} {\bibfnamefont {A.~S.}\ \bibnamefont {Rodin}}, \bibinfo {author} {\bibfnamefont {E.~C.~T.}\ \bibnamefont {O'Farrell}}, \bibinfo {author} {\bibfnamefont {G.}~\bibnamefont {Eda}}, \bibinfo {author} {\bibfnamefont {A.~H.}\ \bibnamefont {Castro~Neto}},\ and\ \bibinfo {author} {\bibfnamefont {B.}~\bibnamefont {{\"O}zyilmaz}},\ }\bibfield  {title} {\bibinfo {title} {Spin--orbit proximity effect in graphene},\ }\href {https://doi.org/10.1038/ncomms5875} {\bibfield
  {journal} {\bibinfo  {journal} {Nature Communications}\ }\textbf {\bibinfo {volume} {5}},\ \bibinfo {pages} {4875} (\bibinfo {year} {2014})}\BibitemShut {NoStop}%
\bibitem [{\citenamefont {Wang}\ \emph {et~al.}(2015)\citenamefont {Wang}, \citenamefont {Ki}, \citenamefont {Chen}, \citenamefont {Berger}, \citenamefont {MacDonald},\ and\ \citenamefont {Morpurgo}}]{Wang2015}%
  \BibitemOpen
  \bibfield  {author} {\bibinfo {author} {\bibfnamefont {Z.}~\bibnamefont {Wang}}, \bibinfo {author} {\bibfnamefont {D.-K.}\ \bibnamefont {Ki}}, \bibinfo {author} {\bibfnamefont {H.}~\bibnamefont {Chen}}, \bibinfo {author} {\bibfnamefont {H.}~\bibnamefont {Berger}}, \bibinfo {author} {\bibfnamefont {A.~H.}\ \bibnamefont {MacDonald}},\ and\ \bibinfo {author} {\bibfnamefont {A.~F.}\ \bibnamefont {Morpurgo}},\ }\bibfield  {title} {\bibinfo {title} {Strong interface-induced spin--orbit interaction in graphene on {WS}$_2$},\ }\href {https://doi.org/10.1038/ncomms9339} {\bibfield  {journal} {\bibinfo  {journal} {Nature Communications}\ }\textbf {\bibinfo {volume} {6}},\ \bibinfo {pages} {8339} (\bibinfo {year} {2015})}\BibitemShut {NoStop}%
\bibitem [{\citenamefont {Yang}\ \emph {et~al.}(2016)\citenamefont {Yang}, \citenamefont {Tu}, \citenamefont {Kim}, \citenamefont {Wu}, \citenamefont {Wang}, \citenamefont {Alicea}, \citenamefont {Wu}, \citenamefont {Bockrath},\ and\ \citenamefont {Shi}}]{Yang2016}%
  \BibitemOpen
  \bibfield  {author} {\bibinfo {author} {\bibfnamefont {B.}~\bibnamefont {Yang}}, \bibinfo {author} {\bibfnamefont {M.-F.}\ \bibnamefont {Tu}}, \bibinfo {author} {\bibfnamefont {J.}~\bibnamefont {Kim}}, \bibinfo {author} {\bibfnamefont {Y.}~\bibnamefont {Wu}}, \bibinfo {author} {\bibfnamefont {H.}~\bibnamefont {Wang}}, \bibinfo {author} {\bibfnamefont {J.}~\bibnamefont {Alicea}}, \bibinfo {author} {\bibfnamefont {R.}~\bibnamefont {Wu}}, \bibinfo {author} {\bibfnamefont {M.}~\bibnamefont {Bockrath}},\ and\ \bibinfo {author} {\bibfnamefont {J.}~\bibnamefont {Shi}},\ }\bibfield  {title} {\bibinfo {title} {Tunable spin–orbit coupling and symmetry-protected edge states in graphene/{WS}$_2$},\ }\href {https://doi.org/10.1088/2053-1583/3/3/031012} {\bibfield  {journal} {\bibinfo  {journal} {2D Materials}\ }\textbf {\bibinfo {volume} {3}},\ \bibinfo {pages} {031012} (\bibinfo {year} {2016})}\BibitemShut {NoStop}%
\bibitem [{\citenamefont {V\"olkl}\ \emph {et~al.}(2017)\citenamefont {V\"olkl}, \citenamefont {Rockinger}, \citenamefont {Drienovsky}, \citenamefont {Watanabe}, \citenamefont {Taniguchi}, \citenamefont {Weiss},\ and\ \citenamefont {Eroms}}]{Volkl2017}%
  \BibitemOpen
  \bibfield  {author} {\bibinfo {author} {\bibfnamefont {T.}~\bibnamefont {V\"olkl}}, \bibinfo {author} {\bibfnamefont {T.}~\bibnamefont {Rockinger}}, \bibinfo {author} {\bibfnamefont {M.}~\bibnamefont {Drienovsky}}, \bibinfo {author} {\bibfnamefont {K.}~\bibnamefont {Watanabe}}, \bibinfo {author} {\bibfnamefont {T.}~\bibnamefont {Taniguchi}}, \bibinfo {author} {\bibfnamefont {D.}~\bibnamefont {Weiss}},\ and\ \bibinfo {author} {\bibfnamefont {J.}~\bibnamefont {Eroms}},\ }\bibfield  {title} {\bibinfo {title} {Magnetotransport in heterostructures of transition metal dichalcogenides and graphene},\ }\href {https://doi.org/10.1103/PhysRevB.96.125405} {\bibfield  {journal} {\bibinfo  {journal} {Phys. Rev. B}\ }\textbf {\bibinfo {volume} {96}},\ \bibinfo {pages} {125405} (\bibinfo {year} {2017})}\BibitemShut {NoStop}%
\bibitem [{\citenamefont {Wakamura}\ \emph {et~al.}(2018)\citenamefont {Wakamura}, \citenamefont {Reale}, \citenamefont {Palczynski}, \citenamefont {Gu\'eron}, \citenamefont {Mattevi},\ and\ \citenamefont {Bouchiat}}]{Wakamura2018}%
  \BibitemOpen
  \bibfield  {author} {\bibinfo {author} {\bibfnamefont {T.}~\bibnamefont {Wakamura}}, \bibinfo {author} {\bibfnamefont {F.}~\bibnamefont {Reale}}, \bibinfo {author} {\bibfnamefont {P.}~\bibnamefont {Palczynski}}, \bibinfo {author} {\bibfnamefont {S.}~\bibnamefont {Gu\'eron}}, \bibinfo {author} {\bibfnamefont {C.}~\bibnamefont {Mattevi}},\ and\ \bibinfo {author} {\bibfnamefont {H.}~\bibnamefont {Bouchiat}},\ }\bibfield  {title} {\bibinfo {title} {Strong anisotropic spin-orbit interaction induced in graphene by monolayer {${\mathrm{WS}}_{2}$}},\ }\href {https://doi.org/10.1103/PhysRevLett.120.106802} {\bibfield  {journal} {\bibinfo  {journal} {Phys. Rev. Lett.}\ }\textbf {\bibinfo {volume} {120}},\ \bibinfo {pages} {106802} (\bibinfo {year} {2018})}\BibitemShut {NoStop}%
\bibitem [{\citenamefont {Zihlmann}\ \emph {et~al.}(2018)\citenamefont {Zihlmann}, \citenamefont {Cummings}, \citenamefont {Garcia}, \citenamefont {Kedves}, \citenamefont {Watanabe}, \citenamefont {Taniguchi}, \citenamefont {Sch\"onenberger},\ and\ \citenamefont {Makk}}]{Zihlmann2018}%
  \BibitemOpen
  \bibfield  {author} {\bibinfo {author} {\bibfnamefont {S.}~\bibnamefont {Zihlmann}}, \bibinfo {author} {\bibfnamefont {A.~W.}\ \bibnamefont {Cummings}}, \bibinfo {author} {\bibfnamefont {J.~H.}\ \bibnamefont {Garcia}}, \bibinfo {author} {\bibfnamefont {M.}~\bibnamefont {Kedves}}, \bibinfo {author} {\bibfnamefont {K.}~\bibnamefont {Watanabe}}, \bibinfo {author} {\bibfnamefont {T.}~\bibnamefont {Taniguchi}}, \bibinfo {author} {\bibfnamefont {C.}~\bibnamefont {Sch\"onenberger}},\ and\ \bibinfo {author} {\bibfnamefont {P.}~\bibnamefont {Makk}},\ }\bibfield  {title} {\bibinfo {title} {Large spin relaxation anisotropy and valley-zeeman spin-orbit coupling in {${\mathrm{WSe}}_{2}$}/graphene/$h$-{BN} heterostructures},\ }\href {https://doi.org/10.1103/PhysRevB.97.075434} {\bibfield  {journal} {\bibinfo  {journal} {Phys. Rev. B}\ }\textbf {\bibinfo {volume} {97}},\ \bibinfo {pages} {075434} (\bibinfo {year} {2018})}\BibitemShut {NoStop}%
\bibitem [{\citenamefont {Island}\ \emph {et~al.}(2019)\citenamefont {Island}, \citenamefont {Cui}, \citenamefont {Lewandowski}, \citenamefont {Khoo}, \citenamefont {Spanton}, \citenamefont {Zhou}, \citenamefont {Rhodes}, \citenamefont {Hone}, \citenamefont {Taniguchi}, \citenamefont {Watanabe}, \citenamefont {Levitov}, \citenamefont {Zaletel},\ and\ \citenamefont {Young}}]{Island2019}%
  \BibitemOpen
  \bibfield  {author} {\bibinfo {author} {\bibfnamefont {J.~O.}\ \bibnamefont {Island}}, \bibinfo {author} {\bibfnamefont {X.}~\bibnamefont {Cui}}, \bibinfo {author} {\bibfnamefont {C.}~\bibnamefont {Lewandowski}}, \bibinfo {author} {\bibfnamefont {J.~Y.}\ \bibnamefont {Khoo}}, \bibinfo {author} {\bibfnamefont {E.~M.}\ \bibnamefont {Spanton}}, \bibinfo {author} {\bibfnamefont {H.}~\bibnamefont {Zhou}}, \bibinfo {author} {\bibfnamefont {D.}~\bibnamefont {Rhodes}}, \bibinfo {author} {\bibfnamefont {J.~C.}\ \bibnamefont {Hone}}, \bibinfo {author} {\bibfnamefont {T.}~\bibnamefont {Taniguchi}}, \bibinfo {author} {\bibfnamefont {K.}~\bibnamefont {Watanabe}}, \bibinfo {author} {\bibfnamefont {L.~S.}\ \bibnamefont {Levitov}}, \bibinfo {author} {\bibfnamefont {M.~P.}\ \bibnamefont {Zaletel}},\ and\ \bibinfo {author} {\bibfnamefont {A.~F.}\ \bibnamefont {Young}},\ }\bibfield  {title} {\bibinfo {title} {Spin--orbit-driven band inversion in bilayer graphene by the van der waals proximity effect},\ }\href
  {https://doi.org/10.1038/s41586-019-1304-2} {\bibfield  {journal} {\bibinfo  {journal} {Nature}\ }\textbf {\bibinfo {volume} {571}},\ \bibinfo {pages} {85} (\bibinfo {year} {2019})}\BibitemShut {NoStop}%
\bibitem [{\citenamefont {Wang}\ \emph {et~al.}(2016)\citenamefont {Wang}, \citenamefont {Ki}, \citenamefont {Khoo}, \citenamefont {Mauro}, \citenamefont {Berger}, \citenamefont {Levitov},\ and\ \citenamefont {Morpurgo}}]{Wang2016}%
  \BibitemOpen
  \bibfield  {author} {\bibinfo {author} {\bibfnamefont {Z.}~\bibnamefont {Wang}}, \bibinfo {author} {\bibfnamefont {D.-K.}\ \bibnamefont {Ki}}, \bibinfo {author} {\bibfnamefont {J.~Y.}\ \bibnamefont {Khoo}}, \bibinfo {author} {\bibfnamefont {D.}~\bibnamefont {Mauro}}, \bibinfo {author} {\bibfnamefont {H.}~\bibnamefont {Berger}}, \bibinfo {author} {\bibfnamefont {L.~S.}\ \bibnamefont {Levitov}},\ and\ \bibinfo {author} {\bibfnamefont {A.~F.}\ \bibnamefont {Morpurgo}},\ }\bibfield  {title} {\bibinfo {title} {Origin and magnitude of `designer' spin-orbit interaction in graphene on semiconducting transition metal dichalcogenides},\ }\href {https://doi.org/10.1103/PhysRevX.6.041020} {\bibfield  {journal} {\bibinfo  {journal} {Phys. Rev. X}\ }\textbf {\bibinfo {volume} {6}},\ \bibinfo {pages} {041020} (\bibinfo {year} {2016})}\BibitemShut {NoStop}%
\bibitem [{\citenamefont {Masseroni}\ \emph {et~al.}(2024)\citenamefont {Masseroni}, \citenamefont {Gull}, \citenamefont {Panigrahi}, \citenamefont {Jacobsen}, \citenamefont {Fischer}, \citenamefont {Tong}, \citenamefont {Gerber}, \citenamefont {Niese}, \citenamefont {Taniguchi}, \citenamefont {Watanabe}, \citenamefont {Levitov}, \citenamefont {Ihn}, \citenamefont {Ensslin},\ and\ \citenamefont {Duprez}}]{Masseroni2024}%
  \BibitemOpen
  \bibfield  {author} {\bibinfo {author} {\bibfnamefont {M.}~\bibnamefont {Masseroni}}, \bibinfo {author} {\bibfnamefont {M.}~\bibnamefont {Gull}}, \bibinfo {author} {\bibfnamefont {A.}~\bibnamefont {Panigrahi}}, \bibinfo {author} {\bibfnamefont {N.}~\bibnamefont {Jacobsen}}, \bibinfo {author} {\bibfnamefont {F.}~\bibnamefont {Fischer}}, \bibinfo {author} {\bibfnamefont {C.}~\bibnamefont {Tong}}, \bibinfo {author} {\bibfnamefont {J.~D.}\ \bibnamefont {Gerber}}, \bibinfo {author} {\bibfnamefont {M.}~\bibnamefont {Niese}}, \bibinfo {author} {\bibfnamefont {T.}~\bibnamefont {Taniguchi}}, \bibinfo {author} {\bibfnamefont {K.}~\bibnamefont {Watanabe}}, \bibinfo {author} {\bibfnamefont {L.}~\bibnamefont {Levitov}}, \bibinfo {author} {\bibfnamefont {T.}~\bibnamefont {Ihn}}, \bibinfo {author} {\bibfnamefont {K.}~\bibnamefont {Ensslin}},\ and\ \bibinfo {author} {\bibfnamefont {H.}~\bibnamefont {Duprez}},\ }\bibfield  {title} {\bibinfo {title} {Spin-orbit proximity in {MoS}$_2$/bilayer graphene heterostructures},\
  }\href {https://doi.org/10.1038/s41467-024-53324-z} {\bibfield  {journal} {\bibinfo  {journal} {Nature Communications}\ }\textbf {\bibinfo {volume} {15}},\ \bibinfo {pages} {9251} (\bibinfo {year} {2024})}\BibitemShut {NoStop}%
\bibitem [{\citenamefont {Zhang}\ \emph {et~al.}(2025)\citenamefont {Zhang}, \citenamefont {Shavit}, \citenamefont {Ma}, \citenamefont {Han}, \citenamefont {Siu}, \citenamefont {Mukherjee}, \citenamefont {Watanabe}, \citenamefont {Taniguchi}, \citenamefont {Hsieh}, \citenamefont {Lewandowski}, \citenamefont {von Oppen}, \citenamefont {Oreg},\ and\ \citenamefont {Nadj-Perge}}]{zhang2024}%
  \BibitemOpen
  \bibfield  {author} {\bibinfo {author} {\bibfnamefont {Y.}~\bibnamefont {Zhang}}, \bibinfo {author} {\bibfnamefont {G.}~\bibnamefont {Shavit}}, \bibinfo {author} {\bibfnamefont {H.}~\bibnamefont {Ma}}, \bibinfo {author} {\bibfnamefont {Y.}~\bibnamefont {Han}}, \bibinfo {author} {\bibfnamefont {C.~W.}\ \bibnamefont {Siu}}, \bibinfo {author} {\bibfnamefont {A.}~\bibnamefont {Mukherjee}}, \bibinfo {author} {\bibfnamefont {K.}~\bibnamefont {Watanabe}}, \bibinfo {author} {\bibfnamefont {T.}~\bibnamefont {Taniguchi}}, \bibinfo {author} {\bibfnamefont {D.}~\bibnamefont {Hsieh}}, \bibinfo {author} {\bibfnamefont {C.}~\bibnamefont {Lewandowski}}, \bibinfo {author} {\bibfnamefont {F.}~\bibnamefont {von Oppen}}, \bibinfo {author} {\bibfnamefont {Y.}~\bibnamefont {Oreg}},\ and\ \bibinfo {author} {\bibfnamefont {S.}~\bibnamefont {Nadj-Perge}},\ }\bibfield  {title} {\bibinfo {title} {Twist-programmable superconductivity in spin--orbit-coupled bilayer graphene},\ }\href {https://doi.org/10.1038/s41586-025-08959-3}
  {\bibfield  {journal} {\bibinfo  {journal} {Nature}\ }\textbf {\bibinfo {volume} {641}},\ \bibinfo {pages} {625} (\bibinfo {year} {2025})}\BibitemShut {NoStop}%
\bibitem [{\citenamefont {Seiler}\ \emph {et~al.}(2025)\citenamefont {Seiler}, \citenamefont {Zhumagulov}, \citenamefont {Zollner}, \citenamefont {Yoon}, \citenamefont {Urbaniak}, \citenamefont {Geisenhof}, \citenamefont {Watanabe}, \citenamefont {Taniguchi}, \citenamefont {Fabian}, \citenamefont {Zhang},\ and\ \citenamefont {Weitz}}]{Seiler2024}%
  \BibitemOpen
  \bibfield  {author} {\bibinfo {author} {\bibfnamefont {A.~M.}\ \bibnamefont {Seiler}}, \bibinfo {author} {\bibfnamefont {Y.}~\bibnamefont {Zhumagulov}}, \bibinfo {author} {\bibfnamefont {K.}~\bibnamefont {Zollner}}, \bibinfo {author} {\bibfnamefont {C.}~\bibnamefont {Yoon}}, \bibinfo {author} {\bibfnamefont {D.}~\bibnamefont {Urbaniak}}, \bibinfo {author} {\bibfnamefont {F.~R.}\ \bibnamefont {Geisenhof}}, \bibinfo {author} {\bibfnamefont {K.}~\bibnamefont {Watanabe}}, \bibinfo {author} {\bibfnamefont {T.}~\bibnamefont {Taniguchi}}, \bibinfo {author} {\bibfnamefont {J.}~\bibnamefont {Fabian}}, \bibinfo {author} {\bibfnamefont {F.}~\bibnamefont {Zhang}},\ and\ \bibinfo {author} {\bibfnamefont {R.~T.}\ \bibnamefont {Weitz}},\ }\bibfield  {title} {\bibinfo {title} {Layer-selective spin-orbit coupling and strong correlation in bilayer graphene},\ }\href {https://doi.org/10.1088/2053-1583/add74a} {\bibfield  {journal} {\bibinfo  {journal} {2D Materials}\ }\textbf {\bibinfo {volume} {12}},\ \bibinfo {pages}
  {035009} (\bibinfo {year} {2025})}\BibitemShut {NoStop}%
\bibitem [{\citenamefont {Tiwari}\ \emph {et~al.}(2022)\citenamefont {Tiwari}, \citenamefont {Jat}, \citenamefont {Udupa}, \citenamefont {Narang}, \citenamefont {Watanabe}, \citenamefont {Taniguchi}, \citenamefont {Sen},\ and\ \citenamefont {Bid}}]{Tiwari2022}%
  \BibitemOpen
  \bibfield  {author} {\bibinfo {author} {\bibfnamefont {P.}~\bibnamefont {Tiwari}}, \bibinfo {author} {\bibfnamefont {M.~K.}\ \bibnamefont {Jat}}, \bibinfo {author} {\bibfnamefont {A.}~\bibnamefont {Udupa}}, \bibinfo {author} {\bibfnamefont {D.~S.}\ \bibnamefont {Narang}}, \bibinfo {author} {\bibfnamefont {K.}~\bibnamefont {Watanabe}}, \bibinfo {author} {\bibfnamefont {T.}~\bibnamefont {Taniguchi}}, \bibinfo {author} {\bibfnamefont {D.}~\bibnamefont {Sen}},\ and\ \bibinfo {author} {\bibfnamefont {A.}~\bibnamefont {Bid}},\ }\bibfield  {title} {\bibinfo {title} {Experimental observation of spin-split energy dispersion in high-mobility single-layer graphene/{${\mathrm{WSe}}_{2}$} heterostructures},\ }\href {https://doi.org/10.1038/s41699-022-00348-y} {\bibfield  {journal} {\bibinfo  {journal} {npj 2D Materials and Applications}\ }\textbf {\bibinfo {volume} {6}},\ \bibinfo {pages} {68} (\bibinfo {year} {2022})}\BibitemShut {NoStop}%
\bibitem [{\citenamefont {Holleis}\ \emph {et~al.}(2025)\citenamefont {Holleis}, \citenamefont {Patterson}, \citenamefont {Zhang}, \citenamefont {Vituri}, \citenamefont {Yoo}, \citenamefont {Zhou}, \citenamefont {Taniguchi}, \citenamefont {Watanabe}, \citenamefont {Berg}, \citenamefont {Nadj-Perge},\ and\ \citenamefont {Young}}]{Holleis2024}%
  \BibitemOpen
  \bibfield  {author} {\bibinfo {author} {\bibfnamefont {L.}~\bibnamefont {Holleis}}, \bibinfo {author} {\bibfnamefont {C.~L.}\ \bibnamefont {Patterson}}, \bibinfo {author} {\bibfnamefont {Y.}~\bibnamefont {Zhang}}, \bibinfo {author} {\bibfnamefont {Y.}~\bibnamefont {Vituri}}, \bibinfo {author} {\bibfnamefont {H.~M.}\ \bibnamefont {Yoo}}, \bibinfo {author} {\bibfnamefont {H.}~\bibnamefont {Zhou}}, \bibinfo {author} {\bibfnamefont {T.}~\bibnamefont {Taniguchi}}, \bibinfo {author} {\bibfnamefont {K.}~\bibnamefont {Watanabe}}, \bibinfo {author} {\bibfnamefont {E.}~\bibnamefont {Berg}}, \bibinfo {author} {\bibfnamefont {S.}~\bibnamefont {Nadj-Perge}},\ and\ \bibinfo {author} {\bibfnamefont {A.~F.}\ \bibnamefont {Young}},\ }\bibfield  {title} {\bibinfo {title} {Nematicity and orbital depairing in superconducting bernal bilayer graphene},\ }\href {https://doi.org/10.1038/s41567-024-02776-7} {\bibfield  {journal} {\bibinfo  {journal} {Nature Physics}\ }\textbf {\bibinfo {volume} {21}},\ \bibinfo {pages} {444}
  (\bibinfo {year} {2025})}\BibitemShut {NoStop}%
\bibitem [{\citenamefont {F{\"u}l{\"o}p}\ \emph {et~al.}(2021)\citenamefont {F{\"u}l{\"o}p}, \citenamefont {M{\'a}rffy}, \citenamefont {Zihlmann}, \citenamefont {Gmitra}, \citenamefont {T{\'o}v{\'a}ri}, \citenamefont {Szentp{\'e}teri}, \citenamefont {Kedves}, \citenamefont {Watanabe}, \citenamefont {Taniguchi}, \citenamefont {Fabian}, \citenamefont {Sch{\"o}nenberger}, \citenamefont {Makk},\ and\ \citenamefont {Csonka}}]{Fulop2021}%
  \BibitemOpen
  \bibfield  {author} {\bibinfo {author} {\bibfnamefont {B.}~\bibnamefont {F{\"u}l{\"o}p}}, \bibinfo {author} {\bibfnamefont {A.}~\bibnamefont {M{\'a}rffy}}, \bibinfo {author} {\bibfnamefont {S.}~\bibnamefont {Zihlmann}}, \bibinfo {author} {\bibfnamefont {M.}~\bibnamefont {Gmitra}}, \bibinfo {author} {\bibfnamefont {E.}~\bibnamefont {T{\'o}v{\'a}ri}}, \bibinfo {author} {\bibfnamefont {B.}~\bibnamefont {Szentp{\'e}teri}}, \bibinfo {author} {\bibfnamefont {M.}~\bibnamefont {Kedves}}, \bibinfo {author} {\bibfnamefont {K.}~\bibnamefont {Watanabe}}, \bibinfo {author} {\bibfnamefont {T.}~\bibnamefont {Taniguchi}}, \bibinfo {author} {\bibfnamefont {J.}~\bibnamefont {Fabian}}, \bibinfo {author} {\bibfnamefont {C.}~\bibnamefont {Sch{\"o}nenberger}}, \bibinfo {author} {\bibfnamefont {P.}~\bibnamefont {Makk}},\ and\ \bibinfo {author} {\bibfnamefont {S.}~\bibnamefont {Csonka}},\ }\bibfield  {title} {\bibinfo {title} {Boosting proximity spin--orbit coupling in graphene/{${\mathrm{WSe}}_{2}$} heterostructures via
  hydrostatic pressure},\ }\href {https://doi.org/10.1038/s41699-021-00262-9} {\bibfield  {journal} {\bibinfo  {journal} {npj 2D Materials and Applications}\ }\textbf {\bibinfo {volume} {5}},\ \bibinfo {pages} {82} (\bibinfo {year} {2021})}\BibitemShut {NoStop}%
\bibitem [{\citenamefont {Rao}\ \emph {et~al.}(2023)\citenamefont {Rao}, \citenamefont {Kang}, \citenamefont {Xue}, \citenamefont {Ye}, \citenamefont {Feng}, \citenamefont {Watanabe}, \citenamefont {Taniguchi}, \citenamefont {Wang}, \citenamefont {Liu},\ and\ \citenamefont {Ki}}]{Rao2023}%
  \BibitemOpen
  \bibfield  {author} {\bibinfo {author} {\bibfnamefont {Q.}~\bibnamefont {Rao}}, \bibinfo {author} {\bibfnamefont {W.-H.}\ \bibnamefont {Kang}}, \bibinfo {author} {\bibfnamefont {H.}~\bibnamefont {Xue}}, \bibinfo {author} {\bibfnamefont {Z.}~\bibnamefont {Ye}}, \bibinfo {author} {\bibfnamefont {X.}~\bibnamefont {Feng}}, \bibinfo {author} {\bibfnamefont {K.}~\bibnamefont {Watanabe}}, \bibinfo {author} {\bibfnamefont {T.}~\bibnamefont {Taniguchi}}, \bibinfo {author} {\bibfnamefont {N.}~\bibnamefont {Wang}}, \bibinfo {author} {\bibfnamefont {M.-H.}\ \bibnamefont {Liu}},\ and\ \bibinfo {author} {\bibfnamefont {D.-K.}\ \bibnamefont {Ki}},\ }\bibfield  {title} {\bibinfo {title} {Ballistic transport spectroscopy of spin-orbit-coupled bands in monolayer graphene on {WSe}$_2$},\ }\href {https://doi.org/10.1038/s41467-023-41826-1} {\bibfield  {journal} {\bibinfo  {journal} {Nature Communications}\ }\textbf {\bibinfo {volume} {14}},\ \bibinfo {pages} {6124} (\bibinfo {year} {2023})}\BibitemShut {NoStop}%
\bibitem [{\citenamefont {Eich}\ \emph {et~al.}(2018)\citenamefont {Eich}, \citenamefont {Herman}, \citenamefont {Pisoni}, \citenamefont {Overweg}, \citenamefont {Kurzmann}, \citenamefont {Lee}, \citenamefont {Rickhaus}, \citenamefont {Watanabe}, \citenamefont {Taniguchi}, \citenamefont {Sigrist}, \citenamefont {Ihn},\ and\ \citenamefont {Ensslin}}]{Eich2018}%
  \BibitemOpen
  \bibfield  {author} {\bibinfo {author} {\bibfnamefont {M.}~\bibnamefont {Eich}}, \bibinfo {author} {\bibfnamefont {F.~c.~v.}\ \bibnamefont {Herman}}, \bibinfo {author} {\bibfnamefont {R.}~\bibnamefont {Pisoni}}, \bibinfo {author} {\bibfnamefont {H.}~\bibnamefont {Overweg}}, \bibinfo {author} {\bibfnamefont {A.}~\bibnamefont {Kurzmann}}, \bibinfo {author} {\bibfnamefont {Y.}~\bibnamefont {Lee}}, \bibinfo {author} {\bibfnamefont {P.}~\bibnamefont {Rickhaus}}, \bibinfo {author} {\bibfnamefont {K.}~\bibnamefont {Watanabe}}, \bibinfo {author} {\bibfnamefont {T.}~\bibnamefont {Taniguchi}}, \bibinfo {author} {\bibfnamefont {M.}~\bibnamefont {Sigrist}}, \bibinfo {author} {\bibfnamefont {T.}~\bibnamefont {Ihn}},\ and\ \bibinfo {author} {\bibfnamefont {K.}~\bibnamefont {Ensslin}},\ }\bibfield  {title} {\bibinfo {title} {Spin and valley states in gate-defined bilayer graphene quantum dots},\ }\href {https://doi.org/10.1103/PhysRevX.8.031023} {\bibfield  {journal} {\bibinfo  {journal} {Phys. Rev. X}\ }\textbf {\bibinfo
  {volume} {8}},\ \bibinfo {pages} {031023} (\bibinfo {year} {2018})}\BibitemShut {NoStop}%
\bibitem [{\citenamefont {Banszerus}\ \emph {et~al.}(2018)\citenamefont {Banszerus}, \citenamefont {Frohn}, \citenamefont {Epping}, \citenamefont {Neumaier}, \citenamefont {Watanabe}, \citenamefont {Taniguchi},\ and\ \citenamefont {Stampfer}}]{Banszerus2018}%
  \BibitemOpen
  \bibfield  {author} {\bibinfo {author} {\bibfnamefont {L.}~\bibnamefont {Banszerus}}, \bibinfo {author} {\bibfnamefont {B.}~\bibnamefont {Frohn}}, \bibinfo {author} {\bibfnamefont {A.}~\bibnamefont {Epping}}, \bibinfo {author} {\bibfnamefont {D.}~\bibnamefont {Neumaier}}, \bibinfo {author} {\bibfnamefont {K.}~\bibnamefont {Watanabe}}, \bibinfo {author} {\bibfnamefont {T.}~\bibnamefont {Taniguchi}},\ and\ \bibinfo {author} {\bibfnamefont {C.}~\bibnamefont {Stampfer}},\ }\bibfield  {title} {\bibinfo {title} {Gate-defined electron--hole double dots in bilayer graphene},\ }\href {https://doi.org/10.1021/acs.nanolett.8b01303} {\bibfield  {journal} {\bibinfo  {journal} {Nano Letters}\ }\textbf {\bibinfo {volume} {18}},\ \bibinfo {pages} {4785} (\bibinfo {year} {2018})}\BibitemShut {NoStop}%
\bibitem [{\citenamefont {Kraft}\ \emph {et~al.}(2018)\citenamefont {Kraft}, \citenamefont {Krainov}, \citenamefont {Gall}, \citenamefont {Dmitriev}, \citenamefont {Krupke}, \citenamefont {Gornyi},\ and\ \citenamefont {Danneau}}]{Kraft2018}%
  \BibitemOpen
  \bibfield  {author} {\bibinfo {author} {\bibfnamefont {R.}~\bibnamefont {Kraft}}, \bibinfo {author} {\bibfnamefont {I.~V.}\ \bibnamefont {Krainov}}, \bibinfo {author} {\bibfnamefont {V.}~\bibnamefont {Gall}}, \bibinfo {author} {\bibfnamefont {A.~P.}\ \bibnamefont {Dmitriev}}, \bibinfo {author} {\bibfnamefont {R.}~\bibnamefont {Krupke}}, \bibinfo {author} {\bibfnamefont {I.~V.}\ \bibnamefont {Gornyi}},\ and\ \bibinfo {author} {\bibfnamefont {R.}~\bibnamefont {Danneau}},\ }\bibfield  {title} {\bibinfo {title} {Valley subband splitting in bilayer graphene quantum point contacts},\ }\href {https://doi.org/10.1103/PhysRevLett.121.257703} {\bibfield  {journal} {\bibinfo  {journal} {Phys. Rev. Lett.}\ }\textbf {\bibinfo {volume} {121}},\ \bibinfo {pages} {257703} (\bibinfo {year} {2018})}\BibitemShut {NoStop}%
\bibitem [{\citenamefont {Overweg}\ \emph {et~al.}(2018{\natexlab{a}})\citenamefont {Overweg}, \citenamefont {Eggimann}, \citenamefont {Chen}, \citenamefont {Slizovskiy}, \citenamefont {Eich}, \citenamefont {Pisoni}, \citenamefont {Lee}, \citenamefont {Rickhaus}, \citenamefont {Watanabe}, \citenamefont {Taniguchi}, \citenamefont {Fal’ko}, \citenamefont {Ihn},\ and\ \citenamefont {Ensslin}}]{Overweg2018Nano}%
  \BibitemOpen
  \bibfield  {author} {\bibinfo {author} {\bibfnamefont {H.}~\bibnamefont {Overweg}}, \bibinfo {author} {\bibfnamefont {H.}~\bibnamefont {Eggimann}}, \bibinfo {author} {\bibfnamefont {X.}~\bibnamefont {Chen}}, \bibinfo {author} {\bibfnamefont {S.}~\bibnamefont {Slizovskiy}}, \bibinfo {author} {\bibfnamefont {M.}~\bibnamefont {Eich}}, \bibinfo {author} {\bibfnamefont {R.}~\bibnamefont {Pisoni}}, \bibinfo {author} {\bibfnamefont {Y.}~\bibnamefont {Lee}}, \bibinfo {author} {\bibfnamefont {P.}~\bibnamefont {Rickhaus}}, \bibinfo {author} {\bibfnamefont {K.}~\bibnamefont {Watanabe}}, \bibinfo {author} {\bibfnamefont {T.}~\bibnamefont {Taniguchi}}, \bibinfo {author} {\bibfnamefont {V.}~\bibnamefont {Fal’ko}}, \bibinfo {author} {\bibfnamefont {T.}~\bibnamefont {Ihn}},\ and\ \bibinfo {author} {\bibfnamefont {K.}~\bibnamefont {Ensslin}},\ }\bibfield  {title} {\bibinfo {title} {Electrostatically induced quantum point contacts in bilayer graphene},\ }\href {https://doi.org/10.1021/acs.nanolett.7b04666} {\bibfield
  {journal} {\bibinfo  {journal} {Nano Letters}\ }\textbf {\bibinfo {volume} {18}},\ \bibinfo {pages} {553} (\bibinfo {year} {2018}{\natexlab{a}})}\BibitemShut {NoStop}%
\bibitem [{\citenamefont {Lee}\ \emph {et~al.}(2020)\citenamefont {Lee}, \citenamefont {Knothe}, \citenamefont {Overweg}, \citenamefont {Eich}, \citenamefont {Gold}, \citenamefont {Kurzmann}, \citenamefont {Klasovika}, \citenamefont {Taniguchi}, \citenamefont {Wantanabe}, \citenamefont {Fal'ko}, \citenamefont {Ihn}, \citenamefont {Ensslin},\ and\ \citenamefont {Rickhaus}}]{Lee2020}%
  \BibitemOpen
  \bibfield  {author} {\bibinfo {author} {\bibfnamefont {Y.}~\bibnamefont {Lee}}, \bibinfo {author} {\bibfnamefont {A.}~\bibnamefont {Knothe}}, \bibinfo {author} {\bibfnamefont {H.}~\bibnamefont {Overweg}}, \bibinfo {author} {\bibfnamefont {M.}~\bibnamefont {Eich}}, \bibinfo {author} {\bibfnamefont {C.}~\bibnamefont {Gold}}, \bibinfo {author} {\bibfnamefont {A.}~\bibnamefont {Kurzmann}}, \bibinfo {author} {\bibfnamefont {V.}~\bibnamefont {Klasovika}}, \bibinfo {author} {\bibfnamefont {T.}~\bibnamefont {Taniguchi}}, \bibinfo {author} {\bibfnamefont {K.}~\bibnamefont {Wantanabe}}, \bibinfo {author} {\bibfnamefont {V.}~\bibnamefont {Fal'ko}}, \bibinfo {author} {\bibfnamefont {T.}~\bibnamefont {Ihn}}, \bibinfo {author} {\bibfnamefont {K.}~\bibnamefont {Ensslin}},\ and\ \bibinfo {author} {\bibfnamefont {P.}~\bibnamefont {Rickhaus}},\ }\bibfield  {title} {\bibinfo {title} {Tunable valley splitting due to topological orbital magnetic moment in bilayer graphene quantum point contacts},\ }\href
  {https://doi.org/10.1103/PhysRevLett.124.126802} {\bibfield  {journal} {\bibinfo  {journal} {Phys. Rev. Lett.}\ }\textbf {\bibinfo {volume} {124}},\ \bibinfo {pages} {126802} (\bibinfo {year} {2020})}\BibitemShut {NoStop}%
\bibitem [{\citenamefont {Overweg}\ \emph {et~al.}(2018{\natexlab{b}})\citenamefont {Overweg}, \citenamefont {Knothe}, \citenamefont {Fabian}, \citenamefont {Linhart}, \citenamefont {Rickhaus}, \citenamefont {Wernli}, \citenamefont {Watanabe}, \citenamefont {Taniguchi}, \citenamefont {S\'anchez}, \citenamefont {Burgd\"orfer}, \citenamefont {Libisch}, \citenamefont {Fal'ko}, \citenamefont {Ensslin},\ and\ \citenamefont {Ihn}}]{Overweg2018PRL}%
  \BibitemOpen
  \bibfield  {author} {\bibinfo {author} {\bibfnamefont {H.}~\bibnamefont {Overweg}}, \bibinfo {author} {\bibfnamefont {A.}~\bibnamefont {Knothe}}, \bibinfo {author} {\bibfnamefont {T.}~\bibnamefont {Fabian}}, \bibinfo {author} {\bibfnamefont {L.}~\bibnamefont {Linhart}}, \bibinfo {author} {\bibfnamefont {P.}~\bibnamefont {Rickhaus}}, \bibinfo {author} {\bibfnamefont {L.}~\bibnamefont {Wernli}}, \bibinfo {author} {\bibfnamefont {K.}~\bibnamefont {Watanabe}}, \bibinfo {author} {\bibfnamefont {T.}~\bibnamefont {Taniguchi}}, \bibinfo {author} {\bibfnamefont {D.}~\bibnamefont {S\'anchez}}, \bibinfo {author} {\bibfnamefont {J.}~\bibnamefont {Burgd\"orfer}}, \bibinfo {author} {\bibfnamefont {F.}~\bibnamefont {Libisch}}, \bibinfo {author} {\bibfnamefont {V.~I.}\ \bibnamefont {Fal'ko}}, \bibinfo {author} {\bibfnamefont {K.}~\bibnamefont {Ensslin}},\ and\ \bibinfo {author} {\bibfnamefont {T.}~\bibnamefont {Ihn}},\ }\bibfield  {title} {\bibinfo {title} {Topologically nontrivial valley states in bilayer graphene quantum
  point contacts},\ }\href {https://doi.org/10.1103/PhysRevLett.121.257702} {\bibfield  {journal} {\bibinfo  {journal} {Phys. Rev. Lett.}\ }\textbf {\bibinfo {volume} {121}},\ \bibinfo {pages} {257702} (\bibinfo {year} {2018}{\natexlab{b}})}\BibitemShut {NoStop}%
\bibitem [{\citenamefont {Fogler}\ and\ \citenamefont {McCann}(2010)}]{Fogler2010}%
  \BibitemOpen
  \bibfield  {author} {\bibinfo {author} {\bibfnamefont {M.~M.}\ \bibnamefont {Fogler}}\ and\ \bibinfo {author} {\bibfnamefont {E.}~\bibnamefont {McCann}},\ }\bibfield  {title} {\bibinfo {title} {Comment on ``screening in gated bilayer graphene''},\ }\href {https://doi.org/10.1103/PhysRevB.82.197401} {\bibfield  {journal} {\bibinfo  {journal} {Phys. Rev. B}\ }\textbf {\bibinfo {volume} {82}},\ \bibinfo {pages} {197401} (\bibinfo {year} {2010})}\BibitemShut {NoStop}%
\bibitem [{\citenamefont {Naimer}\ \emph {et~al.}(2021)\citenamefont {Naimer}, \citenamefont {Zollner}, \citenamefont {Gmitra},\ and\ \citenamefont {Fabian}}]{Naimer2021}%
  \BibitemOpen
  \bibfield  {author} {\bibinfo {author} {\bibfnamefont {T.}~\bibnamefont {Naimer}}, \bibinfo {author} {\bibfnamefont {K.}~\bibnamefont {Zollner}}, \bibinfo {author} {\bibfnamefont {M.}~\bibnamefont {Gmitra}},\ and\ \bibinfo {author} {\bibfnamefont {J.}~\bibnamefont {Fabian}},\ }\bibfield  {title} {\bibinfo {title} {Twist-angle dependent proximity induced spin-orbit coupling in graphene/transition metal dichalcogenide heterostructures},\ }\href {https://doi.org/10.1103/PhysRevB.104.195156} {\bibfield  {journal} {\bibinfo  {journal} {Phys. Rev. B}\ }\textbf {\bibinfo {volume} {104}},\ \bibinfo {pages} {195156} (\bibinfo {year} {2021})}\BibitemShut {NoStop}%
\bibitem [{\citenamefont {Zollner}\ \emph {et~al.}(2023)\citenamefont {Zollner}, \citenamefont {Jo\~ao}, \citenamefont {Nikoli\ifmmode~\acute{c}\else \'{c}\fi{}},\ and\ \citenamefont {Fabian}}]{Zollner2023}%
  \BibitemOpen
  \bibfield  {author} {\bibinfo {author} {\bibfnamefont {K.}~\bibnamefont {Zollner}}, \bibinfo {author} {\bibfnamefont {S.~a.~M.}\ \bibnamefont {Jo\~ao}}, \bibinfo {author} {\bibfnamefont {B.~K.}\ \bibnamefont {Nikoli\ifmmode~\acute{c}\else \'{c}\fi{}}},\ and\ \bibinfo {author} {\bibfnamefont {J.}~\bibnamefont {Fabian}},\ }\bibfield  {title} {\bibinfo {title} {Twist- and gate-tunable proximity spin-orbit coupling, spin relaxation anisotropy, and charge-to-spin conversion in heterostructures of graphene and transition metal dichalcogenides},\ }\href {https://doi.org/10.1103/PhysRevB.108.235166} {\bibfield  {journal} {\bibinfo  {journal} {Phys. Rev. B}\ }\textbf {\bibinfo {volume} {108}},\ \bibinfo {pages} {235166} (\bibinfo {year} {2023})}\BibitemShut {NoStop}%
\bibitem [{\citenamefont {Korm\'anyos}\ \emph {et~al.}(2014)\citenamefont {Korm\'anyos}, \citenamefont {Z\'olyomi}, \citenamefont {Drummond},\ and\ \citenamefont {Burkard}}]{Kormanyos2014}%
  \BibitemOpen
  \bibfield  {author} {\bibinfo {author} {\bibfnamefont {A.}~\bibnamefont {Korm\'anyos}}, \bibinfo {author} {\bibfnamefont {V.}~\bibnamefont {Z\'olyomi}}, \bibinfo {author} {\bibfnamefont {N.~D.}\ \bibnamefont {Drummond}},\ and\ \bibinfo {author} {\bibfnamefont {G.}~\bibnamefont {Burkard}},\ }\bibfield  {title} {\bibinfo {title} {Spin-orbit coupling, quantum dots, and qubits in monolayer transition metal dichalcogenides},\ }\href {https://doi.org/10.1103/PhysRevX.4.011034} {\bibfield  {journal} {\bibinfo  {journal} {Phys. Rev. X}\ }\textbf {\bibinfo {volume} {4}},\ \bibinfo {pages} {011034} (\bibinfo {year} {2014})}\BibitemShut {NoStop}%
\bibitem [{\citenamefont {Knothe}\ and\ \citenamefont {Fal'ko}(2018)}]{Knothe2018}%
  \BibitemOpen
  \bibfield  {author} {\bibinfo {author} {\bibfnamefont {A.}~\bibnamefont {Knothe}}\ and\ \bibinfo {author} {\bibfnamefont {V.}~\bibnamefont {Fal'ko}},\ }\bibfield  {title} {\bibinfo {title} {Influence of minivalleys and berry curvature on electrostatically induced quantum wires in gapped bilayer graphene},\ }\href {https://doi.org/10.1103/PhysRevB.98.155435} {\bibfield  {journal} {\bibinfo  {journal} {Phys. Rev. B}\ }\textbf {\bibinfo {volume} {98}},\ \bibinfo {pages} {155435} (\bibinfo {year} {2018})}\BibitemShut {NoStop}%
\bibitem [{\citenamefont {Fang}\ and\ \citenamefont {Stiles}(1968)}]{Fang1968}%
  \BibitemOpen
  \bibfield  {author} {\bibinfo {author} {\bibfnamefont {F.~F.}\ \bibnamefont {Fang}}\ and\ \bibinfo {author} {\bibfnamefont {P.~J.}\ \bibnamefont {Stiles}},\ }\bibfield  {title} {\bibinfo {title} {Effects of a tilted magnetic field on a two-dimensional electron gas},\ }\href {https://doi.org/10.1103/PhysRev.174.823} {\bibfield  {journal} {\bibinfo  {journal} {Phys. Rev.}\ }\textbf {\bibinfo {volume} {174}},\ \bibinfo {pages} {823} (\bibinfo {year} {1968})}\BibitemShut {NoStop}%
\bibitem [{\citenamefont {Pallecchi}\ \emph {et~al.}(2002)\citenamefont {Pallecchi}, \citenamefont {Heyn}, \citenamefont {Lohse}, \citenamefont {Kramer},\ and\ \citenamefont {Hansen}}]{Pallecchi2002}%
  \BibitemOpen
  \bibfield  {author} {\bibinfo {author} {\bibfnamefont {I.}~\bibnamefont {Pallecchi}}, \bibinfo {author} {\bibfnamefont {C.}~\bibnamefont {Heyn}}, \bibinfo {author} {\bibfnamefont {J.}~\bibnamefont {Lohse}}, \bibinfo {author} {\bibfnamefont {B.}~\bibnamefont {Kramer}},\ and\ \bibinfo {author} {\bibfnamefont {W.}~\bibnamefont {Hansen}},\ }\bibfield  {title} {\bibinfo {title} {Magnetocapacitance of quantum wires: Effect of confining potential on one-dimensional subbands and suppression of exchange enhanced g factor},\ }\href {https://doi.org/10.1103/PhysRevB.65.125303} {\bibfield  {journal} {\bibinfo  {journal} {Phys. Rev. B}\ }\textbf {\bibinfo {volume} {65}},\ \bibinfo {pages} {125303} (\bibinfo {year} {2002})}\BibitemShut {NoStop}%
\bibitem [{\citenamefont {Gall}\ \emph {et~al.}(2022)\citenamefont {Gall}, \citenamefont {Kraft}, \citenamefont {Gornyi},\ and\ \citenamefont {Danneau}}]{Gall2022}%
  \BibitemOpen
  \bibfield  {author} {\bibinfo {author} {\bibfnamefont {V.}~\bibnamefont {Gall}}, \bibinfo {author} {\bibfnamefont {R.}~\bibnamefont {Kraft}}, \bibinfo {author} {\bibfnamefont {I.~V.}\ \bibnamefont {Gornyi}},\ and\ \bibinfo {author} {\bibfnamefont {R.}~\bibnamefont {Danneau}},\ }\bibfield  {title} {\bibinfo {title} {Spin and valley degrees of freedom in a bilayer graphene quantum point contact: Zeeman splitting and interaction effects},\ }\href {https://doi.org/10.1103/PhysRevResearch.4.023142} {\bibfield  {journal} {\bibinfo  {journal} {Phys. Rev. Res.}\ }\textbf {\bibinfo {volume} {4}},\ \bibinfo {pages} {023142} (\bibinfo {year} {2022})}\BibitemShut {NoStop}%
\bibitem [{\citenamefont {Janak}(1969)}]{Janak1969}%
  \BibitemOpen
  \bibfield  {author} {\bibinfo {author} {\bibfnamefont {J.~F.}\ \bibnamefont {Janak}},\ }\bibfield  {title} {\bibinfo {title} {$g$ factor of the two-dimensional interacting electron gas},\ }\href {https://doi.org/10.1103/PhysRev.178.1416} {\bibfield  {journal} {\bibinfo  {journal} {Phys. Rev.}\ }\textbf {\bibinfo {volume} {178}},\ \bibinfo {pages} {1416} (\bibinfo {year} {1969})}\BibitemShut {NoStop}%
\bibitem [{\citenamefont {Ando}\ and\ \citenamefont {Uemura}(1974)}]{Ando1974}%
  \BibitemOpen
  \bibfield  {author} {\bibinfo {author} {\bibfnamefont {T.}~\bibnamefont {Ando}}\ and\ \bibinfo {author} {\bibfnamefont {Y.}~\bibnamefont {Uemura}},\ }\bibfield  {title} {\bibinfo {title} {Theory of oscillatory g factor in an mos inversion layer under strong magnetic fields},\ }\href {https://doi.org/10.1143/JPSJ.37.1044} {\bibfield  {journal} {\bibinfo  {journal} {Journal of the Physical Society of Japan}\ }\textbf {\bibinfo {volume} {37}},\ \bibinfo {pages} {1044} (\bibinfo {year} {1974})}\BibitemShut {NoStop}%
\bibitem [{\citenamefont {Masseroni}(2024)}]{MasseroniphD2024}%
  \BibitemOpen
  \bibfield  {author} {\bibinfo {author} {\bibfnamefont {M.}~\bibnamefont {Masseroni}},\ }\bibfield  {title} {\bibinfo {title} {Electronic transport experiments in 2d materials with spin-orbit coupling},\ }\href@noop {} {\bibfield  {journal} {\bibinfo  {journal} {Doctoral Thesis ETH}\ } (\bibinfo {year} {2024})}\BibitemShut {NoStop}%
\bibitem [{\citenamefont {Hou}\ \emph {et~al.}(2022)\citenamefont {Hou}, \citenamefont {Wang}, \citenamefont {Ma}, \citenamefont {Feng}, \citenamefont {Chen},\ and\ \citenamefont {Filleter}}]{Hou2022}%
  \BibitemOpen
  \bibfield  {author} {\bibinfo {author} {\bibfnamefont {Y.}~\bibnamefont {Hou}}, \bibinfo {author} {\bibfnamefont {G.}~\bibnamefont {Wang}}, \bibinfo {author} {\bibfnamefont {C.}~\bibnamefont {Ma}}, \bibinfo {author} {\bibfnamefont {Z.}~\bibnamefont {Feng}}, \bibinfo {author} {\bibfnamefont {Y.}~\bibnamefont {Chen}},\ and\ \bibinfo {author} {\bibfnamefont {T.}~\bibnamefont {Filleter}},\ }\bibfield  {title} {\bibinfo {title} {Quantification of the dielectric constant of {MoS}$_2$ and {${\mathrm{WSe}}_{2}$} nanosheets by electrostatic force microscopy},\ }\href {https://doi.org/https://doi.org/10.1016/j.matchar.2022.112313} {\bibfield  {journal} {\bibinfo  {journal} {Materials Characterization}\ }\textbf {\bibinfo {volume} {193}},\ \bibinfo {pages} {112313} (\bibinfo {year} {2022})}\BibitemShut {NoStop}%
\bibitem [{\citenamefont {McCann}\ and\ \citenamefont {Koshino}(2013)}]{mccann2013}%
  \BibitemOpen
  \bibfield  {author} {\bibinfo {author} {\bibfnamefont {E.}~\bibnamefont {McCann}}\ and\ \bibinfo {author} {\bibfnamefont {M.}~\bibnamefont {Koshino}},\ }\bibfield  {title} {\bibinfo {title} {The electronic properties of bilayer graphene},\ }\href {https://doi.org/10.1088/0034-4885/76/5/056503} {\bibfield  {journal} {\bibinfo  {journal} {Reports on Progress in Physics}\ }\textbf {\bibinfo {volume} {76}},\ \bibinfo {pages} {056503} (\bibinfo {year} {2013})}\BibitemShut {NoStop}%
\bibitem [{\citenamefont {Zollner}\ and\ \citenamefont {Fabian}(2021)}]{Zollner2021}%
  \BibitemOpen
  \bibfield  {author} {\bibinfo {author} {\bibfnamefont {K.}~\bibnamefont {Zollner}}\ and\ \bibinfo {author} {\bibfnamefont {J.}~\bibnamefont {Fabian}},\ }\bibfield  {title} {\bibinfo {title} {Bilayer graphene encapsulated within monolayers of {${\mathrm{WS}}_{2}$} or {${\mathrm{Cr}}_{2}{\mathrm{Ge}}_{2}{\mathrm{Te}}_{6}$}: Tunable proximity spin-orbit or exchange coupling},\ }\href {https://doi.org/10.1103/PhysRevB.104.075126} {\bibfield  {journal} {\bibinfo  {journal} {Phys. Rev. B}\ }\textbf {\bibinfo {volume} {104}},\ \bibinfo {pages} {075126} (\bibinfo {year} {2021})}\BibitemShut {NoStop}%
\end{thebibliography}
\end{document}